%% file: EXOFASTv2.tex
\documentclass[twocolumn]{aastex62}

\usepackage{amsmath}
\usepackage{graphicx}
\usepackage{natbib}
\usepackage[scaled]{helvet}
\usepackage{epsfig}
\usepackage{url}
\usepackage{textcomp}
\usepackage{mathpazo}
\usepackage{xspace}
\usepackage{hyperref}
\usepackage{apjfonts}
\usepackage{mathrsfs}
\usepackage{verbatim}
\usepackage{color}
\usepackage[normalem]{ulem}

\bibpunct{(}{)}{;}{a}{}{,}
\newcommand{\bjdtdb}{\ensuremath{\rm {BJD_{TDB}}}}
\newcommand{\feh}{\ensuremath{\left[{\rm Fe}/{\rm H}\right]}}

\newcommand{\teff}{\ensuremath{T_{\rm eff}}}
\newcommand{\logk}{\ensuremath{\log K}}
\newcommand{\logg}{\ensuremath{\log g_*}}
\newcommand{\loggp}{\ensuremath{\log g_{\rm{P}}}}
\newcommand{\ecosw}{\ensuremath{e\cos{\omega_{*}}}}
\newcommand{\esinw}{\ensuremath{e\sin{\omega_{*}}}}
\newcommand{\secosw}{\ensuremath{\sqrt{e}\cos{\omega_{*}}}}
\newcommand{\sesinw}{\ensuremath{\sqrt{e}\sin{\omega_{*}}}}
\newcommand{\msun}{\ensuremath{\,M_\Sun}}
\newcommand{\rsun}{\ensuremath{\,R_\Sun}}
\newcommand{\lsun}{\ensuremath{\,L_\Sun}}

\newcommand{\rhostar}{\ensuremath{\,\rho_*}}
\newcommand{\mj}{\ensuremath{\,M_{\rm J}}}
\newcommand{\mearth}{\ensuremath{\,M_{\Earth}}}
\newcommand{\mplanet}{\ensuremath{\,M_{\rm P}}}
\newcommand{\rplanet}{\ensuremath{\,R_{\rm P}}}
\newcommand{\msini}{\ensuremath{\,M_{\rm P}\sin{i}}}
\newcommand{\cosi}{\ensuremath{\cos{i}}}

\newcommand{\rj}{\ensuremath{\,R_{\rm J}}}
\newcommand{\re}{\ensuremath{\,R_{\rm \Earth}}\xspace}
\newcommand{\me}{\ensuremath{\,M_{\rm \Earth}}\xspace}
\newcommand{\fave}{\langle F \rangle}
\newcommand{\fluxcgs}{10$^9$ erg s$^{-1}$ cm$^{-2}$}

\newcommand{\kepler}{{\it Kepler}}
\newcommand{\ktwo}{{\it K2}}
\newcommand{\tess}{{\it TESS}}
\newcommand{\gaia}{{\it Gaia}}
\newcommand{\spitzer}{{\it Spitzer}}
\newcommand{\corot}{{\it CoRoT}}

\newcommand{\mstar}{\ensuremath{M_{*}}}
\newcommand{\rstar}{\ensuremath{R_{*}}}
\newcommand{\ar}{\ensuremath{a/R_*}}
\newcommand{\vsini}{\ensuremath{V\sin{I_*}}}
\newcommand{\exofasttwo}{{\tt EXOFASTv2}}
\newcommand{\exofast}{{\tt EXOFAST}}
\newcommand{\multifast}{{\tt MULTIFAST}}

\newcommand{\Rnom}{\hbox{$\mathcal{R}^{\rm N}_{\odot}$}}

\newcommand{\Lnom}{\hbox{$\mathcal{L}^{\rm N}_{\odot}$}}

\newcommand{\GMnom}{\hbox{$\mathcal{(GM)}^{\rm N}_{\odot}$}}
\newcommand{\GMJnom}{\hbox{$\mathcal{(GM)}^{\rm N}_{\rm J}$}}
\newcommand{\GMEnom}{\hbox{$\mathcal{(GM)}^{\rm N}_{\rm E}$}}
\newcommand{\MJnom}{\hbox{$\mathcal{M}^{\rm N}_{\rm J}$}}
\newcommand{\MEnom}{\hbox{$\mathcal{M}^{\rm N}_{\rm E}$}}
\newcommand{\ReJnom}{\hbox{$\mathcal{R}^{\rm N}_{e\rm J}$}}

\newcommand{\ReEnom}{\hbox{$\mathcal{R}^{\rm N}_{e\rm E}$}}

\newcommand{\um}{\ensuremath{\mu m}}

\begin{document}

\title{EXOFASTv2: A public, generalized, publication-quality exoplanet modeling code}

\author[0000-0003-3773-5142]{Jason D.\ Eastman}
\affiliation{Center for Astrophysics \textbar \ Harvard \& Smithsonian, 60 Garden St, Cambridge, MA 02138, USA}

\author[0000-0001-8812-0565]{Joseph E.\ Rodriguez}
\affiliation{Center for Astrophysics \textbar \ Harvard \& Smithsonian, 60 Garden St, Cambridge, MA 02138, USA}

\author[0000-0002-0802-9145]{Eric Agol}
\affiliation{Virtual Planetary Laboratory, University of Washington, Seattle, WA, USA}
\affiliation{Department of Astronomy, University of Washington, Seattle, WA, USA}

\author[0000-0002-3481-9052]{Keivan G.\ Stassun}
\affiliation{Department of Physics and Astronomy, Vanderbilt University, 6301 Stevenson Center Ln., Nashville, TN 37235, USA}
\affiliation{Department of Physics, Fisk University, 1000 17th Avenue North, Nashville, TN 37208, USA}

\author[0000-0002-9539-4203]{Thomas G.\ Beatty}
\affiliation{Department of Astronomy and Steward Observatory, University of Arizona, Tucson, AZ 85721, USA}

\author[0000-0001-7246-5438]{Andrew Vanderburg}
\affiliation{Department of Astronomy, The University of Texas at Austin, 2515 Speedway, Stop C1400, Austin, TX 78712}
\affiliation{NASA Sagan Fellow}

\author[0000-0003-0395-9869]{B.\ Scott Gaudi}
\affiliation{Department of Astronomy, The Ohio State University, 140 West 18th Avenue, Columbus, OH 43210, USA}	

\author[0000-0001-6588-9574]{Karen A.\ Collins}
\affiliation{Center for Astrophysics \textbar \ Harvard \& Smithsonian, 60 Garden St, Cambridge, MA 02138, USA}

\author[0000-0002-0296-3826]{Rodrigo Luger}
\affil{Center~for~Computational~Astrophysics, Flatiron~Institute, New~York, NY}

\correspondingauthor{Jason D. Eastman}
\email{jason.eastman@cfa.harvard.edu}

\shorttitle{\exofasttwo}
\shortauthors{Eastman et al.}

\begin{abstract}

We present the next generation public exoplanet fitting software, \exofasttwo. It is capable of fitting an arbitrary number of planets, radial velocity data sets, astrometric data sets, and/or transits observed with any combination of wavelengths. We model the star simultaneously in the fit and provide several state-of-the-art ways to constrain its properties, including taking advantage of the now-ubiquitous all-sky catalog photometry and Gaia parallaxes. \exofasttwo \ can model the star by itself, too. Multi-planet systems are modeled self-consistently with the same underlying stellar mass that defines their semi-major axes through Kepler's law and the planetary period. Transit timing, duration, and depth variations can be modeled with a simple command line option.

We explain our methodology and rationale as well as provide an improved version of the core transit model that is both 25\% faster and more accurate. We highlight several potential pitfalls in exoplanet modeling, including the handling of eccentricity in transit-only fits, that the standard exoplanet convention for $\omega$ uses a left-handed coordinate system, contrary to most modern textbooks, how to avoid an important degeneracy when allowing negative companion masses, and a widely unappreciated, potential 10-minute ambiguity in the reported transit times.

\exofasttwo \ is available at \url{https://github.com/jdeast/EXOFASTv2}. The code is written in IDL, and includes an executable that can be run freely and legally without an IDL license or any knowledge of the language. Extensive documentation and tutorials are included in the distribution for a variety of example fits. Advanced amateurs and undergrads have successfully performed sophisticated global fits of complex planetary systems with \exofasttwo. It is therefore a powerful tool for education and outreach as well as the broader professional community.

\end{abstract}


\section{Introduction}

From the first discovery of a Hot Jupiter around a main sequence star \citep{Mayor:1995}, the field of exoplanets has exploded in a relatively short amount of time, with nearly 4000 planets confirmed today, largely discovered from the highly successful \kepler \ mission. With \tess \ ramping up, the availability of precision radial velocity instrumentation increasing dramatically, \gaia \ poised to announce the first astrometric planetary detections, and several new missions on the horizon, the field has a promising future.

At the same time, we understand that our knowledge of the individual planets is limited by our understanding of their host stars, and that planetary transits provide a unique opportunity to directly constrain the host star's density \citep{Seager:2003}. New stellar models have improved accuracy and range, while all sky photometric surveys, galactic dust maps, and \gaia \ parallaxes empirically constrain the bolometric luminosity and therefore stellar radius for the vast majority of RV and transit detected exoplanet hosts. 

With this flood of complex -- but complementary -- data, modeling a single system is a major endeavor requiring an expertise in stars, radial velocities, transits, and even potentially astrometry. Often, the approach has been for the respective experts to model each component separately and combine the results in some fashion. This is a labor intensive effort that is forced to ignore the often immense complementarity of the data sets, and one that does not scale well with the enormous numbers of planets already in hand, much less the flood that of discoveries that continues to come.

A global model of a stellar system that simultaneously includes many data sets, all planets, and the star simultaneously offers significant advantages over a piece-wise approach that is common today. Most obviously, a piece-wise model often fits the same parameter multiple times, resulting in degraded precision, a requirement for iteration and priors, and mathematically (and sometimes statistically) inconsistent parameter values. While some parameters are only constrained by one data set (e.g., the planetary radius from the transit), and some data sets constrain some parameters so much better than others that the constraint can be safely ignored in inferior data sets (e.g., $T_C$ from transits), there exists a significant grey area where the whole is greater than the sum of its parts, or inhomogeneous data quality can change which data set is the dominant constraint in non-intuitive ways.

The most obvious example of the synergy between data sets is in the eccentricity and argument of periastron. The transit duration depends critically on $e$ and $\omega_{*}$. While the transit duration eliminates a significant fraction of the joint $e$-$\omega_{*}$ parameter space, the degeneracy between them effectively means most values of $e$ and $\omega_{*}$, individually, are allowed. The radial velocity curve can break that degeneracy, but the precision can be greatly improved by including the degenerate constraint from the transit, especially with few RV data points (see \S \ref{sec:ecc}).

As planetary systems become more abundant and more complex, the existence of flexible, standard, trusted tools to determine their properties and uncertainties provides powerful advantages to the community. First and foremost, such tools significantly lower the barrier of entry and reduce the time and effort required to create a sophisticated model of a planetary system. 

Standardized and heavily used tools benefit from community feedback and error reporting, generally resulting in more robust, well-tested, better documented, easier to use, and validated code. That has certainly been the case for \exofast \ and \exofasttwo. Finally, standardized tools mean that a consistent set of parameters are reported with a consistent set of underlying and well-documented assumptions, enabling a more robust comparative and statistical analysis of exoplanet populations.

Many exoplanet modeling programs have existed in the public domain and offer various advantages and disadvantages. Hans J. Deeg summarized the status of many public tools related to exoplanets\footnote{\url{https://owncloud.ll.iac.es/index.php/s/5iKfRHf25YUEVwB}}, but we will focus on just the codes that model both transits and radial velocities: \exofast \ \citep{Eastman:2013}, {\tt exonailer} \citep{Espinoza:2016}, the Transit Light Curve Modeller ({\tt TLCM}) \citep{Smith:2017,Csizmadia:2019}, and {\tt Pyaneti} \citep{Barragan:2019}. More recent and promising contributions in the combined RV and transit modeling space include {\tt exoplanet}\footnote{\url{https://github.com/dfm/exoplanet}}, {\tt Juliet} \citep{Espinoza:2018a,Espinoza:2018b}, and {\tt allesfitter} \citep{Gunther:2019}.

{\tt Exonailer}, {\tt Pyaneti}, {\tt Juliet}, {\tt allesfitter}, and {\tt exoplanet} are all written in Python, a language that is quickly becoming the standard in the exoplanet community and astronomy as a whole, which is a significant advantage to many potential users more comfortable with Python and who wish to customize the code for their purposes. Being in Python also allows these codes to build on other well-tested, lower-level, public codes including {\tt BATMAN} \citep{Kreidberg:2014}, {\tt EMCEE} \citep{ForemanMackey:2012}, {\tt ellc} \citep{Maxted:2016}, {\tt dynesty} \citep{Speagle:2019}, {\tt celerite} \citep{ForemanMackey:2017,ForemanMackey:2018}, and {\tt George} \citep{ForemanMackey:2015}. {\tt exoplanet} in particular shows significant promise in terms of runtime, utilizing PyMC3 \citep{Salvatier:2016}, an implementation of a Hamiltonian MCMC to achieve factors of 100-1000 times faster convergence than the differential evolution or affine invariant samplers more commonly used today. {\tt Juliet}, {\tt exoplanet} and {\tt allesfitter} allow the inclusion of Gaussian Processes, a promising method to model correlated noise in both transit and radial velocity data. {\tt allesfitter} has the unique advantage of being able to model star spots and stellar flares. 

In contrast, \exofast, \exofasttwo, and {\tt TLCM} are all written in IDL. {\tt TLCM} can be run with the open source compiler GDL (GNU Data Language), and has the unique advantage of being able to model RVs from both components of a binary star simultaneously, and includes ellipsoidal and beaming effects in the light curve.

But in general, the sophistication and flexibility of \exofasttwo \ is unmatched. It has already been used for many published planets -- often without our involvement -- including Kepler/K2 systems \citep{Rodriguez:2018a, Yu:2018, Rodriguez:2018b, Canas:2019a}, KELT systems \citep{Johns:2019, Rodriguez:2019}, Qatar systems \citep{Alsubai:2019a, Alsubai:2019b}, and about half of the currently-published TESS planet discovery papers \citep[e.g.][]{Canas:2019b, Winters:2019, Huber:2019}. 

No other global code constrains the star simultaneously with isochrones or SEDs, as we do, described in \S \ref{sec:star}, fits the limb darkening simultaneously constrained by a lookup table (\S \ref{sec:limbdarkening}), or is flexible enough to simultaneously model a wide variety of real data sets or the most interesting multi-planet systems with relatively simple changes to command line options or configuration files, as described in \S \ref{sec:multiplanet}.

While many public codes fit just transits (e.g., {\tt TAP} \citep{Gazak:2012}, {\tt BATMAN} \citep{Kreidberg:2014}) or just RVs (e.g., {\tt Systemic} \citep{Meschiari:2012}, {\tt RadVel} \citep{Fulton:2018}), we are unaware of another public code that models the Doppler Tomography signal (our implementation is described in \S \ref{sec:dt}) or the stellar and planetary astrometric signal (see \S \ref{sec:astrometry}) at all. We discuss how the code weights each data set to determine a global constraint in \S \ref{sec:combining}.

We know of no other code that integrates an exoplanet mass-radius relation (see \S \ref{sec:chen}). Some codes allow for negative planet masses to avoid the Lucy-Sweeney-like bias for planet mass, but do not address the degeneracy that results (see \S \ref{sec:negmass}). No other public code we are aware of easily allows variations between transit timing, depth, and duration (\S \ref{sec:ttvs}), allows an arbitrary number of additive or multiplicative detrending parameters in the transit light curve (\S \ref{sec:detrending}), or allows simultaneous fitting of the dilution in the transit light curve (\S \ref{sec:dilute}). Only a couple codes handle the significant smearing effect of long exposures (\S \ref{sec:longcadence}). We also allow uniform and/or Gaussian priors on all fitted or derived parameters, as discussed in \S \ref{sec:priors}.

We outline our global model parameterization in \S \ref{sec:parameterization} and go through a detailed derivation of the RV, astrometric, and transit models in \S \ref{sec:planetpath}, while clarifying an inconsistent use of the argument of periastron in the literature that is likely to lead to confusion, especially when incorporating astrometric data sets in the future.

In the age of \kepler, \ktwo, and \tess, it is critical that we be able to model a lightcurve without RVs. While this is most typically done by assuming the planetary orbits are circular \citep[e.g.,][]{Thompson:2018,Mayo:2018}, such an assumption requires us to break the link between the transit and the stellar density lest an eccentric orbit bias the stellar parameters. However, that link is actually a powerful, albeit degenerate, constraint on the eccentricity, particularly when Gaia DR2 allows us to derive precise densities for most host stars. We discuss the major complications this introduces and how we deal with it in \S \ref{sec:ecc}.

The available public and private codes use inconsistent definitions of the transit time, which can differ by as much as 10 minutes and is discussed in detail in \S \ref{sec:tc}. 

While the model is generated from the projected planetary path using a full Keplerian orbit, the code reports some approximate values (\S \ref{sec:approx}). We use the same constants consistently throughout the code, and summarize their exact values in \S \ref{sec:constants} for transparency and reproducibility. In \S \ref{sec:runtime} we discuss the runtime of the code. \exofasttwo \ can be run without an IDL license (see \S \ref{sec:license}). 

Section \ref{sec:underhood} discusses options and enhancements to the core MCMC algorithm, which are general to any optimization problem. Most important, the code can now easily perform parallel tempering (\S \ref{sec:paralleltempering}) to robustly sample multi-modal distributions or enable the code to navigate complex likelihood surfaces, including more robustly finding the optimal global solution. It also has a more robust calculation of the burn-in (\S \ref{sec:burnin}), and we explain the importance of strict convergence criteria and when the user may wish to relax it to improve the runtime (\S \ref{sec:convergence}).

Finally, \S \ref{sec:walkthrough} walks through an example fit of HAT-P-3. We compare its results to the results of the original \exofast \ \citep{Eastman:2013}, which are significantly improved due to the SED+\gaia \ constraint on the stellar radius. Similar improvements can be expected for most planets characterized before or without constraints from \gaia \ DR2 \citep{Gaia:2018}. Section \ref{sec:explain} describes the outputs of \exofasttwo, including a table of all output parameters and a detailed explanation of their bounds, how they were derived, and their underlying assumptions. Section \ref{sec:troubleshooting} describes general steps one might take to troubleshoot problematic fits, and \S \ref{sec:tess} explains how we do our automated fits for TESS Objects of Interest available on the ExoFOP-TESS.

In many sections throughout the paper, we identify important caveats and likely areas for future improvement.

This work would not have been possible without the work of countless members of the community. We ask that those who use this code use the acknowledgement in \S \ref{sec:acknowledgement} as a guide when citing this work to recognize their foundational effort.

\section{Stellar Mass and Radius}
\label{sec:star}

Significant changes have been made to allow several additional options for constraining the stellar mass and radius. Because ultimately the properties we care most about (e.g., the radius and mass of a planet, \rplanet, \mplanet) depend upon the stellar properties, and because the transit alone provides a valuable, independent constraint on the stellar density through its duration \citep{Seager:2003} and its \teff, \logg, and \feh \ through the limb darkening, both \exofast \ and \exofasttwo \ always require the user to fit the stellar properties along with the planetary fit. 

In order to do this, the original \exofast \ used the \logg, \teff, \feh, and the empirical \citet{Torres:2010} relations to derive the stellar mass and radius. A prior was required on \teff \ and \feh. A major disadvantage to this was that the Torres relation does not apply to low mass stars, making the original \exofast \ inapplicable in the regime that is a major focus of \ktwo, \tess, and the exoplanet community as a whole.

Now with \exofasttwo, we provide several different methods, described in the following subsections, to robustly constrain the stellar parameters. Which method the user chooses depends on the availability of information and the stellar type, but trying multiple ways to constrain the stellar properties can also serve as an important cross check on potential systematic errors. So as to not unnecessarily limit its use, the code does not require that any of the following methods be used and places no restrictions on the combinations of allowed methods, but we urge extreme caution when straying -- even slightly -- from the following five methods to constrain the stellar properties. The code is unforgiving if the stellar parameters are unconstrained and it is likely to generate non-intuitive error messages. Conversely, a common mistake users make is to overconstrain the stellar properties, supplying priors derived from the same or similar methods used in the global fit, resulting in unjustifiably small uncertainties.

\begin{itemize}

\item By default, \exofasttwo \ now uses the MIST stellar evolutionary models \citep{Dotter:2016} (see \S \ref{sec:mist}).

\item By setting the {\tt NOMIST} and {\tt YY} keywords, the user can disable the MIST model constraint and instead use the Yonsie Yale (YY) stellar evolutionary models \citep{Yi:2001} (see \S \ref{sec:yy}).

\item By setting the {\tt NOMIST} and {\tt TORRES} keywords, we recover the behavior of the original \exofast, using the empirical \citet{Torres:2010} relations to derive the stellar mass and radius (see \S \ref{sec:torres}).

\item The user can supply a list of broad band photometry of the host star, along with priors on the parallax and extinction. Then \exofasttwo \ will fit an SED (see \S \ref{sec:sed}).

\item The user can set the {\tt NOMIST} flag and supply a direct Gaussian or uniform prior on \mstar \ and/or \rstar \ (see \S \ref{sec:starprior}).

\end{itemize}

The code is able to use any combination of these methods, however, we strongly advise against using more than one of the first three, as there is a significant risk of double counting information, resulting in underestimated uncertainites. Similarly, the user must be careful when using priors (\S \ref{sec:starprior}) to ensure that the information is not derived from a similar technique. With clean, broadband photometry (not contaminated by an unresolved companion), the best constraints can typically be obtained by combining the SED fit (which largely constrains the stellar radius) with another method, like the MIST stellar tracks, to pin down the stellar mass. While in principle, the stellar density constraint from a transit plus the stellar radius constraint from the SED allows one to derive the stellar mass, the mass uncertainty is typically too large to be an independent check on the stellar models ($\gtrsim 10\%$), even with precise RVs to constrain the eccentricity. In the age of \gaia, \kepler, \ktwo, and \tess, the known stellar density can often be used to constrain the transit duration better than the light curve, effectively removing a free parameter from the transit fit (or it can be used to constrain the planetary eccentricity from a transit alone -- see \S \ref{sec:ecc}). Part of the beauty of a global model is that we need not know beforehand what data set will constrain which parameters best -- the global model will automatically balance all constraints to return the best possible parameters.

The stellar effective temperature, \teff\, and metallicity, \feh\, are still typically best constrained by spectroscopic priors, but can be independently determined from the SED, isochrones, and transit limb darkening when spectroscopic values are either unavailable or not trusted. The stellar surface gravity, \logg, can also be independently determined from the SED, isochrones, and transit, and often much more precisely than the spectroscopic prior, which can have large systematic errors \citep{Torres:2008}. Generally, we advise against supplying a spectroscopic prior on \logg \ when fitting a transit, unless fitting for the period of a single transit or similarly unconstrained global fits.

Because \exofasttwo \ can be used to model a star without a planet, it recreates much of the functionality of codes dedicated to stellar modeling, like {\tt MINESweeper}\footnote{\url{https://github.com/pacargile/MINESweeper}}, though \exofasttwo \ uses a less sophisticated interpolation of the isochrones and bolometric correction grid and we do not fit spectra directly to constrain \teff \ or \feh.

As a brief aside to head off confusion between the various stellar {\it atmospheric} models and stellar {\it evolution} models we use, we clarify that here. There are two theoretical stellar evolution models we can use: MIST (\S \ref{sec:mist}) and YY (\S \ref{sec:yy}), as well as the empirical Torres relation derived from well-separated eclipsing binaries (\S \ref{sec:torres}). These essentially supply a constraint on the stellar mass and radius based on the stellar \logg, \teff, and \feh. 

There are also two stellar atmospheric models implicitly used by the code. The MIST bolometric correction grid, one way of computing the SED (\S \ref{sec:mistsed}), as well as the limb darkening tables (\S \ref{sec:limbdarkening}) used to constrain the limb darkening of the transit light curves, are built from ATLAS \citep{Kurucz:1970,Kurucz:1993}, while the equivalent width summation SED grid, another method to model the SED (\S \ref{sec:ewsed}), is built from NextGen stellar atmospheric models \citep{Allard:2012}.

\subsection{MIST Evolutionary tracks}
\label{sec:mist}

The MESA Isochrones and Stellar Tracks (MIST) are a relatively new set of isochrones that are computed using MESA \citep{Paxton:2011, Paxton:2013, Paxton:2015} to derive stellar tracks from 0.1 \msun \ to 300 \msun, ages from 100,000 years to well beyond the age of the universe (we impose a prior to exclude ages beyond 13.82 Gyr), evolutionary phases from the pre-main sequence to the white dwarf cooling sequence, and metalicities from -4.0 to +0.5 dex \citep{Dotter:2016, Choi:2016}.

Their stellar tracks are computed in a regular grid of initial mass, initial \feh, and ``Equivalent Evolutionary Phase'' (EEP), which describe common phases of evolutionary history (e.g., EEP of 202 corresponds to the zero age main sequence and EEP of 454 corresponds to the terminal age main sequence). The EEPs are chosen so neighboring tracks look as similar as possible at a given EEP, which improves the accuracy of interpolation compared to a regular grid in age.

When MIST models are used, we replace the fitted parameter age with EEP. However, a uniform prior on EEP imposes a nonphysical prior on age that biases the star toward quickly changing regions of its life cycle (i.e., away from the main sequence). In order to correct the uniform EEP prior to a uniform age prior, we weight the acceptance of the step by the interpolated gradient in age, $\partial (EEP)/\partial (Age)$. This often leads to discontinuities in the EEP probability distribution function, which may appear alarming, but is expected.

We also add the initial metallicity, $\feh_0$ as a new fitted parameter. The observed surface iron abundance, \feh, is not constant over the lifetime of the star, and in fact varies by as much as 0.12 dex for the Sun. Initial metallicity and metallicity are highly correlated, but the covariance is linear which is well handled by the differential evolution algorithm we employ.

We convert the surface metallicity at each EEP provided in the MIST tracks into a surface \feh \ using the equation:

\begin{equation}
\feh = \log_{10}{\left(\frac{Z}{X}\right)} - \log_{10}{\left(\frac{Z}{X}\right)}_\Sun
\end{equation}

\noindent where $\left(\frac{Z}{X}\right)_\Sun = 0.0181$ is the metal fraction of the Sun from \citet{Asplund:2009}, on which the MIST models are calibrated. Z is the surface metallicity fraction and X is the surface $^1$H fraction provided for each age in the MIST EEP file. To be clear, we fit both \feh \ and $\feh_0$. We use the $\feh_0$ to define the grid point in the isochrone, then penalize the fit for the difference between the MIST interpolated \feh \ and the MCMC step \feh.

While the mass of the star also changes over its lifetime, most planet hosting stars lose a negligible amount of mass (compared to the uncertainty) during the normal stages of their lives, so we assume that initial mass is the current mass. Very high mass stars, where this assumption is invalid, should not be fitted with MIST tracks.

We repackage the MIST EEP tracks into IDL save files and include only the age, \rstar, \teff, as well as our derived quantities \feh \ and $\partial (EEP)/\partial (Age)$, for each \mstar \ and $\feh_0$ \ in order to reduce the size of the installation and improve load times during the fit. 

We perform a trilinear interpolation in EEP, \mstar, $\feh_0$, \ space using IDL's {\tt INTERPOLATE} procedure to derive \teff, \rstar, and \feh \ at each step in the Markov Chain. Loading a track takes about 10 times longer than the interpolation, so we store them in memory between calls. Because MCMC typically samples relatively few tracks, this once-per-track overhead is negligible compared to the total MCMC runtime.

Once we have the interpolated values for \teff, \rstar, and \feh, we compare them to the corresponding model values at the current MCMC step, and penalize the model for the difference between the two:

\begin{equation}
\begin{tabular}{lcl}
    $\chi^2$ & += & $\left[\left(R_{*,{\rm MCMC}} - R_{*,{\rm MIST}}\right)/\left(\sigma_{R,{\rm MIST}}\right)\right]^2 + $ \\
             &    & $\left[\left(T_{{\rm eff, MCMC}} - T_{{\rm eff,MIST}}\right)/\left(\sigma_{T_{\rm eff},MIST}\right)\right]^2 + $ \\
             &    & $\left[\left(\feh_{{\rm MCMC}} - \feh_{{\rm MIST}}\right)/\left(\sigma_{\feh,MIST}\right)\right]^2$,

\end{tabular}
\end{equation}

\noindent where $\sigma_{R,{\rm MIST}}$, $\sigma_{T_{\rm eff},MIST}$, and $\sigma_{\feh,MIST}$ are the assumed MIST model uncertainties corresponding to each parameter. They are percent errors that vary as a function of stellar mass equal to 10\% at 0.1 \msun, 3\% at 1 \msun, 5\% at 10 \msun, and smoothly varying in between as a quadratic in $\log{\mstar}$. This relatively simple penalty enforces the physics of stellar evolution encoded in the MIST models, including the complex covariances between EEP, \mstar, \feh, age, \teff, and \rstar. It is important to understand that, while the penalty is only on \rstar, \teff, and \feh, the MCMC algorithm is free to modify \mstar, $\feh_0$, and/or EEP to match a well-constrained \rstar, \teff, or \feh.

Because we use the models to guide the stellar parameters rather than define them, additional information from any source (e.g., priors, transit duration, limb darkening, parallax, SED) can dramatically improve the stellar mass and radius, and therefore the planetary mass and radius. In this way, a degeneracy in the evolutionary models may be broken by the transit density constraint, or the transit density-eccentricity degeneracy may be constrained by the stellar evolutionary models. 

A critical caveat is that enforcing an external constraint means the fit also inherits any of its shortcomings or assumptions. While the MIST models represent the current state-of-the-art in our theoretical understanding of stellar astrophysics, there is much we do not understand about stars and there are many approximations we must use. The MIST models we use assume the star is not rotating, has solar scaled abundances (i.e., $[\alpha/{\rm Fe}]=0$), is limited to $\feh \leq 0.5$ dex, and are fundamentally derived from a 1D stellar evolution model. Further, all theoretical stellar models have known problems reproducing the observations in detail for low-mass stars. Even for the Sun (using the grid point at $EEP=354.1661$, $\feh_0=0.01837$, $\mstar=1 \msun$, corresponding to $\feh=0$, age $=4.603$ Gyr), our interpolation does not recreate the exact solar values (\rstar=1.02\rsun, \teff=5824), which is due to the fact that the \citet{Asplund:2009} proto-solar abundances are mildly inconsistent with the MIST grids (Dotter, 2018, Priv. Comm.), though this systematic error is well within the assumed 3\% model uncertainty. See \citet{Choi:2016} for a more in depth discussion of the MIST models. 

Especially for users not used to modeling stars, it can be difficult to define appropriate starting conditions for the star to ensure the initial guess is defined, much less values that will not get stuck in a locally optimized solution far from the correct one (e.g., on the wrong side of the turn-off). By default, \exofasttwo \ assumes Sun-like values for any parameter not supplied ($\rstar=\rsun$, $\mstar=\msun$, $\teff=5778$ K, $\feh=0$, $\feh_0=0$, Age $=4.603$ Gyr, EEP $=354.1661$). If the user changes some starting parameters, but not all consistently, it is easy for the initial global model to be rejected as nonphysical with relatively little feedback. For example, if the user selects a high mass star, it will die long before 4.6 Gyr. Conversely, if the user changes the EEP to be at the turnoff (454), but then selects a low-mass star, its age will exceed the age of the universe and the star will be rejected as nonphysical.

If a roughly self-consistent set of parameters (for at least mass, radius, \teff, and age) are not already known, the user may wish to fit the star independently first, with parallel tempering enabled (see \S \ref{sec:paralleltempering}) to better explore the complex parameter space of the MIST models.

When MIST tracks are used, a plot of the mass track is created, an example of which can be seen later in the text: Figure \ref{fig:mcmcmist} for HAT-P-3b.

\subsection{YY Evolutionary tracks}
\label{sec:yy}

Using the YY models \citep{Yi:2001} is conceptually the same as using the MIST models, but our implementation of the YY models are nearly 20 years out of date. While more recent updates to the YY isochrones include low-mass stars \citep{Spada:2013}, we have not included those updates, and so our implementation of YY isochrones does not apply to low mass ($\mstar \lesssim 0.5 \msun$) stars. The YY model implementation is well tested, routinely being used for KELT discoveries, and is described in \citet{Eastman:2016}, but largely made obsolete by the introduction of the MIST models. We retain it as a useful check on the systematic errors of different models.

There are a few differences between MIST and YY worth noting. When using the YY isochrones, we step directly in Age and \feh; EEP and $\feh_0$ are not fitted or derived. We also assume a constant $3\%$ uncertainty regardless of the stellar mass, which is too low for non-solar type stars. When YY isochrones are used, figures similar to Figure \ref{fig:mcmcmist} are generated, but using the YY mass track.

\subsection{Torres Relations}
\label{sec:torres}

\citet{Torres:2010} identified an ensemble of well-separated (non-interacting) eclipsing binaries, for which a robust dynamical mass and geometric radius could be derived. They defined an empirical relation that mapped \logg, \teff, and \feh \ to the stellar mass and radius. This relation is fast and is much better behaved owing to its smooth likelihood surface, but the sample is relatively sparse and does not include low mass stars, so it should not be used for extreme systems or low mass ($\mstar \lesssim 0.5 \msun$) stars. No age is fitted or derived, and no stellar track is created when using the \citet{Torres:2010} relations. When using the \citet{Torres:2010} relations, we recover the behavior of the original \exofast. A more detailed description of this can be found in \citet{Eastman:2013}.

\subsection{SED fitting}
\label{sec:sed}

We now optionally include Spectral Energy Distribution (SED) fitting within the global model with the user's choice of two different methods described below. The SED model, which is essentially a measure of the star's bolometric flux, is relatively insensitive to large (0.5 dex) changes in \logg \ and \feh, making the constraint largely independent of stellar models \citep{Stassun:2016}. It is also therefore a poor constraint on \logg \ and \feh, but a moderate to strong constraint on \teff \ and the V-band extinction, $A_V$, (both of which set the shape of the SED), and a strong constraint on the stellar radius divided by the distance squared, $\left(R_*/d\right)^2$ (which sets the normalization of the model). When spectroscopic priors on the stellar parameters are non-existent or not trusted, the SED model is a powerful tool to independently constrain \teff, especially if optical photometry is available. 

If the broad band photometry is not supplied, the distance, extinction, and SED error scaling are automatically excluded from the fit. If the stellar temperature is known and a parallax is not, the only value of fitting the SED is in determining a relatively imprecise photometric distance. However, when coupled with an accurate prior on the parallax, or the astrometry directly, (e.g., from \gaia), the SED model imposes a tight constraint on the stellar radius that is only loosely dependent on stellar models and can dramatically improve the precision of the fundamental stellar and planetary parameters.

Guided by the findings from \citet{Stassun:2018}, we recommend adding 0.082 mas to the reported Gaia DR2 parallax, and adding the 0.033 mas uncertainty in the offset in quadrature with the Gaia-reported uncertainties.

One must be careful to only include trustworthy sources of the apparent magnitudes with accurate uncertainties that are correctly predicted by the stellar models. For example, SDSS magnitudes from the Sloan Digital Sky Survey show strong systematic errors and their quoted photon-limited uncertainties are typically far too precise for the bright exoplanet hosts we often fit. SDSS broad band apparent magnitudes should be used with extreme caution. The stellar atmospheres are usually poorly modeled in the UV Galex bands, making their inclusion suspect. We fit a scaling term to the uncertainties of all photometric bands so as to not unfairly weight the SED model relative to other constraints (e.g., a spectroscopic prior on \teff), but results will be significantly improved by carefully inspecting the photometry and removing outliers. {\tt MKSED} is a program distributed with \exofasttwo \ that will automatically query the most trusted photometric catalogs, apply a reasonable systematic error floor based on our experience, and generate the SED file in the required format for \exofasttwo \ to fit the SED.

Regardless of the method, the user must understand a fundamental limitation of SED fitting. An SED fit should not be used with broad band photometry that is blended with a nearby star, since the extra flux from the companion will make the star appear brighter and therefore larger than it is. The user must either forego the SED fit or manually deblend the broadband photometry prior to inputting it into \exofasttwo. Future enhancements may add the ability to model multiple SEDs to account for this, which would also enable us to self-consistently constrain the dilution in the light curves (see \S \ref{sec:dilute}). Since it can be difficult to know if there is an unresolved stellar companion, it is wise to ensure the radius implied by the SED makes sense. An unidentified companion is a likely reason the best-fit model does not lie close to the stellar model tracks and can cause catastrophic failures of the entire fit.

When fitting an SED, \exofasttwo \ automatically outputs a publication-quality figure of the SED fit, as shown in Figure \ref{fig:mcmcsed} for HAT-P-3b. When using the MIST bolometric correction grid, we generate a similar plot, except without the atmosphere, since it does not use the stellar atmosphere directly. The input magnitudes are a mix of AB and Vega systems to match the system most commonly reported by VizieR.

\subsubsection{Equivalent width summation}
\label{sec:ewsed}

With the equivalent width summation method, the NextGen stellar atmospheric models \citep{Allard:2012} are tri-linearly interpolated at each step in  \logg, \teff, and \feh. We then apply the reddening with an $R_V=3.1$ law, normalize it by $\left(R_*/d\right)^2$, and sum the flux over each band's equivalent width. The grid of stellar models are coarsely sampled, by 0.5 dex, in \logg \ and \feh \ and more finely sampled in \teff \ ($\sim$100 K).  

Summing over the equivalent width of each bandpass ignores their detailed shape. For complex or wide bandpasses, this can be a source of systematic error that has not been thoroughly investigated. For this reason, we do not support the relatively wide \gaia \ bands. In principle, \gaia's space-based, all sky blue bandpass could be a particular help in constraining \teff \ for most planet hosts, which ultimately limits the precision of the derived \rstar. 

\subsubsection{MIST bolometric correction grid}
\label{sec:mistsed}

The MIST team has supplied pre-computed bolometric corrections in a grid of \logg, \teff, \feh, and $A_V$\footnote{\url{http://waps.cfa.harvard.edu/MIST/model_grids.html}}, based on the ATLAS/SYNTHE stellar atmospheres \citep{Kurucz:1970,Kurucz:1993} and the detailed shape for each of a large number of standard filters in a way that is consistent with their stellar models (Conroy et al., in prep). Compared to the NextGen grids, the MIST grids are somewhat finer in \feh \ (typically 0.25 dex around solar metallicities), similarly sampled in \logg, and more coarsely sampled in \teff \ (typically 250 K around solar type stars). We have repackaged their grids and added them to our distribution for efficient use with \exofasttwo.

When using the MIST grid to fit the SED, we linearly interpolate their 4D grid at the current step values for \logg, \teff, \feh, and $A_V$ to derive a bolometric correction, $BC_x$, for each observed band, $x$. Then, using the derived step's stellar luminosity and distance modulus, $\mu$, we derive an apparent magnitude ($m_x=-2.5\log{\left(L/L_\Sun\right)} + 4.74 + \mu - BC_x$), which we compare to the supplied catalog magnitudes. 

We have yet to perform a detailed comparison between the two methods of fitting the SED, and in particular, our implementation of the MIST grid method is not well studied. We may have a more concrete recommendation in the future. Still, we expect the MIST grid to perform better primarily because the \gaia \ data are often the only reliable source of visible band photometry. The self-consistency between the SED and the MIST stellar tracks (which are used by default to constrain the star) is an added bonus, in addition to the precise handling of the detailed shape of each bandpass. However, the differently sampled stellar atmosphere grid and effectively much courser sampling of the reddening means there is no obvious winner. In the only head-to-head comparison we did, for HAT-P-3b, the SED photometric error scaling was 1-sigma lower using the MIST grid, meaning the model is a better match to the data. 

While we expected the MIST grid method to be significantly faster, in the few test cases we performed, models using the MIST grids required four times as many model evaluations to converge (for reasons not currently understood), making the total runtime far slower, despite the fact that a single SED model calculation is typically twice as fast.

\subsection{Stellar priors}
\label{sec:starprior}

\exofasttwo \ allows the user to supply Gaussian or uniform priors on any fitted or derived parameter. If available, the user may wish to include, for example, a prior on the Age from gyrochronology or a prior on the density from asteroseismology in conjunction with one of the above methods. Or, if for some reason none of the above methods are trusted, the user can disable them and supply a Gaussian prior directly on the stellar mass and/or radius from a variety of other sources, such as the \citet{Mann:2015} relations, or other stellar models like the \citet{Baraffe:1998} or Parsec \citep{Bressan:1993} models for low mass stars. The user should exercise extreme caution when supplying priors on stellar parameters in conjunction with any other method to verify that the priors are derived in a completely independent manner. The vast majority of stellar parameters are derived using something similar or identical to the above methods, and using both a prior and its underlying method would double count the constraint, resulting in underestimated stellar parameter uncertainties, which would trickle down to underestimated planetary uncertainties.

For example, we note that \gaia \ typically reports a \teff \ and \rstar. These reported values are based on an SED fit of only the \gaia \ photometry and parallax. Priors on \teff \ and \rstar \ from \gaia \ should never be used in conjunction with an SED fit within \exofasttwo \ (\S \ref{sec:sed}). Furthermore, the \gaia-only-derived \teff \ and \rstar \ are often far inferior to what we can determine from our SED fit using additional, readily-available photometry.

When applying a prior directly on \rstar \ and \mstar from some external method, the user must be careful about the other stellar parameters. In particular, \feh \ and \teff \ are still fit and constrain the limb darkening of a transit model. Therefore, the user must also either impose realistic priors on  \feh \ and \teff \ or fix them to some arbitrary value, disable the claret limb darkening constraint, and ignore any derived quantities that depend on them (e.g., $T_{\rm eq}$). 

\section{Stellar Limb darkening}
\label{sec:limbdarkening}

As in the original \exofast, we apply priors on the quadratic limb darkening by interpolating the \citet{Claret:2011} limb darkening models at each step in \logg, \teff, and \feh. When \exofast \ was originally published, we were concerned that potential systematic errors in the quadratic limb darkening tables may bias the stellar parameters if the limb darkening could be constrained from the light curve. In particular, we were worried because the values were derived with relatively simple assumptions, and that more sophisticated 3D hydrodynamic simulations would be better \citep[e.g.][]{Hayek:2012}, or that the quadratic form we assume for expediency would break down and, at some point, the non-linear law would be required before accurate stellar parameters could be inferred. However, subsequent tests have shown that our assumed systematic errors in the limb darkening parameters are accurate and still small enough that they can meaningfully constrain the stellar properties. Still, more accurate limb darkening models with smaller systematic uncertainties are likely to provide even better constraints. Users can disable this model constraint using the {\tt NOCLARET} flag and supply their own priors, decoupled from \logg, \teff, and \feh \ if they believe theirs to be superior. In particular, for low-mass stars, the ATLAS stellar atmospheres from which these tables were derived are suspect. In such cases, we recommend imposing a wide uniform prior ($\pm 0.15$) around the value predicted by the tables, and setting the {\tt NOCLARET} flag.

We impose limits on the limb darkening parameters that only allow steps within the physical bounds identified by \citet{Kipping:2013} for any band: $u_1 + u_2 < 1$, $u_1 > 0$, and $u_1 + 2u_2 > 0$.

These tables have been updated with \tess \ bands from \citet{Claret:2017}. Only supported bands are allowed, which are currently the Johnson/Cousins U, B, V, R, I, J, H, and K; the Sloan u', g', r', i', and z'; the Stromgren u, b, v, and y; the \spitzer \ 3.6\um, 4.5\um, 5.8\um, and 8.0\um; and \kepler, \tess, and \corot. Should the user wish to model a transit in a different band, they should either pick the closest supported equivalent band, or disable the limb darkening table interpolation using the {\tt NOCLARET} flag and either apply their own prior or let the data constrain it.

Using a non-linear or more complex limb darkening law, or including gravity darkening is typically not required to adequately model the transit, even for most \kepler \ light curves. And, with current IDL implementations, it is $\sim 1000 \times$ slower than the quadratic light curve outlined by \citet{Mandel:2002}, optimized for speed and accuracy by \citet{Eastman:2013}, and optimized again by \citet{Agol:2019} that we use. Improvements using techniques described in the codes for {\tt BATMAN} \citep{Kreidberg:2014} or {\tt STARRY} \citep{Luger:2018} could help model transit light curves for extreme stars or extremely precise light curves in a tractable amount of time, but we have not yet explored that possibility.

\subsection{Improved handling of quadratic limb-darkening}

The \citet{Mandel:2002} expressions contain the evaluation of complete elliptic integrals which turn out to be subject to numerical instability near the regions where the projected separation, $z$, equals zero ($z=0$) or equals the occultor radius ($z=R_p/R_*$), and near the points of contact ($z=\vert 1-R_p/R_*\vert$ and $z=1+R_p/R_*$). Although these instabilities cause errors in a very a small region of parameter space, and thus are rarely encountered, it is still advantageous to use formulae which are numerically stable. 

There are three issues which cause numerical instability:  division by the projected separation, $1/z$, division by the difference between the projected separation and occultor radius, $1/(R_p/R_*-z)$, and divergence of the complete elliptic integrals.  The first two of these can be eliminated analytically with a transformation of the parameters of the complete elliptic integrals. The nature of the third instability is that the complete elliptic integrals of the first and third kinds, $K(k)$ and $\Pi(n,k)$, can diverge logarithmically as $k \rightarrow 1$.  When the full transit expression is computed analytically, these divergences cancel out;  however, round-off errors in the numerical evaluation of these elliptic integrals can cause these cancellations to be imperfect when $k \approx 1$.  This can be fixed by transforming the elliptic integrals to another form of elliptic integral, the generalized complete elliptic integral \citep{Bulirsch1969}, which evaluates the sums and difference of these elliptic integrals simultaneously, giving an analytic integral which does not diverge.  The evaluation of the generalized complete elliptic integral uses a similar approach as the function used in the original \exofast \ for the complete elliptic integral of the third kind  \citep{Bulirsch1965a,Bulirsch1965b} based on the method of \citet{Bartky1938}.  In addition to these improvements, we use an improved computation of the uniform limb-darkening term which uses a more accurate formulation of the area of overlap of two circles, and we no longer use Hasting's approximation for the complete elliptic integrals of the first and second kinds, but we evaluate these along with the generalized complete elliptic integral.

The transformations and new expressions are described at length in \citet{Agol:2019}, and we have implemented these expressions in IDL to replace the former version of the quadratic limb-darkening routine used in the original version of \exofast, {\tt EXOFAST\_OCCULTQUAD}. The updated code used in \exofasttwo \ is now called {\tt EXOFAST\_OCCULTQUAD\_CEL}.

The new expressions we find to be accurate over the expected range of use ($10^{-2} \le R_p/R_* \le 10^2$, $0 \le z \le 1+R_p/R_*$), and to eliminate the numerical instabilities in problematic regions. While the difference for a typical Hot Jupiter near the contact points of the primary transit is small ($\sim$12 ppm), the maximum error in the old formulae during the secondary eclipse of an Earth-like planet near the contact points is extremely large (20\%). Previously, when such errors would be encountered, the practical effect would be to unfairly (but slightly) bias the transit time to move the data off the precise times of contact.

Additionally, while testing the new code, we found cases with the old code that were misidentified due to rounding errors and then entered an infinite loop in the iterative {\tt ELLPIC\_BULIRSCH} code. While such an error was only encountered when the limb of the star and planet were within $10^{-13}$ of one another and was never knowingly encountered in the billions of subsequent evaluations from countless MCMC runs over the past decade, the error will instantly and irrecoverably hang any fit that encounters it. We have fixed that in the original code. The new code never encountered it, highlighting the importance of numerical stability.

Finally, because of the new code's numerical stability near edge cases, we can safely eliminate the expensive check for these cases. This results in code that is 25\% faster than the previous code. Since the transit model calculation is a substantial fraction of the runtime of total code for fits with a large number of transit points, this improvement has a noticeable benefit to the total runtime of a typical fit.

We refer the interested reader to \citet{Agol:2019} for more details of the transformation and implementation.

\section{Multi-planet systems}
\label{sec:multiplanet}
\exofasttwo \ now handles an arbitrary number of planets around the same star, using Kepler's law, the mass of the system, the radius of the star, and the planetary period to derive (rather than fit) \ar \ for each planet at each step of the fit. This means that the stellar densities implied by the transit light curve model for each planet are mathematically identical. We also transform the input Solar System Barycentric time stamps to the target system's barycenter to create the self consistent stellar model and reconcile the transit, eclipse, and RV timing of multiple planets \citep{Eastman:2010}. 

No non-Keplerian motion is included and long-term stability is not considered. If timing variations are to be fit (e.g., due to Non-Keplerian interactions, reflex motion, or tidal decay), the user should set the {\tt TTV} flag (see \S \ref{sec:ttvs}).
 
We exclude eccentricities where they would cause the planets to cross into each others' Hill spheres. When we do not have masses, we use the estimate of \citet{Chen:2017} as described in \S \ref{sec:chen} to calculate the Hill Sphere. If the \citet{Chen:2017} relation is suppressed during the fit, the companion mass and therefore the Hill radius is assumed to be zero.

The eccentricity exclusion is simply based on the apastron and periastron distances plus or minus the hill radius. We do not consider stable orbital resonances or account for mutual inclination in this exclusion. For multiple transiting planets, which must have a small mutual inclination, the effect is small. Also, it is important to note that transits alone cannot distinguish between mutual inclinations of $i_b$ - $i_c$, or $180-i_b$ - $i_c$, so the mutual inclination of two planets, both with a measured inclination of $89^\circ$ could be 0 or 2 degrees even assuming the longitudes of ascending nodes are aligned, while even more significant mutual inclinations are allowed if the nodes are misaligned.

However, mutual inclination may be important for systems in which one or more fitted planets does not transit. For example, systems like Neptune and Pluto would be a priori excluded because their projected orbits cross, even though their physical orbits do not and they are dynamically stable. Quickly and precisely calculating the Minimum Orbit Intersection Distance (MOID) is non-trivial \citep{Dybczynski:1986}. A nice analytical approximation was done by \citet{Bonanno:2000}, but we have not implemented it. If the user is concerned about excluding dynamically allowed orbits that have overlapping projected orbits or orbital resonances, this constraint can be disabled by setting the {\tt ALLOWORBITCROSSING} flag.

\exofasttwo \ naturally handles simultaneous, but non-overlapping, transits without TTVs  (see \S \ref{sec:ttvs} for explanation of the latter limitation). We have not yet implemented support for planets that block other planets, which would require an upgrade of the core transit light curve model code to something like the code developed by \citet{Pal:2012} or \citet{Kipping:2011}. If mutual eclipses exist in the data, they should be excluded from the fit so they do not bias the results. In certain systems, the lack of mutual eclipses might break the $i$ and $180-i$ inclination degeneracy, at least for some of the planets. The inclination can be modeled over the entire $0 < i < 180$ degree range by setting the {\tt I180} boolean array for which planet(s) should be allowed to be fit for the whole range. Without astrometry, at least one planet must be artificially constrained to $0 < i < 90$, and the true inclination of all planets might be flipped.

\section{Doppler Tomography and Rossiter-McLaughlin}
\label{sec:dt}
Radial velocities during a transit across the star enables a measurement of the projected ``spin-orbit alignment'' angle, $\lambda$, between the spin axis of the star and the orbital axis of the planet. Because the star is rotating, part of the stellar surface is approaching the observer (blue-shifted) and part of the stellar surface is receding from the observer (red-shifted)\footnote{assuming the star is not in the unlikely, pole-on orientation}. For isolated stars where one can see both the approaching and receding stellar surfaces equally, this has the effect of broadening the observed stellar lines symmetrically. However, when a planet transits the star and blocks a part of the stellar surface, it also obscures the corresponding blue-shifted or red-shifted area of the stellar surface. This creates an unequal broadening of the stellar lines, and results in a distortion of the line profile that is often seen as anomalous radial velocity measurements during a planetary transit. These anomalous in-transit velocities are said to be caused by the Rossiter-McLaughlin (RM) effect \citep{Rossiter:1924,McLaughlin:1924}, and it was initially seen in eclipsing binary systems. 

The advent of extremely precise, high-resolution spectrographs allows us to measure and model the underlying line distortions themselves, using a technique dubbed Doppler Tomography \citep[DT,][]{collierDT2010}. In DT observations it is possible to see a traveling ``Doppler shadow'' in the stellar lines caused by the transiting planet obscuring sequential portions of the stellar surface \citep{Gaudi:2007}. The location at any given time of this shadow, relative to the line centers, is determined by the radial velocity of the stellar surface underneath the planet, and so given a measured stellar rotation velocity DT observations are capable of re-constructing the two-dimensional path of the planet across the stellar disk. This, in turn, allows us to determine the spin-orbit angle $\lambda$ between the on-sky projected rotation axis of the star and the orientation of the planetary orbit.

\exofasttwo\ allows the inclusion of either RM modeling or DT modeling. The inclusion of an RM fit adds the parameters $\lambda$ (star-planet spin-orbit angle) and $V\sin{I_*}$ (projected stellar rotation velocity) to our model, and is somewhat crude, relying on the \citet{Ohta:2005} approximation. This breaks down for very hot or rapidly rotating stars, but has the advantage that the signal is trivially treated as an anomalous radial velocity signal during transit.

Our DT model is more time-intensive than the RM model, but as mentioned it has the advantage of fitting the underlying planetary Doppler shadow directly -- rather than implicitly fitting the line distortions via their induced velocity changes in an RM model. Including a DT dataset will add the parameters $\lambda$, $V\sin{I_*}$, $V_{\mathrm{line}}$, and $\sigma_{DT}$ to our model, where $\lambda$ and $V\sin{I_*}$ are defined as above. $\sigma_{DT}$ is an error scaling term to ensure that the DT model is weighted appropriately relative to other constraints and $V_{\mathrm{line}}$ is the average line-width for the star without any rotational broadening, in units of velocity. $V_{\mathrm{line}}$ therefore includes effects from macroturbulence, thermal, and pressure broadening and is typically 2 to 10 $\mathrm{km\ s}^{-1}$.  

As input to the DT model, we expect that users will construct a properly formatted input file outside of \exofasttwo\ that is a two-dimensional \textsc{fits} array containing only the observations of the planetary shadow in velocity-space. To do so, for each observation in a DT dataset one should calculate the cross-correlation function (CCF) using a \textit{non-rotating} stellar template. The use of a non-rotating template is critical, since it makes the resulting CCFs effectively a composite line-profile for the entire spectrum. One should then subtract the out-of-transit averaged CCF from each observation, and also subtract the systemic velocity from each CCF's velocity coordinates. The resulting subtracted-CCFs should then be stacked to create a two-dimensional dataset with velocity on the x-axis and time on the y-axis. Ideally, this stacked 2D CCF will now only contain a signal from the planetary Doppler shadow.

The DT model in \exofasttwo\ is designed to only fit for this planetary shadow, and will not account for residual stellar contributions. To do so, we first determine the projected position of the planet on the stellar disk by using Equations 7 and 8 from \citet{collierDT2010}, and the perpendicular distance of the planet to the stellar rotation axis using Equation 10 from \citet{collierDT2010}. We note that there is a typo in Equation 10 from \citet{collierDT2010} as written: the two terms should be added together, rather than subtracted. Together with a value for $V\sin{I_*}$, the perpendicular distance of the planet to the stellar rotation axis allows us to calculate the center of the planetary Doppler shadow (in velocity-space) at any given time.

We next determine the shape of the planetary Doppler shadow. To do so, we first calculate the planetary shadow for the case of a non-rotating star, which we assume is a Gaussian with a velocity width of
\begin{equation}
    \sigma_\mathrm{non-rot} = \sqrt{V_{\mathrm{line}}^2 + V_{\mathrm{spec}}^2}, 
\end{equation}
where $V_{\mathrm{line}}$ is the average line-width for the star without any rotational broadening and $V_{\mathrm{spec}}$ is the line broadening in velocity-space caused by the finite spectral resolution of the observing spectrograph. As mentioned previously, $V_{\mathrm{line}}$ is a free parameter in an \exofasttwo\ DT fit, while $V_{\mathrm{spec}}$ is fixed to be $\mathrm{R}_{\mathrm{spec}}/c$. Note that if the measured $V\sin{I_*} < (\sigma_\mathrm{non-rot}/5)$, we judge that stellar rotation will be insignificant in setting the shape of the observed Doppler shadow, and so we use a Gaussian shadow with a width of $\sigma_\mathrm{non-rot}$. For a high-resolution spectrograph of $\mathrm{R}_{\mathrm{spec}}>30,000$ the typical value of $\sigma_\mathrm{non-rot}$ is a few $\mathrm{km\ s}^{-1}$, but at lower spectral resolutions the instrumental broadening can dominate both the inherent line width and the rotational broadening.

Assuming that stellar rotation is significant, we next calculate the rotational kernel for the stellar surface directly underneath the planet. To do so we use Equation 18.14 from \citet{gray2005} and assume that the limb-darkening parameter $\varepsilon=0$. The assumption of no limb-darkening is justified by the fact that here we are only considering the portion of the stellar surface obscured by the planet: whatever the global limb-darkening properties of the stellar disk, the small region obscured by the planet can be approximated as uniformly illuminated. Similarly, the velocity-width of the rotation kernel caused by the planet's obscuration \citep[$\nu_\mathrm{L}$ in the notation from][]{gray2005} is then
\begin{equation}
    \nu_\mathrm{L} = V\sin{I_*}\,\frac{R_P}{R_*}.
\end{equation}
To determine the final shape of the planetary Doppler shadow we then convolve this rotation kernel with a Gaussian with width $\sigma_\mathrm{non-rot}$.

We then calculate the amplitude of the planetary shadow. To do so, we note that the integral of the Doppler shadow at any given time must equal what we would see as the transit depth, at that time, for a photometric measurement using the passband of the observing spectrograph. We therefore normalize the Doppler shadow so that its area matches that of a $V$-band limb-darkened transit model for the planet in question. This additionally allows the \exofasttwo\ DT model to appropriately treat the rapidly changing Doppler shadow amplitude during transit ingress and egress. Note that we have specified the $V$-band for this amplitude calculation, to correspond to the wavelengths covered by most optical spectrographs. Observers modeling DT data at much different wavelengths will need to alter this bandpass in the \exofasttwo\ code itself to appropriately reflect their observations.

Finally, we subtract the resulting DT model from the DT observations and calculate the resulting $\chi^2$ fit statistic. We estimate the ``error'' on the DT observations by taking the standard deviation of the entire 2D data array, and scaling this so that $\chi^2/\mathrm{dof}=1$ when we subtract the median of the 2D data array. When we calculate the $\chi^2$ for a given DT model, we further scale the resulting $\chi^2$ value by the number of independent velocity measurements in the data. We do this to account for the fact that typical CCFs super-sample a spectrograph's actual instrumental resolution, and so the number of points in an individual CCF is not the actual number of independent velocity measurements. We calculate the number of independent velocities as
\begin{equation}
    N_\mathrm{indep} = \frac{V_{\mathrm{spec}}}{\Delta V_{\mathrm{CCF}}},
\end{equation}
where $\Delta V_{\mathrm{CCF}}$ is the velocity step of the CCFs. We then scale the $\chi^2$ from the DT model so that $\chi^2_\mathrm{DT} = \chi^2_\mathrm{raw}/N_\mathrm{indep}$. 

\section{Astrometry}
\label{sec:astrometry}

\exofasttwo \ is capable of fitting a full astrometric solution (including position, proper motion, parallax, and all orbital elements of each planet) to an arbitrary number of astrometric data sets, with an arbitrary number of planets (including 0), and with or without transits or RVs. 

While no planet has ever been definitively discovered via astrometry, \gaia \ Data Release 4, expected at the end of 2022, promises to change that in dramatic fashion, with estimated yields of 15,000 to 100,000 Jupiter-mass planets, depending on the planet occurrence rate and the ultimate mission duration \citep{Perryman:2014}. Importantly, the low end of their estimate assumes the nominal 5-year mission, which has since been extended at least two years\footnote{\url{http://sci.esa.int/director-desk/60943-extended-life-for-esas-science-missions/}}, and it excludes long-period planets that will not have at least a full orbit of observations. 

Nearly every currently known transit and RV detected exoplanet will have epoch astrometry from \gaia. While many previously known planets will have undetectable astrometric signals, of the nearly 2000 planets with sufficient information in the exoplanet archive\footnote{using the \citet{Chen:2017} mass-radius relation to translate radii to masses where missing}, over 500 (primarily RV-detected planets) will have astrometric signals larger than 1 $\mu as$ and periods less than 10 years, 25 of which are transiting.

Unlike transits, astrometry does not require an unlikely (edge-on) geometry for detection, and unlike RVs, we can measure the inclination and determine the true planet mass. While the baseline required for observation increases with semi-major axis, the expected signal grows, so astrometry probes a somewhat overlapping, but largely complementary sample of planets. DR4 will include partial data sets on countless systems, where a joint analysis can be particularly valuable. Combined with, e.g., a single transit from \tess, \kepler, or \ktwo, we will be able to determine significantly more about the system than with either data set alone. Since the transit probability scales as 1/$a$ and the astrometric signal scales as $a$, the number of transiting, long-period planets detectable by \gaia \ is roughly constant as a function of $a$. However, the relatively short \tess, \ktwo, and \kepler \ dwell times means the ones we detect as transiting falls off as 1/Period -- 0.7\% for a true Jupiter analog in a \tess \ sector. Nevertheless, if there is 1 long-period, $\sim$\mj \ planet per star, we can expect to find a few that transit within 200 pc necessary for a detection in \gaia.

The astrometric signal does not depend on \vsini \ or the stellar type to nearly the same degree as radial velocities or transits, allowing us to significantly expand the census of exoplanets around different types of stars. It does, however, have a strong selection effect against distant stars.

Astrometry allows us to resolve the $i$ and $180^\circ-i$ \ inclination degeneracy for transiting planets and measure the longitude of the node, $\Omega$ (and so properly constrain the mutual inclination of multi-planet systems). Astrometry alone cannot differentiate between the longitude of Ascending and Descending nodes ($\Omega$ and $\Omega + \pi$) -- essentially when the planet is coming toward us or moving away from us. However, when combined with either RVs or transits, that ambiguity is resolved. In addition, with an astrometric-only data set on multiple planets, the longitude relative to one fixed planet can be measured. Therefore, when astrometry is included, the inclination is reported from 0 to 180 degrees, and the impact parameter may be negative. When astrometry alone is fitted, $\Omega$ is reported from 0 to 180 degrees for the first planet, and its interpretation is the longitude of the node. All subsequent planets report $\Omega$ from 0 to 360 degrees, but there is a 50-50 chance all reported $\Omega$s are the longitude of the {\it descending} node. If RVs or transits are also fit, $\Omega$ is reported from 0 to 360 degrees for all planets and they are unambiguously the longitude of {\it ascending} node.

Astrometry provides an additional constraint on most of the orbital elements, providing independent confirmation of the physical nature of the signal and complementary constraints on the orbital parameters. That is, RVs measure the velocity in the Z direction, and astrometry measures the positions and velocities in X and Y directions. As a side note, \gaia \ DR4 will also supply a handful of low-precision ($\sim200$ m/s) RVs for many exoplanet targets, which will also help identify companions and false positives.

Even in the majority of cases where we expect the astrometric signal to be too weak for a detection, the all-sky nature of \gaia \ means it will provide robust upper limits on the mass and eccentricity for all detected bodies, as well as constrain the presence of outer bodies in the system, which are often invoked to explain the migration of Hot Jupiters. Finally, astrometry will greatly reduce the number of allowed false positive scenarios for transit and RV detected planets.

Fitting the astrometry directly avoids the need to supply a prior on the parallax to constrain the SED (\S \ref{sec:sed}). Supplied priors are assumed to be Gaussian and uncorrelated, but fitting the original data set removes that limitation. Correlations may exist which bias the parallax, especially if a companion is not identified and accounted for by the \gaia \ pipeline. This will be particularly important for data sets with less than 1 complete orbit, or marginal astrometric detections, neither of which will be modeled by the \gaia \ team.

The precise target coordinates and proper motion are important for determining the barycentric correction to the observed radial velocities \citep{Wright:2014}, and the absolute radial velocity can be used to determine the 3-space motion in the galaxy and constrain the age and population of the star through its kinematics, though we do not currently impose any such constraints within the fit.

The price of adding the astrometric data set to an RV or transit detection is a slight increase in the runtime to calculate the astrometric model and a handful of additional free parameters ($\alpha_{\rm ICRS}, \delta_{\rm ICRS}, \mu_\alpha, \mu_\delta, \Omega$). Computing the astrometric model is actually a subset of calculating the transit or RV models (see \S \ref{sec:planetpath}), and the data sets of at most few hundred points are small relative to \kepler, \ktwo, and \tess \ data sets. The extra parameters are robustly measured and largely independent of the planetary parameters. Therefore, it should not add significantly to either the calculation of a single model or the convergence time. In fact, to the extent the astrometric constraint can break degeneracies (e.g., for eccentricity or inclination), its inclusion may actually decrease the convergence time and therefore the total runtime.

The input format has been written around the expected format for \gaia \ -- that is, absolute astrometry as a function of \bjdtdb. Often, for non-\gaia \ data sets, we are given relative astrometry from a given epoch or separation and position angle as a function of time. At the moment, these formats are unsupported and require the user to convert them to something approximating the absolute astrometry. Further, we assume the companion is dark (e.g., a typical exoplanet). A companion that contributes significant flux to the system will bias the observed photocenter away from the position of host star and bias the measured semi-major axis. The solar system barycentric coordinates of the observer are required inputs to the astrometric model. When that observer is Earth (geocenter), the code can automatically calculate it based on JPL's DE405 ephemeris. When that observer is \gaia, the code can perform an automated telnet session to HORIZONS to calculate it. The latter requires {\tt horizons.exp} (included with \exofasttwo) and {\tt expect} to be in the user's path.

The astrometric model, at the time of publication, has not been rigorously tested. The stellar motion model follows the procedure described in \citet{Lindegren:2018}, which, as they state, does not correctly apply the contribution from the absolute radial velocity, and so does not apply to the nearest stars. From that, we approximate the contribution from each of the planetary orbits as an independent effect, simply added to the signal from the proper motion and parallax. This is another approximation that introduces an error that we believe is small, but has not been thoroughly investigated.

Generally, we expect the details of the implementation of the astrometric model within \exofasttwo, as well as the output files, to evolve leading up to and somewhat beyond \gaia \ DR4 (in $\sim$2022), and those changes are not likely to be backward compatible. Users relying on the astrometric capability should be prepared for that possibility over the next few years, and for that reason we do not provide specific examples of the current outputs here. 

While these are major caveats meaning any results including an astrometric fit currently should not be trusted without independent verification, its current implementation handles all the behind-the-scenes bookkeeping and can accurately recover the parameters used to generate a simulated data set (generated with the same, imperfect model). This is the majority of effort required to implement the astrometric model and eliminates the need for users to have an advanced understanding of the internals of \exofasttwo \ if they wish to improve upon it. Users wishing to validate the astrometric code, develop a more sophisticated astrometric model, or expand the allowed astrometric inputs are encouraged to contact the authors, but should only have to modify the {\tt EXOFAST\_ASTROM} and {\tt READASTROM} codes. At the moment, the astrometric modeling of \exofasttwo \ should be seen as a good starting point and a promise of what is to come.

\section{Combining data sets}
\label{sec:combining}

In the original \exofast \ (and in the customized version for KELT discoveries), we fit each data set individually first, scaled the uncertainties such that the probability of getting the $\chi^2$ that we did was 50\% (roughly $\chi^2_\nu = 1$), and then did a global fit of all data sets. This is a simplification that makes it easy to combine data sets because the $\chi^2$ from each data set can simply be added to one another, and the likelihood is proportional to $\exp{(-\chi^2/2)}$ (see \S 2.1 and \S 5.1 of \citet{Eastman:2013}).

However, this method ignores the uncertainty in the error scaling, it requires each data set to be able to independently and robustly constrain the model (meaning, for example, each RV instrument must have at least 5 data points for a circular orbit or 7 for an eccentric orbit), and the pre-fitting is both critical to get right (otherwise the errors will be inflated too much and the dataset unfairly de-weighted) and time consuming.

\exofasttwo \ fits the error scaling alongside of each data set (transit, radial velocity, SED, Doppler tomography, astrometry) to dynamically and properly weight each data set according to its true likelihood. This requires a more sophisticated treatment of the underlying likelihood, which includes the previously ignored (no longer constant) term in front.

We also note that, in addition to the user supplied data sets, we can potentially apply prior penalties from the user, MIST models (see \S \ref{sec:mist}), YY models (see \S \ref{sec:yy}), the Torres relations (see \S \ref{sec:torres}), SED models (see \S \ref{sec:sed}), limb darkening tables (see \S \ref{sec:limbdarkening}), exoplanet mass-radius relations (see \S \ref{sec:chen}), and for the ephemeris when TTVs are fit (see \S \ref{sec:ttvs}). And, we still want to be able to correct our priors using a Jacobian transformation (see \S 2.3 of \citet{Eastman:2013}). These constraints must all be weighted correctly, lest, for example, an overestimated likelihood in the MIST models unfairly dominates the transit-derived stellar density constraint.

In order to do this, we now compute the appropriately normalized likelihood, $\mathscr{L}$, of each constraint and multiply them together. However, in order to maintain backward compatibility with {\tt EXOFAST\_DEMC}, the ``$\chi^2$'' we calculate is actually
$-2\ln{\mathscr{L}}$. Throughout this paper and the code, we use the terms ``minimizing $\chi^2$'' and ``maximizing likelihood'' interchangeably, as a relic of \exofast's history. In all cases, the more precise term and intended meaning is ``maximizing likelihood.''

\section{Planetary Mass and Radius}
\label{sec:chen}
The original \exofast \ could fit either a radial velocity data set, a transit data set, or both. However, if a transit data set was not supplied, the planetary radius was not fit and all derived parameters were not estimated or displayed. Similarly, if a radial velocity data set was not supplied, \msini \ was not fit. And if both were not fit, the true planetary mass, \mplanet, and all derived parameters (e.g., planetary density) were not displayed.

\exofasttwo \ can now use the exoplanet mass-radius relations (and uncertainties) of \citet{Chen:2017} to derive a mass or radius of the planet. Notably, the RV semi-amplitude is included in the set of derived quantities. Therefore, when RVs are not fit, \exofasttwo \ can be used to estimate the RV semi-amplitude and therefore the amount of follow up effort required to obtain a true mass measurement.

While \citet{Chen:2017} include {\tt Forecaster}, a python package to derive the mass given the radius, marginalized over their posteriors, the code is too slow (and incompatible with IDL) to use directly. Instead, we take the median values for each of their fitted parameters (normalization, power-law slope, break point, and fractional dispersion), and add a $\chi^2$ penalty at each step in the MCMC chain.

By default, the relation is only applied if RVs and/or astrometry, or just transits are fit for a given planet (so the relation does not bias parameters that can be measured). If no RVs or astrometry is fit, the mass-radius relation constrains the mass. If no transit is fit, the mass-radius relation constrains the radius (while marginalizing over the planetary inclination, which is uniformly distributed in \cosi). This default behavior can be overridden at the command line if desired by specifying {\tt CHEN}, a boolean array corresponding to each of the planets. A ``1'' tells the code to use the relation and a ``0'' tells the code not to use relation for each of the corresponding planets.

For grazing transits, it is often the case that highly grazing, large companions are indistinguishable from barely grazing, smaller planets. However, RVs and the mass-radius relation is a natural way to realistically bound that degeneracy, as we did in \citet{Rodriguez:2018b} for the grazing inner-most planet, K2-266b.

While this constraint is extremely useful, it is important to distinguish between an extrapolated mass with large potential systematics and a directly measured mass using RVs or astrometry. The value of the latter will never be diminished, though masses estimated through the mass-radius relation may be sufficient for many purposes and may alleviate some of the crippling demand on precision RV instrumentation.

However, a major downside to this method is that, when used, all derived quantities inherit the assumptions and systematic errors of the \citet{Chen:2017} mass-radius relations. We warn the reader that this relation is relatively new, the low-mass ``planets'' that make up this relation are dominated by solar system bodies (and therefore may be biased by the solar environment), and has remarkably tight fractional dispersions -- $\sim4\%$ at $R_P < 2 \re$. Since all detection methods used in the \citet{Chen:2017} sample become easier with larger masses and they expressly exclude marginal ($<3 \sigma$) detections, it is likely that the relation is biased toward higher masses -- that is, the sample where nature scatters the planet mass low may avoid detection and artificially increase the inferred mass at a given radius while decreasing the observed scatter. Therefore, it is likely that the uncertainties on all quantities derived from the \citet{Chen:2017} mass-radius relation are underestimated, and values are overestimated at the low mass end. In addition, at brown-dwarf masses, the relation lacks sufficient data and does not actually reproduce the observed masses and radii of brown dwarfs. 

Despite these caveats and the existence of other relations that address some of these concerns \citep[e.g.,][]{Weiss:2014, Wolfgang:2016}, the \citet{Chen:2017} relation is currently the only mass radius relation with a continuous function spanning the entire range of potential companion masses, which is essential for integration into a generalized code like \exofasttwo.

Finally, even a perfect mass-radius relation is fundamentally multi-valued -- that is, a companion near a Jupiter radius can have up to three corresponding masses, and can create multi-modal distributions that are difficult to properly sample (though see \S \ref{sec:paralleltempering}).

Even with those significant caveats, we feel its advantages far outweigh its disadvantages. The impact on eccentricity due to the hill sphere that relies on the mass is typically small, and no other derived quantities impact the quality of the fit. It is mostly useful for a ballpark estimate of fundamental parameters that are often critical to estimate for follow up.

\section{Negative Planet Masses}
\label{sec:negmass}
Previously, \exofast \ fit in \logk \ to impose a more realistic planet mass prior that favors smaller planets. That is still the default in \exofasttwo. However, for a marginally detected planet or for calculating an upper limit, there is an infinite volume of nearly identical likelihood to explore at the small planet mass end (as \logk \ approaches negative infinity), requiring the user to supply some arbitrary cutoff to allow convergence (we impose a lower limit of $\logk \geq -6$), and imposing a Lucy-Sweeney-like bias on the inferred planet mass from the boundary at zero \citep{Lucy:1971}. We now optionally allow the user to fit in K directly, allowing it to be negative, similar to other codes like {\tt RadVel} \citep{Fulton:2018}. This is not physical and requires some special handling and non-obvious choices for several parameters and the global model. 

When the MCMC steps to a negative K, the RV model is inverted. This allows a smooth transition from positive to negative values of K. In terms of the RV model, this is identical to a phase shift of $\omega_*$ by $\pi$, which would swap the primary transit and occultation times. However, we do not change the transit time accordingly, which would effectively reject such steps when a transit is fit and re-introduce the Lucy-Sweeney like bias in K. This means the model we generate when K is negative is expressly non-physical. 

This degeneracy is broken when a transit or astrometry constrains $T_C$. When only RVs are fit, there is a discrete degeneracy between $K$, $\omega_*$ and $-K$, $\omega_* + \pi$, which, if properly sampled, would always lead to a median value for $K$ of 0. With typical MCMC samplers and a significant mass or eccentricity, it is practically impossible to sample this discrete degeneracy. However, for marginally detected planets (when this is most important), or with parallel tempering enabled (see \S \ref{sec:paralleltempering}), this is a serious risk that can unfairly make the detection appear less significant than it is and cause the efficiency of the MCMC to plummet. In such situations, the user should provide a loose, uniform bound on $T_C \pm P/4$ that will prevent the fit from swapping the primary transit and secondary occultation times, breaking the degeneracy similar to supplying transit or astrometric data, while not excluding viable solutions. If the RV data is so marginal that those bounds are encountered, the RV data set provides nothing more than an upper limit and this degeneracy is not relevant.

When K is negative, the planet mass is then calculated as if K were positive, then negative sign is applied. In this way, we can assess the significance of low-mass planets without imposing a bias by the physical boundary at zero. If there is a small tail that goes negative, we likely have a detection and we can be assured its mass is not biased high by the assumption that it is physical. If a significant fraction of the PDF is negative, the RVs probably represent an upper limit or systematic error, or it was fit without a transit and the starting value for $T_C$ was off by $P/2$.

The planetary surface gravity, \loggp \ is a derived quantity that is undefined for negative masses. Its reported median value and confidence interval (and all summary plots) excludes all negative entries, and the code will issue a warning.

When deriving the semi-major axis, $a$, and the hill sphere, we sum the planet and host mass as is, even if the planet mass is negative. If the total system mass is negative, we reject the step, since the semi-major axis would be undefined. For most systems where the companion might be consistent with a negative mass, the impact on $a$ from the companion should be negligible. To the extent it is not, and if a transit is also fit, it may eliminate some negative mass solutions through its impact on the transit duration.

When using the \citet{Chen:2017} exoplanet mass-radius relation, our input mass is always the greater of $10^{-10} \mearth$ (the size of a small asteroid) or the model planetary mass, since the \citet{Chen:2017} relation is not defined for negative masses. This limit was chosen arbitrarily to be so small it will never be hit with a real system. Practically speaking, if the \citet{Chen:2017} relation is used at all, it will effectively (and rightly) exclude negative masses.

\section{Transit timing, depth, and inclination variations}
\label{sec:ttvs}

The original \exofast \ assumed a strictly linear ephemeris. The modified version of \exofast \ used to fit most of the KELT planets, internally dubbed \multifast, allowed Transit Timing Variations (TTVs), but they completely decoupled the ephemeris from the transits, which required an iteration to first do the fit, fit a linear ephemeris to the transit times, and then impose that ephemeris as a prior to redo the fit. Further, no ability to allow Transit Depth Variations (T$\delta$Vs) or duration variations (TDVs) was included in either \exofast \ or \multifast.

\exofasttwo \ now enables the user to fit a non-physically motivated fudge factor to the transit times, transit durations, or transit depths. This allows the code to generate unbiased transit times and impact parameters for external analysis. When TTVs are enabled, the code fits a linear ephemeris to the transit times and adds a penalty for the deviation of the step's linear ephemeris from the best-fit linear ephemeris of the transit times, which automatically feeds the transit-based ephemeris constraint into the rest of the global model (e.g., RV curve, transit duration) while the fitted TTV parameter allows complete freedom to ensure the ephemeris does not bias the inferred transit times. 

An N-body code could replace the planet path calculation (see \ref{sec:planetpath}), but its serial nature is not well suited to IDL, and would dramatically increase the runtime of the code. Even then, the origin of TTVs and TDVs may also be due to astrophysical systematics (e.g., star spots), observational systematics (e.g., clouds) or different physical processes (e.g., tides, unmodeled planets). Eventually, we may add a {\tt TTVFast}-like model to include the effects of non-Keplerian motion and provide a constraint on the planetary masses \citep{Deck:2014}, but the utility of this fudge factor will never disappear due to the potential for unmodeled planets or non-astrophysical origins of transit variations. 

\begin{figure}[!htbp]
  \begin{center}
    \includegraphics[width=3.5in]{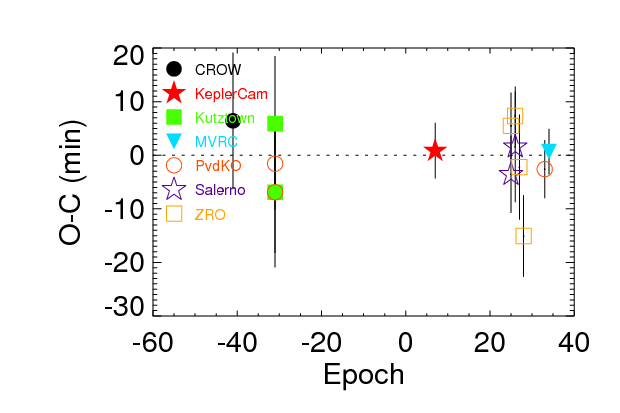} 
    \caption{The Transit Timing Variations (TTVs), refitting the discovery data of KELT-12b \citep{Stevens:2017} with \exofasttwo. Each symbol is a different telescope, and they are plotted as a function of epoch number. This publication-quality figure is a direct output of \exofasttwo. The telescope names and legends are automatically generated based on the filenames of the supplied transits.}
    \label{fig:ttv}
  \end{center}
\end{figure}

When TTVs are fitted, the results are plotted with the linear ephemeris subtracted, as in Figure \ref{fig:ttv}. When any variations are fitted, the ancillary output table containing the times of minimum projected separation, eclipse times, impact parameters, and transit depths for each transit with supplied data, is critical to include in any publication.

There are a few limitations and caveats with how we handle transit variations of which the user must be aware. For TDVs, we assume the root cause is a change in the orbital inclination. However, changes in $a$, $e$, or $\omega_*$ may also impact the observed transit duration. We do not currently include the ability to model duration variations as anything other than a change in inclination. Similarly, for T$\delta$Vs, we assume the root cause is a change in the planetary radius. This may be the correct physical model when the optical depth of the planetary atmosphere varies as a function of bandpass, but less so when the depth variations are due to something like variability in the spot covering fraction of the star. Generally, variations will be small and the potential systematic errors introduced by using the wrong physical interpretation will be negligible, but this is something for users to keep in mind.

Because variations could be due to {\it observational} systematic errors, they are fit to each transit data file (not to each transit epoch). Therefore, if users wish to model TTVs, TDVs, and/or T$\delta$Vs, each transit must be in a separate input file. There are two downsides to this implementation. First, simultaneous transits of two or more planets cannot be modeled with any variations, since each planet would be forced to have the same variation. Second, we also fit a new baseline flux ($F_0$) and added variance ($\sigma^2$) to each file, adding many additional free parameters that can make the fit less robust and take longer.

Ideally, we would allow the user to fit variations by epoch, by input file, or both, and epoch variations could be enabled without the need to generate separate light curves for each. However, this would require a major architectural change to the code, and that is unlikely to be implemented. A more likely improvement we may include in the future is to add the ability to link parameters in the prior file so the fitted baselines and error terms could be combined across files where they are expected to be similar. Until then, the user may wish to fix the baseline flux and added variance for each transit file to the the median value found from a fit without any variations, essentially ignoring the uncertainty in those parameters. While not ideal, the effect should be minimal.

If the user wishes to model a multi-planet system, some planets with variations, and some without, the variation parameter must be fixed to zero for each data file where variations are not desired. Transits for any number of planets without variations can be combined into a single data file provided the baseline flux and added variance are approximately equal for all data.

\section{Detrending}
\label{sec:detrending}

Like the original, \exofasttwo \ can fit an arbitrary number of detrending parameters for each transit, simply by including additional columns in the transit file to detrend against. The number of detrending parameters can vary with each transit. The original only included additive detrending, which is appropriate for things like sky background. Additive detrending is a decent approximation to multiplicative detrending in most regimes, but does introduce a systematic error proportional to its magnitude times the depth of the event for multiplicative effects like airmass or meridian flips. We now allow an arbitrary number of additive ($C_j$) or multiplicative ($M_j$) detrending parameters, which are distinguished by an optional header in the transit file. Any detrending column that has a corresponding header beginning with a capital ``M'' is modeling with multiplicative detrending. Otherwise (or if left blank), all detrending columns are modeled as an additive effect.

The lightcurve model at the $i$th time, with detrending, is then:

\begin{equation}
    Model_i = \left(Transit_i + \sum_{j=1}^{n_a} C_j d_{i,j}\right)\left(F_0 + \sum_{j=1}^{n_m}M_j d_{i,j}\right)
\end{equation}

\noindent where ``Transit'' is the quadratically limb darkened transit model given in equation \ref{eq:modellcall}, $F_0$ is the baseline flux, $n_a$ is the number of additive detrending parameters, $n_m$ is the number of multiplicative detrending parameters, and $d_{i,j}$ is the detrending value corresponding to the $j$th detrending parameter at time $i$.

Before fitting each user-supplied detrending column, we zero-average it and normalize it by the maximum value in the column. This ensures the values in each column are in the range -1 to 1 and the value of each coefficient roughly corresponds to the same magnitude of correction. Previously, normalizing the detrending columns was not done, making it difficult for our minimization algorithms to find small coefficients corresponding to large numbers like sky background. This generally makes the fits better behaved, but the user should not use starting values on these coefficients identified with \exofast.

\subsection{Keplerspline}

In addition to the detrending above, we allow the user to request a spline detrending of any subset of lightcurves. After detrending as above, we multiply our model by a spline fit to the residuals using {\tt KEPLERSPLINE} from \citet{Vanderburg:2014}. As the name would imply, it was written with \kepler \ data in mind and is intended to remove long-term stellar variability and instrumental artifacts from long, continuous data sets. It also works well for \tess \ data. The user may specify the knot spacing of the spline fit, which should be much larger than the expected transit signal (by default, 0.75 days), and break points where separate spline fits are applied to account for expected discontinuities in the data (e.g., momentum dumps in \tess). The break points are specified as an optional second header line of the transit file as a white-space delimited list of Julian dates. Gaps in the data larger than the knot spacing are automatically identified as break points.

Including this step inside the fitting allows the user to include the uncertainty introduced by the lightcurve flattening, which is useful for assessing the significance of weak signals and the impact the flattening procedure may have on the planet parameters. This is especially important when a transit coincides with a discontinuity in the lightcurve and is therefore particularly sensitive to the flattening method. However, it is also slow. Not only is the procedure itself computationally expensive (generally slightly more expensive than the rest of the transit model calculation), but it requires a much longer out-of-transit baseline to robustly compute.

\section{Dilution}
\label{sec:dilute}

The light curve may be blended with nearby stars, diluting the transit signal and making the planet appear smaller than it actually is. This dilution can be modeled, recovering the intrinsic planet properties, with the command line option {\tt FITDILUTE}, a string array that specifies which bands should include a blending term. The dilution of a light curve is highly degenerate with the transit depth, so it is wise to include a prior on the dilution in each band derived from high resolution imaging. Otherwise, the uncertainty in the (unblended) transit depth -- and the corresponding planet radius -- will be large.

While a negative dilution is non-physical, many light curves have the dilution already corrected (e.g., the TESS and Kepler light curves) and an over-corrected dilution is possible. Even if the light curve has not had a dilution correction applied, we do not want to bias the fitted dilution to positive-only values. Therefore, we allow negative values for the dilution, which effectively take flux away from the star, making the transit depth deeper than measured.

Future enhancements may add the ability to model multiple SEDs which would also enable us to self-consistently constrain the dilution in the light curves, but currently there is no connection between the dilution term here (used only for the transit light curve), and any potential dilution in the broadband photometry used to constrain the SED model.

Note that the dilution is often given as a contrast ratio, $C$, between the combined flux of all contaminating sources, $F_2$, and the flux of the target star, $F_1$, $C=F_2/F_1$. However, the dilution in \exofasttwo, $D$, is the fractional contribution from the neighboring stars, $D=F_2/(F_1+F_2)$. We can convert between the two by computing $D = C/(1+C)$. 

\section{Long exposure times}
\label{sec:longcadence}

Originally inspired by Kepler long cadence data, we now allow the input of an arbitrary exposure time and the number of model data points to average over that span in order to account for the smearing of the light curve due to long exposure times. 

We also include a {\tt LONGCADENCE} flag for Kepler long cadence data that automatically sets the exposure time to 29.425 minutes and averages 10 model points for each data point. We allow a boolean array for these values so the user can simultaneously fit a mix of long cadence data with, for example, high cadence ground-based follow up. 

The \tess \ postage stamps are 2 minutes and need no integration, but the full frame images during the prime mission are 30 minutes. Since \tess \ is likely to change its full frame exposure time in the extended mission, providing a similar ``longcadence'' flag for \tess \ might lead to confusion, and the difference between the 29.425 minute \kepler \ integrations and the 30 minute full frame \tess \ integrations is $\pm 40$ ppm for a typical Hot Jupiter -- an error that is comparable to \tess's systematic floor and four times the numerical integration error with the default settings and a typical hot Jupiter. For the highest accuracy, users modeling \tess \ data should specify the \tess \ exposure times -- 30 minutes for the full frame images in the primary mission -- and the number of data points to average (we recommend 10) explicitly, and avoid the use of the {\tt LONGCADENCE} flag. 

We use a simple Riemann sum to integrate the model over the exposure. Trapezoidal integration is less accurate, and Simpson integration requires twice as many model evaluations for the same resolution. For the same number of model evaluations, a Simpson integration was less accurate than a Riemann sum, too.

In the original \exofast, our evaluated model points were equally spaced between the beginning and end of the exposure, but that biased the integration toward the boundaries between exposures, which get double counted by the exposures on either side. Now, following the recommendation of \citet{Kipping:2010} and equivalent to a midpoint Riemann sum, our evaluated model points are equally spaced between the center of the exposure $\pm ((N-1)/2N)*T$, where $N$ is the number of sampled points (10 by default) and $T$ is the exposure time. This spacing yields uniform model evaluations between adjacent exposures, assuming no overhead (i.e., the exposure cadence is equal to the exposure time).

\section{Priors}
\label{sec:priors}
Priors can now be specified on any fitted or derived parameters within the configuration file. The configuration file simply lists the parameter name and one to four values after it. If only one value is specified, the default starting value for that parameter is overridden, but no likelihood penalty is imposed. Tweaking the starting values for a handful of parameters is typically required and is the least invasive way to fix a fit that is not working.

The second value, if supplied, is the uncertainty for a Gaussian prior. During each step of the fit, a $\chi^2$ penalty is imposed equal to $((value-prior)/uncertainty)^2$. If the uncertainty is negative, it is ignored and no penalty is imposed. If the uncertainty is zero, the parameter is fixed at the supplied value. However, fixing values is generally not recommended because the uncertainties of any covariant parameter would then be underestimated. It is typically strongly preferred to include an accurate uncertainty. If third and fourth values are supplied, they are lower and upper bounds, respectively. All models that venture beyond those bounds are given a zero likelihood and therefore are immediately discarded. Lower and upper bounds equal to "-Inf" or "Inf", respectively, are ignored. Therefore, one can impose Gaussian or uniform priors on any fitted or derived parameter from this configuration file. The user can never override the hard boundaries within \exofasttwo, which are typically conservative physical boundaries and sometimes limits of the applicability of certain models (and summarized in Table \ref{tab:explain}). For example, the inherent boundary that $0 \leq e < 1-\frac{R_*+R_P}{a}$ only allows physical eccentricities where the star and planet would not collide during periastron. User supplied bounds of -Inf and 1 would have no impact on the fit at all. The configuration file supports comments to help the user experiment with several different options. 

Both uniform and Gaussian priors should be physically and independently motivated. Priors derived from the data themselves (e.g., a period prior from a BLS search of the light curve being fit), or from fits of similar analysis (e.g., a prior on the stellar properties derived from an SED fit while including an SED model) will double count the constraint and underestimate the final uncertainties. The exception to this rule is, especially when using parallel tempering, it is wise to include wide, uniform priors on $T_C$ and $Period$ (based on the data) to prevent the hotter chains from exploring an infinite volume of parameter space, which may slow or even prevent convergence. The user should always check the posteriors of such fits to ensure the boundary did not impact the posterior. Imposing tighter uniform bounds on some parameters may be desired when fitting low signal to noise transits that may only be marginally significant (otherwise, the fit may not work at all). However, the user must understand that doing so artificially inflates the significance of the transit itself, and an independent method must be used to evaluate whether the transit signal is real or just a systematic blip.

If the user is applying a boundary or especially a starting value on an uncommon derived parameter (e.g., Safronov Number), it is wise to confirm it is being applied as expected. There are many parameters to derive and many (sometimes conflicting) ways to derive them. Many such combinations have not been included, but are trivial to add on a case by case basis. When in doubt, and if possible, apply priors or change the starting values directly on the fitted parameter, which is guaranteed to be propagated correctly.

\section{Parameterization}
\label{sec:parameterization}

\exofasttwo \ uses mostly the same parameterizations of the transit, radial velocity, and the star as the original \exofast, with a couple notable exceptions. Instead of stepping in $\log(\ar)$, we now step in $\log(\mstar)$. This allows us to more generally handle the addition of multiple planets, since the stellar mass, stellar radius, Kepler's law, and the planetary period are enough to uniquely derive each planet's \ar. If each planet had its own fitted \ar, the system would no longer be self-consistent, nor would we be able to harness the combined stellar density constraint from all planet transits. Additionally, instead of stepping in \logg, we now step in the EEP of the star, which constrains the age of the star (see \S \ref{sec:mist}). If Yonsie Yale stellar evolutionary models are used \citep{Yi:2001}, we step directly in age (see \S \ref{sec:yy}).

The major downside of physical parameterizations is that they are often highly covariant, and this is no exception. For example, almost by construction, strong covariances exist between the stellar parameters and the orbital inclination (both of which impact the duration of the transit). This tends to make the AMOEBA minimization take a long time, but it is still a small fraction of the total runtime. Fortunately, as long as the covariances are linear (not curved or discrete), the DEMC algorithm \citep{terBraak:2006} naturally adapts the step sizes and directions to efficiently explore the complicated covariances between parameters. Somewhat more problematic, AMOEBA often has difficulty with these covariances, especially as the number of fitted parameters increases. In order to get a well-behaved fit, it is often necessary to start very close to the best-fit. The MCMC can often make up for small deficiencies in the AMOEBA fit, but it will converge slower. We have included a code called {\tt MKPRIOR}, which recreates the user's prior file based on the best-fit model found by a preliminary MCMC run. This can be used iteratively to start at the best-fit value and may help converge faster than letting it run uninterrupted.

Several difficulties arise with periodic parameters and MCMC algorithms. There are two common periodic parameters: angles (e.g., $\omega_*$, $\lambda$) and times (e.g., $T_C$). Formally, periodic parameters can never converge, because the likelihood is identical at integer multiples of the period. Even in practice, it is sometimes possible for multiple chains to get stuck in separate minima, making it practically impossible to converge. This can be solved by modifying the step such that it is always within one period. Note that this is not the same as rejecting steps outside of the period, since the latter would set up an artificial bi-modal distribution and make it difficult to sample both sides of the (artificial) boundary. 

For periodic parameters with likely values that straddle that artificial boundary, further complications arise. The reported values are the median of the distribution and the 68\% confidence interval encloses 34\% of the probability on either side of the median. However, the choice of the boundary can shift the median value arbitrarily. Further, the convergence criteria become less reliable when there is significant power that is widely separated.

Applying priors to periodic parameters is also problematic. If the prior is near a periodic boundary, the application of the prior must have some knowledge of that boundary, lest values on the opposite side of the boundary get unfairly penalized. For example, an angle in the interval [-180,180) degrees with a prior of $-175 \pm 10$ should penalize the value of $175$ with a $\chi^2$ penalty of 1, since 175 degrees is only 10 degrees from -175 degrees. However, an implementation ignorant of the periodicity would see the difference as 350 degrees and impose a penalty of $\chi^2$ penalty of 1225, strongly disfavoring it.

Thus, any angular parameter must be flagged as such in order to be subject to these special rules. This is done by setting its units to ``radians'' or ``degrees'' and is already handled transparently by \exofasttwo \ for all applicable parameters. Note that any directly fitted parameter must be in radians for the proper calculation of the convergence criteria.

All of the same arguments apply to any periodic parameter, such as $T_C$. But in practice, this is almost always constrained to much much better than its periodicity. We apply a bound that $T_C$ must be within half of the current step's period of its starting value. We recommend the user bound the period loosely to avoid convergence problems, especially when using parallel tempering. 

By and large, we avoid these complications with periodic parameters by our choice of parameterization. The inclination is parameterized as $\cos{i}$ and $\omega_*$ is parameterized as \secosw \ and \sesinw, which covers most fits.

However, we now include the option to fit the Rossiter-McLaughlin effect \citep{Rossiter:1924,McLaughlin:1924} or Doppler Tomography (see \S \ref{sec:dt}), both of which include the angular parameter $\lambda$, the projected spin-orbit alignment angle. In order to avoid using $\lambda$ directly, we considered parameterizing it as $\sqrt{\vsini}\cos{\lambda}$ and $\sqrt{\vsini}\sin{\lambda}$, analagous to our parameterization of $e$ and $\omega_*$, but that loses generality with multiple planets, which has one $\vsini$ for the star but a $\lambda$ for each planet.

When fitting astrometry, $\Omega$ is another angular parameter subject to these rules. Technically, the Right Ascension, $\alpha_{\rm ICRS}$, and Declination, $\delta_{\rm ICRS}$, are also angular parameters, but the uncertainty will never be large enough to encounter this type of confusion while still being meaningful to include in the fit. We fit $\alpha_{\rm ICRS}$ and $\delta_{\rm ICRS}$ in degrees.

\section{Derivation of the RV, transit, and astrometric models}
\label{sec:planetpath}

The derivation of the models is similar to the derivation described in \S4.3 of \citet{Eastman:2013} for \exofast. However, the handling of multiple planets and astrometry requires additional complexity. Further, many of the details in that derivation did not follow the standard convention owing to widespread confusion as to what exactly the standard convention is. To be clear, the exoplanet community has inherited a left-handed coordinate system from the stellar binary community to describe the planetary orbital motion. A left-handed coordinate system is required to self-consistently recreate the textbook definitions of $\omega$, $\Omega$, and the ascending node. The exoplanet community almost always quotes the argument of periastron of the stellar orbit due to each planet in this left-handed coordinate system, but are rarely so explicit. Unfortunately, most modern textbooks describe orbital motion in a right-handed coordinate system, which has become the standard in physics and mathematics more broadly. Doubly unfortunate, the right-handed $\omega$ and $\Omega$ each differ by $\pi$ from their standard left-handed counterparts, requiring some amount of behind-the-scenes fudging to reconcile them with their textbook definitions.

To compute the Cartesian coordinates of an orbit, we start with an orbit in the $X_M-Y_M$ plane, oriented such that periastron is in the $+X_M$ direction, at a distance $r = \frac{a(1-e^2)}{1+\cos{\theta}}$. The object orbits counter-clockwise as time (and the true anomaly, $\theta$) grows. In this model coordinate system, denoted by the subscript ``M'',

\begin{equation}
  \label{eq:xyzm}
  \begin{split}
    X_M =& r\cos{\theta} \\
    Y_M =& r\sin{\theta} \\
    Z_M =& 0.
  \end{split}
\end{equation}

We can transform that orbit into the observed frame through a series of rotations of the three Euler angles: $\omega$, $i$, and $\Omega$. By convention, first by $\omega$ in the $X_M$-$Y_M$ plane:

\begin{equation}
  \label{eq:xyzm1}
  \begin{split}
    X_M\prime =& r\left(\cos{\theta}\cos{\omega} - \sin{\theta} \sin{\omega}\right) \\
    Y_M\prime =& r\left(\cos{\theta}\sin{\omega} - \sin{\theta} \cos{\omega}\right) \\
    Z_M\prime =& 0, 
  \end{split}
\end{equation}

\noindent where the prime denotes an intermediate model coordinate system. Using trigonometric identities, this simplifies to

\begin{equation}
  \label{eq:xyzm1simp}
  \begin{split}
    X_M\prime =& r\cos{\left(\theta+\omega \right)} \\
    Y_M\prime =& r\sin{\left(\theta+\omega \right)} \\
    Z_M\prime =& 0.
  \end{split}
\end{equation}

\noindent Then we rotate by $i$ in the new $Y_M\prime$-$Z_M\prime$ plane, which is relatively simple since $Z_M\prime$ = 0:

\begin{equation}
  \label{eq:xyzm2}
  \begin{split}
    X_M\prime\prime =& r\cos{\left(\theta+\omega \right)} \\
    Y_M\prime\prime =& r\sin{\left(\theta+\omega \right)}\cos{i} \\
    Z_M\prime\prime =& r\sin{\left( \theta+\omega \right)}\sin{i},
  \end{split}
\end{equation}

\noindent where the double prime denotes the second intermediate model coordinate system. Finally, we rotate by $\Omega$ in the new $X_M\prime\prime$-$Y_M\prime\prime$ plane, and we drop the subscripts for our final, observed frame:

\begin{equation}
  \label{eq:xyz}
  \begin{split}
    X =& r\left(\cos{\left(\theta+\omega \right)}\cos{\Omega} - \sin{\left(\theta+\omega \right)}\sin{\Omega}\cos{i}\right) \\
    Y =& r\left(\cos{\left(\theta+\omega \right)}\sin{\Omega} + \sin{\left(\theta+\omega \right)}\cos{\Omega}\cos{i}\right) \\
    Z =& r\sin{\left(\theta+\omega \right)}\sin{i}
  \end{split}
\end{equation}

Equation \ref{eq:xyz} for the Cartesian coordinates of an orbit are derived from a standard application of rotation matrices and versions of this derivation can be found in countless references.

If one wants to compute the stellar orbit, use the stellar parameters and if one wants to compute the planetary orbit, use the planetary parameters. And, if one wants to compute the heliocentric coordinates of the planet (instead of the barycentric coordinates), use $a_{\rm tot}$ and the planetary parameters. The stellar and planetary orbits are simply related:

\begin{equation}
    \label{eq:elements_planet_vs_star}
    \begin{split}
        a_{\rm tot} &= a_* + a_P \\
        a_* &= a_P\frac{m_P}{M_*}\\
        e_* &= e_P \\
        P_* &= P_P \\
        i_* &= i_P\\
        \Omega_* &= \Omega_P\\
        \omega_* &= \omega_P + \pi.
    \end{split}
\end{equation}

Until now, the lack of subscripts in eqs \ref{eq:xyzm} through \ref{eq:xyz} imply generality. As we move forward, we will be explicit about the object by including subscripts wherever they differ. We will drop the object subscripts for $i$, $e$, $P$, and $\Omega$, because they are identical for the star and planet.

While $+X$ is always North and $+Y$ is always East for exoplanets, our convention that $+Z$ grows with distance makes it a left-handed coordinate system and will lead to important differences compared to the derivation described in \citet[e.g.][]{Winn:2010} and \citet{Eastman:2013}. Note, however, that all the equations in \citet{Winn:2010} describing important derived quantities (e.g., transit duration, impact parameter) should be computed using the standard reported value of $\omega_{*}$ derived in the left-handed coordinate system, which is numerically equal to the value of argument of periastron of the planetary orbit in the non-standard right-handed coordinate system which they use in the text\footnote{Since the left-handed $\omega_*$ is equal to the right-handed $\omega_* + \pi$ and $\omega_* = \omega_P + \pi$, the left-handed $\omega_*$ is equal to the right-handed $\omega_P$}.

\exofasttwo \ uses $T_{C,j}$, $\sqrt{e_j}\cos{\omega_{*,j}}$, $\sqrt{e_j}\sin{\omega_{*,j}}$, $\log{P_j}$, $\log{K_j}$, $\log{\mstar}$, and \rstar \ as parameters, where the $j$ subscript denotes the $j$th planet. We note that $\omega_{*,j}$, while a measure of the stellar orbit, has a subscript for each planet. This is not a typo --  we use the argument of periastron of the star's orbit imparted from each planet. This is the convention because $\omega$ is most often constrained from radial velocities, which measures the stellar motion.

The first step is to convert those to the parameters required to generate the model. Using trigonometric identities, we derive the eccentricity:

\begin{equation}
\label{eq:e}
e_j = \left(\sqrt{e_j}\cos{\omega_{*,j}}\right)^2 + \left(\sqrt{e_j}\sin{\omega_{*,j}}\right)^2
\end{equation}

\noindent \ and argument of periastron:

\begin{equation}
\label{eq:omega}
\omega_{*,j} = atan2(\sqrt{e_j}\sin{\omega_{*,j}},\sqrt{e_j}\cos{\omega_{*,j}}).
\end{equation}

\noindent While the arctangent typically only maps angles to the range $-\pi/2$ to $\pi/2$, the mathematical function {\tt atan2} allows us to map the angle to the full range from $-\pi$ to $\pi$ when the sign of both the numerator and denominator are independently known.

While the time of minimum projected separation, $T_{T,j}$ is defined such that $\sqrt{X_P^2 + Y_P^2}$ is minimized while the planet is between its host star and the Earth, solving for that time is computationally expensive. Instead, we, along with many in the community, have adopted the time of conjunction, $T_{C,j}$, as the reference time, defined such that $X_M\prime\prime=0$ (or $X=0$ when $\Omega = 0$). It is critical to understand that this definition of $T_{C,j}$, used by many codes as ``the transit time,'' is not generally equal to $T_{T,j}$. This is discussed at length in section \ref{sec:tc}, but the important takeaway is that, while they are often assumed to be the same, $T_{C,j}$ and $T_{T,j}$ can differ by up to 10 minutes for inclined, eccentric, single-planet systems and they can differ arbitrarily in systems with TTVs.

Solving equation \ref{eq:xyzm2} for the planetary orbit's true anomaly at conjunction, $\theta_{C,j}$, where $X_M\prime\prime=0$ yields two solutions: $\theta_{C,j} = \pi/2 - \omega_{P}$ and $\theta_{C,j} = -\pi/2 - \omega_{P}$ (the transit and occultation). In the standard left-handed coordinate system, when the planet's position along the line of site, $Z_P$, is negative, the planet is between its host star and earth, and so $\theta_{C,j} = -\pi/2 - \omega_{P,j}$ at transit. Because we parameterize the model in terms of $\omega_{*,j}$, we will make that substitution using eq \ref{eq:elements_planet_vs_star}: 

\begin{equation}
    \label{eq:trueanomtc}
    \theta_{C,j} = \pi/2 - \omega_{*,j}.
\end{equation}

We must derive the time of periastron, $T_{P,j}$ to compute the model. We plug the true anomaly into the equation for the eccentric anomaly:

\begin{equation}
  \label{eq:eccenanom0}
  E_{j}(T_{C,j}) =  2\arctan{\left[\sqrt{\frac{1-e_j}{1+e_j}}\tan{\left(\frac{\pi/2 - \omega_{*,j}}{2}\right)}\right]}. 
\end{equation}

\noindent The equation for the mean anomaly in terms of the eccentric anomaly is:

\begin{equation}
  \label{eq:meananom0}
  M_j(T_{C,j}) = E_j(T_{C,j}) - e_j\sin{(E_j(T_{C,j}))},
\end{equation}

\noindent from which we solve for $T_{P,j}$:

\begin{equation}
\label{eq:meananom1}
T_{P,j} = T_{C,j} - M_j(T_{C,j})\frac{P_{j}}{2\pi}.
\end{equation}

Next, we work our way backwards through equations \ref{eq:meananom1}, \ref{eq:meananom0}, and \ref{eq:eccenanom0} to derive the true anomaly for each input time, $t$. However, we must first correct $t$, which is in the solar system barycentric frame (\bjdtdb), to the target's barycentric frame -- the time the photon would have left the barycenter of the target system, $t_j$. This per-planet step is important in reconciling the timing of events happening in vastly different locations. Skipping this step would cause a non-physical offset between primary and secondary eclipses (of $2a_{P,j}/c$) or between the primary transit and the time of transit implied by the RV or astrometric data (of $a_{P,j}/c$). Ignoring this would bias the planetary eccentricity to match the observed data. Even a single transit would be compressed in the first half and stretched in the second half due to the change in arrival times throughout the transit, though this effect is typically negligible. We use the routine {\tt BJD2TARGET} to make this conversion, which only accounts for the light travel time. We ignore relativistic effects in both the solar system (the TCB coordinate system) and the target system, as well as the constant offset from the light travel time between solar system barycenter and the target barycenter. Then, we derive the mean anomaly as a function of the target's barycentric time:

\begin{equation}
\label{eq:meananom2}
M_j(t_j) = 2\pi(t_j-T_{P,j})/P_j.
\end{equation}

\noindent Next, we derive the eccentric anomaly: 

\begin{equation}
  \label{eq:meananom}
  M_j(t_j) = E_j(t_j) - e_j\sin{(E_j(t_j))}.
\end{equation}

Unfortunately, equation \ref{eq:meananom} cannot be inverted analytically. Solving this equation robustly and quickly has been the subject of significant research \citep[e.g.][]{Tommasini:2018} and dominates the total runtime of the transit model calculation. Our approach has not changed since the original \exofast.

Then, we derive the true anomaly:

\begin{equation}
  \label{eq:trueanom0}
  \theta_j(t_j) = 2\arctan{\left[\sqrt{\frac{1+e_j}{1-e_j}}\tan{\left(\frac{E_j(t_j)}{2}\right)}\right]}.
\end{equation}

The true anomaly for each input time, coupled with the RV semi-amplitude, is all that is required for the radial velocity model.

\begin{equation}
    \label{eq:rv}
    RV_{*,j} = K_j\left(\cos{(\theta_j(t_j)+\omega_{*,j})} + e_j\cos{\omega_{*,j}}\right).
\end{equation}

\noindent Then we sum the contribution from each planet, and add the zero point ($\gamma$), and if desired, a slope, $\dot{\gamma}(t-t_0)$, and quadratic term $\Ddot{\gamma}\left(t-t_0\right)^2$, where $t_0$ is chosen as the average of the first and last RV data points' observation time.

For the transit and astrometric models, we must compute the projected paths of the planet across the star and the star across the sky, respectively. Before we continue, we must compute the masses of each planet ($M_{P,j}$) and compute the semi-major axes of both the star's orbit due to that planet ($a_{*,j}$) and the semi-major axis of the planet's orbit ($a_{P,j}$).

To compute the planet mass, we start with model parameter $K_j$, the RV semi-amplitude:

\begin{equation}
  \label{eq:KP}
  K_j = \left(\frac{2 \pi G}{P_j(M_{*} + M_{P,j})^2}\right)^{1/3}\frac{M_{P,j}\sin{i_j}}{\sqrt{1-e_j^2}},
\end{equation}

\noindent and solve the cubic for $M_{P,j}$. Then we solve Kepler's equation for the semi-major axis $a_{\rm {tot}, j}$,

\begin{equation}
\label{eq:kepler-newton}
    a_{\rm {tot}, j} = \left(\frac{GP_j^2(\mstar+M_{P,j})}{4\pi^2}\right)^{1/3},
\end{equation}

\noindent and compute the two components of the semi-major axis separately by solving the system of equations:

\begin{equation}
    \label{eq:com}
    \begin{split}
      a_{\rm {tot},j} &= a_{*,j} + a_{P,j} \\
      a_{*,j} \mstar &= a_{P,j} M_{P,j}.
    \end{split}
\end{equation}

Finally, we can apply equation \ref{eq:xyz} to determine the stellar orbit:

\begin{equation}
  \label{eq:xyzstar}
  \begin{split}
    r_{*,j} =& \frac{a_{*,j}}{d}\frac{(1-e_j^2)}{1+e_j\cos{\theta_j(t)}} \\
    X_{*,j} =& r_{*,j}\left(\cos{\left(\theta_j(t_j)+\omega_{*,j} \right)}\cos{\Omega_{j}}\right. \\
             & \left.- \sin{\left(\theta_j(t_j)+\omega_{*,j} \right)}\sin{\Omega_{j}}\cos{i_j}\right) \\
    Y_{*,j} =& r_{*,j}\left(\cos{\left(\theta_j(t_j)+\omega_{*,j} \right)}\sin{\Omega_{j}}\right. \\
             & \left.+ \sin{\left(\theta_j(t_j)+\omega_{*,j} \right)}\cos{\Omega_{j}}\cos{i_j}\right) \\
    Z_{*,j} =& r_{*,j}\sin{\left(\theta_j(t_j)+\omega_{*,j} \right)}\sin{i_j}, 
  \end{split}
\end{equation}

\noindent where $d$ is the distance to the object. We repeat equation \ref{eq:xyzstar} for each planet and sum the contribution from all planets. This is the total stellar reflex motion, in radians, in each Cartesian direction:

\begin{equation}
  \label{eq:reflex}
    \begin{split}
    X_{*,\rm tot} =& \sum_{j=1}^{N_{\rm Planets}} X_{*,j} \\
    Y_{*,\rm tot} =& \sum_{j=1}^{N_{\rm Planets}} Y_{*,j} \\
    Z_{*,\rm{tot}} =& \sum_{j=1}^{N_{\rm Planets}} Z_{*,j}.
    \end{split}
\end{equation}

\noindent For the astrometric model, we simply add that to the stellar angular motion due to parallax and proper motion. In our approximation, the star is far from the observer, so $Z_{*,{\rm {tot}}}$, the motion of the star in the direction of the observer, is not used.

For the transit model, we now repeat equation \ref{eq:xyz}, using the planetary parameters and $a_{\rm{tot},j}/R_*$ to determine the heliocentric coordinates normalized by the stellar radius. Note that we substitute $\omega_{*}$ for $\omega_{P}$ and flip the signs of $X_{\rm{tot},j}$, $Y_{\rm{tot},j}$, and $Z_{\rm{tot},j}$ to avoid introducing another parameter: 

\begin{equation}
  \label{eq:xyzp}
  \begin{split}
    r_{\rm{tot},j} =& \frac{a_{\rm{tot},j}}{R_*} \frac{(1-e_j^2)}{1+e_j\cos{\theta_j(t_j)}} \\
    X_{\rm{tot},j} =& -r_{\rm{tot},j}\left(\cos{\left(\theta_j(t_j)+\omega_{*,j} \right)}\cos{\Omega_{j}}\right. \\
                    &\left.- \sin{\left(\theta_j(t_j)+\omega_{*,j}\right)}\sin{\Omega_{j}}\cos{i_j}\right) \\
    Y_{\rm{tot},j} =& -r_{\rm{tot},j}\left(\cos{\left(\theta_j(t_j)+\omega_{*,j} \right)}\sin{\Omega_{j}}\right. \\
                    &\left. + \sin{\left(\theta_j(t_j)+\omega_{*,j}\right)}\cos{\Omega_{j}}\cos{i_j}\right) \\
    Z_{\rm{tot},j} =& -r_{\rm{tot},j}\sin{\left(\theta_j(t_j)+\omega_{*,j} \right)}\sin{i_j}.
  \end{split}
\end{equation}

The impact parameter of the planet as a function of target barycentric time, $z_{j}(t_j)$, is equal to 

\begin{equation}
    \label{eq:zhelio}
    z_{j}(t_j) = \sqrt{X_{\rm{tot},j}^2 + Y_{\rm{tot},j}^2}
\end{equation}

\noindent which is the expected input to {\tt EXOFAST\_OCCULTQUAD\_CEL}, along with the normalized planetary radius ($R_{P,j}/R_*$) and the limb darkening. This computes the flux decrement due to the planet occulting the star. When $Z_{\rm{tot},j}$ is negative, the result is the transit model, $M_{T,j}$.

When $Z_{\rm{tot},j}$ is positive, we compute the occultation. However, the geometry is such that the star occults the planet, and so the input planetary path to {\tt EXOFAST\_OCCULTQUAD\_CEL} becomes $\frac{R_*}{R_{P,j}}z_{j}(t_j)$, the input normalized radius becomes $R_*/R_{P,j}$, and input the limb darkening is uniform ($u_1=u_2=0$). The resultant model is the fraction of the planet that is visible to the observer as a function of time, $P_{V,j}$. It is multiplied by the fitted thermal emission parameter, $A_{T,j}$ plus the fitted amplitude of the reflection, $A_{R,j}$, phased by the time of conjunction,\footnote{technically, this is should be phased by $T_{T,j}$, or better yet, the phase offset should be fit, but we currently do neither.} then added to the rest of the model. Therefore, the model light curve, for one planet, $M_j$, is:

\begin{equation}
\label{eq:modellc}    
  M_j = M_{T,j}+P_{V,j}\left(A_{T,j} - A_{R,j}\cos{\left(\frac{2\pi}{P_j}(t_j-T_{C,j})\right)}\right).
\end{equation}

We sum the contributions from all planets to generate the combined lightcurve model, $M$:

\begin{equation}
    \label{eq:modellcall}    
    M = 1+\sum_{j=1}^{N_{\rm Planets}} M_j - 1.
\end{equation}

This is further modified by the detrending, as discussed in \S \ref{sec:detrending}.

\section{Degeneracy between $e$ and $\omega_*$}
\label{sec:ecc}

We parameterize $e$ and $\omega_*$ as \secosw \ and \sesinw, which reduces the covariance between $e$ and $\omega_*$, eliminates the periodic parameter $\omega_*$, and naturally imposes a uniform prior on $e$ and $\omega_*$ \citep{Eastman:2013}. When we fit radial velocities or astrometry, the eccentricity is well constrained and this parameterization is well behaved, though it is worth re-emphasizing that it does not remove the hard boundary at zero, which ensures the models can only scatter upward, biasing the eccentricity to nominally higher values. Therefore, an eccentricity should not be considered significant if it is less than $2.3\sigma$ from zero. This is known as the Lucy-Sweeney bias \citep{Lucy:1971}.

\begin{figure}[!htbp]
  \begin{center}
    \includegraphics[width=3.5in, trim={0.5cm 0cm 1.5cm 0cm}]{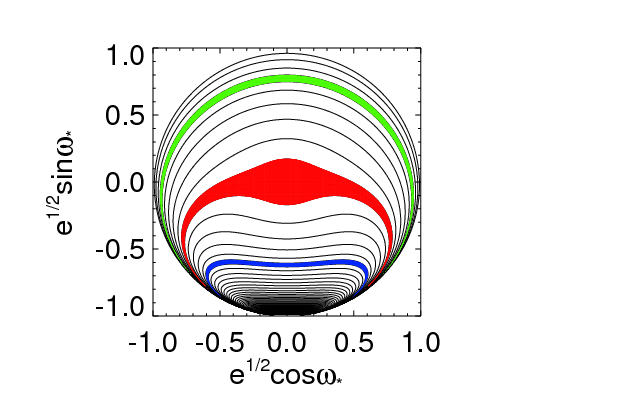} 
    \caption{The covariance between \secosw \ and \sesinw \ for transit-only fits, shown with contours in equal transit duration, only including solutions with $0 \leq e < 1$. The shaded regions denote the typical constraint from the transit duration consistent with a circular orbit (red), half that duration (green), or twice that duration (blue). We see that the vast majority of parameter space is eliminated, including preferentially eliminating high eccentricity solutions. We can also see the diabolically non-linear covariance that is difficult to sample. Differential evolution or Affine invariant samplers will draw a significant number of steps in the unlikely regions inside the contours, severely impacting their efficiency.}
    \label{fig:secoswsesinw}
  \end{center}
\end{figure}

Unfortunately, the \secosw \ and \sesinw \ parameterization has a diabolical covariance when fitting a transit alone, owing to the fact that essentially, we have one constraint (the transit duration) to constrain two parameters ($e$ and $\omega_*$). Figure \ref{fig:secoswsesinw} shows the covariance between \secosw \ and \sesinw \ with contours of constant transit duration. The shaded regions denote the typical constraint from the transit duration consistent with a circular orbit (red), half that duration (green), or twice that duration (blue). This covariance is inefficient for DE-MCMC or the Affine invariant algorithms to sample. As discussed in more detail in \S \ref{sec:underhood}, those algorithms essentially draw two random points from the PDF to define a vector to draw the next step. For linear covariances, that vector is a good means of sampling the covariance. For curved covariances, it often leads to proposed steps in the low-likelihood regions interior to the curve, causing the acceptance rate and therefore the efficiency of the MCMC to plummet ($\lesssim 1\%$ is typical for transit-only fits -- 20$\times$ less than ideal). More sinister, this covariance is particularly difficult to fully explore, and can often pass less strict convergence criteria without properly sampling the tips of the covariance between \secosw \ and \sesinw, biasing the eccentricity low and underestimating the uncertainties in both $e$ and $\omega_*$.

This degeneracy is most often avoided in the literature by assuming the planetary orbit is circular \citep[e.g.,][]{Thompson:2018,Mayo:2018}. This is not ideal for several reasons. First, the degeneracy is not perfect, so in cases of extreme eccentricities or precise data, it introduces small systematic errors in the transit model. Second, for eccentric systems, it either introduces a large systematic error in the stellar parameters or forces us to decouple the planet and star, removing the powerful constraint on the stellar density from Kepler's law. 

Finally, a significant amount of the $e$-$\omega_*$ parameter space is actually excluded by the transit duration. While it is possible to recreate the same transit duration with most eccentricities, transits that are increasingly longer than their circular cousins can only occupy an increasingly narrow range of $\omega_*$. Therefore, fitting a transit consistent with a circular orbit with uniform $e$ and $\omega_*$ priors still tends to favor low eccentricities. Further, many eccentric planets will have a duration that is inconsistent with a circular orbit, and therefore impose a hard lower bound on the allowed eccentricity. This effect was first described by \citet{Ford:2008}, and later explored as the ``photoeccentric effect'' \citep{Dawson:2012}, and is the same basic idea behind ``astrodensity profiling'' \citep{Kipping:2012}. All of these methods use the duration of the transit and the stellar density to constrain the planetary eccentricity. Thanks to \gaia \ DR2, the stellar density is well known for most transiting planet hosts and this technique is now widely applicable.

We can further constrain the eccentricity by excluding solutions where $e \geq 1-(R_*+R_P)/a$. Such solutions would imply the planet collides with its host star during periastron and are obviously not physical. This was an improvement added to the public code shortly after \exofast \ was published. Stricter criteria could be considered for planets inside the Roche limit. Long before a planet collides with its star, tidal forces will be at work. There is no known planet with $e > 1-3R_*/a$. We optionally set the eccentricity of such planets to zero by setting the {\tt TIDES} flag, presuming such planets would be tidally circularized.

Even small effects like the asymmetry between ingress and egress due to the planetary eccentricity is naturally included through the physical model, though it is typically too small to provide a meaningful constraint \citep{Winn:2010}.

Finally, in multi-planet systems, we can further constrain the eccentricity by eliminating solutions with crossing orbits (see \S \ref{sec:multiplanet}).

We therefore marginalize over this degeneracy in our typical transit-only fits, allowing us to use Kepler's law, the stellar parameters, and the transit in the global model and applying all these additional effects to constrain the planetary eccentricity. The resultant constraint on eccentricity is often surprisingly tight.

To make sampling this diabolical degeneracy more efficient, we considered fitting in the duration and marginalizing over $\omega_*$. The transit duration can be parameterized as a unitless scaling factor between the velocity of the planet at the time of transit, $V_e$, divided by the velocity the planet would have if its orbit were circular, $V_c$, which \citet{Winn:2010} showed is approximately equal to

\begin{equation}
\label{eq:vvcirc}
    V_e/V_c = \frac{\sqrt{1-e^2}}{1+\esinw}.
\end{equation}

\begin{figure}[!htbp]
  \begin{center}
    \includegraphics[width=3.5in]{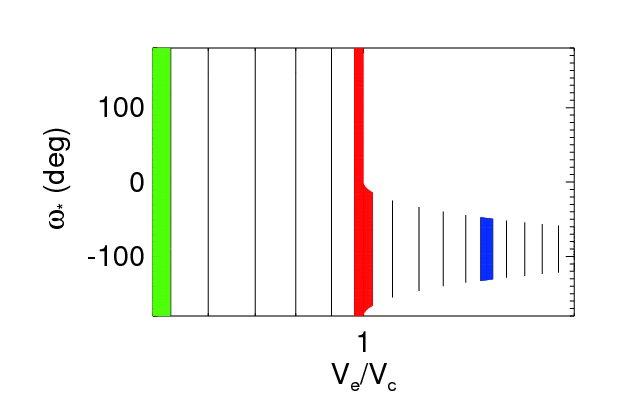} 
    \caption{The covariance between $V_e/V_c$ and $\omega_*$, as shown with contours in equal $V_e/V_c$ for transit-only fits. The shaded regions are the same as Figure \ref{fig:secoswsesinw}. The X-axis is log spaced.}
    \label{fig:vvcirc}
  \end{center}
\end{figure}

\begin{figure}
  \begin{center}
    \includegraphics[width=3.5in]{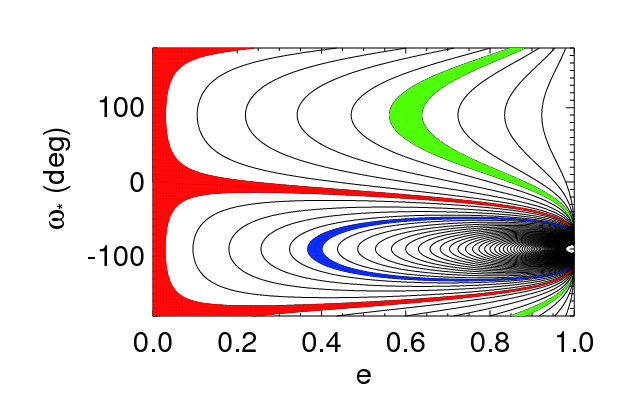} 
    \caption{The covariance between $e$ and $\omega_*$, as shown with contours in equal $V_e/V_c$ for transit-only fits. The shaded regions are the same as Figure \ref{fig:secoswsesinw}.}
    \label{fig:eomega}
  \end{center}
\end{figure}

Figure \ref{fig:vvcirc} shows us the covariance between $V_e/V_c$ and $\omega_*$ is nicely uncorrelated, just as we would like, and so should be efficient to sample. However, when we solve Equation \ref{eq:vvcirc} for $e$, it is a quadratic, so there are two solutions of $e$ for a given $V_e/V_c$ and $\omega_*$. In many cases, both solutions are real and physical, which is better visualized in Figure \ref{fig:eomega}. That could be solved by randomly picking between the two solutions before evaluating the model, if not for the far bigger problem.

The uniform step in $V_e/V_c$ imposes a prior in $e$ that biases the fit toward high eccentricities and $\omega_*=-90^\circ$. Normally, we could correct for such priors by weighting the step probability by Jacobian of the transformation between the two parameterizations:

\begin{equation}
    \label{eq:eccjacobian}
      \partial (V_e/V_c) /\partial e = -\frac{e+\sin{\omega_*}}{\sqrt{1-e^2}(1+e\sin{\omega_*})^2}.
\end{equation}

\noindent However, with the denominator and $e$ always positive and $-1 \leq \sin{\omega_*} \leq 1$, this Jacobian can change signs and the MCMC algorithm will automatically reject any proposed step with a Jacobian of the opposite sign as the first step, unjustly excluding viable regions of parameter space.

Therefore, we abandon this parameterization, though we note early releases of \exofasttwo, from 2018-06-13 to 2018-10-24, used it for transit-only fits. Users who downloaded the code during that period are strongly encouraged to update. The primary effect of such fits is to exclude potentially viable values for $\omega_*$ in eccentric systems ($\omega_* \sim -90^\circ \pm 45^\circ$), which would have downstream effects on the reported eclipse times and transit probabilities. However, since the uncertainty in $\omega_*$ of transit-only fits is typically large (most values are allowed), and $\omega_*$ is not typically a quantity of intrinsic interest, the practical effect is likely minimal.

We explored several other parameterizations for $e$ and $\omega_*$, including $e$ and $\omega_*$ directly; \ecosw \ and \esinw; and $e^{1/4}\cos{\omega_*}$ \ and $e^{1/4}\sin{\omega_*}$, but none was consistently faster than \secosw \ and \sesinw. We also attempted to linearize the covariance by using $\sqrt{1-e^2}$ and $1+\esinw$ (the numerator and denominator in equation \ref{eq:vvcirc}). Like $V_e/V_c$ and $\omega_*$, the fits converged faster, but the Jacobian excluded viable regions of parameter space, so this parameterization was also rejected.

Ultimately, we left the \secosw \ and \sesinw \ parameterization for all fits, accepting the poor efficiency ($\sim$20x longer runtime) for transit-only fits as an acceptable tradeoff for including a proper eccentricity constraint. Not only is the eccentricity interesting in and of itself, but it also enables us to generate a physical global model. While a factor of 20 longer runtime is a steep price to pay, the fundamentally unique insight and smaller uncertainties easily justify it.

While it is strongly discouraged, if the user wishes to disable this feature and force a circular orbit, it is possible with \exofasttwo. They can disable all stellar constraints (set the {\tt NOCLARET} and {\tt NOMIST} flags, remove any SED fit), impose wide, uniform priors on \mstar \ and \rstar, impose appropriate priors on the limb darkening, fix \teff \ and \feh \ at some value, and force a circular orbit. The resultant stellar parameters or anything derived from them should not be trusted, as they would be biased by any non-modeled eccentricity.

\section{Differing definitions of the transit time}
\label{sec:tc}

\citet{Csizmadia:2019} recently noted that the definition of ``transit time'' is not uniformly agreed upon or reported, and that this potential 10-minute difference is measureable for some systems today. \citet{Kipping:2008} noted a similar, $\sim 2$ minute discrepancy due to the asymmetry of ingress/egress. In this section, we reiterate and expand on their explanations and propose the universal adoption of the time of minimum projected separation, $T_T$, as an unambiguous definition for the ``transit time'' applicable to any system.

There are at least three definitions of ``transit time'' currently in use, which can differ for eccentric, inclined systems by as much as 10 minutes from one another, even ignoring the differences between time stamps outlined in \citet{Eastman:2010}. Further, the difference in definitions can be arbitrarily large in systems with transit timing variations.

The exoplanet archive \citep{Akeson:2013} interprets all quoted transit times as the average between first and fourth contact\footnote{\url{https://exoplanetarchive.ipac.caltech.edu/docs/API_exoplanet_columns.html}}. This is also the definition \citet{Kipping:2008} calls $T_{\rm MID,apparent}$. While this is a precise and intuitive definition, it is expensive to calculate and we are unaware of anyone who actually reports this quantity.

The time of inferior conjunction, which we call $T_C$, is the ``transit time'' currently most commonly reported, because it is straight-forward to compute assuming well behaved, Keplerian orbits (see \S \ref{sec:planetpath}). It corresponds to the time at which the planet has a true anomaly of $90^\circ - \omega_*$, and is used by {\tt BATMAN}, {\tt TAP}, \exofast, and up until 2019-07-22, it was the only time reported by \exofasttwo. As discussed below, in certain systems that are known today, it can be significantly offset from an intuitive understanding of the transit time.

The time of minimum projected separation, which we will call $T_T$, is perhaps the most intuitive definition and is used by {\tt TLCM} and {\tt PyTransit} \citep{Parviainen:2015}, which are both built around \citet{Gimenez:2006}. It is also the only time that is well-defined for N-body codes, and is therefore the time assumed by {\tt TTVFast} \citep{Deck:2014}. The major downside to quoting the time of minimum projected separation is that it requires a numerical search to compute the offset between the time of minimum projected separation and the time of conjunction that is necessary to compute the Keplerian orbit. If we were to fit with $T_T$ as a parameter, this numerical search would have to be done at each step in the Markov Chain before we can generate the model (which TLCM and {\tt Pytransit} do). \citet{Csizmadia:2019} outlined a relatively fast, iterative method to calculate the offset (which was first proposed in the eclipsing binary world by \citet{Martynov:1973} and \citet{Gimenez:1983}). However, in the presence of TTVs, the orbital elements themselves vary in time, and so the very meaning of a ``true anomaly'' (and therefore $T_C$ and its offset from $T_T$) is ambiguous and therefore the general applicability of that method is unclear.

The difference between these three times is usually small ($\sim1 s$), but can be measured today in a handful of systems and may be a source of timing error in the literature. Among all planets in the exoplanet archive with sufficient information to calculate it, the worst case difference between the time of conjunction and the time of minimum projected separation is 175 seconds (HATS-41b). Another five planets have discrepancies larger than 30 seconds (CoRoT-16b, CoRoT-10b, HD 89345b, K2-287b, and Kepler-105b). A rough exploration of parameter space around HATS-41b finds that we can make the discrepancy as large as 541 seconds (while still having a grazing transit) by changing its argument of periastron from 136 to -4.7 degrees. There is a linear increase in the difference between $T_C$ and $T_T$ with period and inclination, but this is tempered by the requirement that the planet transit. That is, if the period grows, $a/R_*$ grows and the inclination must shrink for it to still transit.

Every argument made above applies equally to the time of superior conjunction, $T_S$, and the time of minimum projected separation during the secondary occultation, $T_E$.

\exofasttwo \ continues to parameterize the model in terms of the time of conjunction, $T_C$, and this value is displayed in the final output table. For systems with large TTVs, this can create non-intuitive offsets between the starting parameter and the by-eye center of transit that can make it difficult to start the fit. Such systems should be rare, and in such cases, the user can tune their starting values with trial and error. In the future, we may allow the user to specify $T_T$ directly and derive the corresponding model parameter $T_C$.

At the end of the fit, we also compute the times of minimum projected separation using a golden section search at each MCMC step for each transit ($T_T$), occultation ($T_E$), and planet that overlap any supplied data, then summarize them in a separate table. This table also includes the depth of each transit, which can vary with limb darkening if the transits are observed in multiple bands or if depth variations are fit, and it includes the true impact parameter of each transit, which differs slightly from the approximate value reported in the primary table in eccentric, inclined systems, and can differ from transit to transit if duration variations are allowed.

Especially when TTVs are fit or multiple planets are present, this is an important output, since these are not trivially derived and the exact times are critical to constrain the planetary mass and eccentricity with a TTV analysis.

We only use the thinned steps to compute each $T_T$ and $T_E$, so it saves us some time, but this is expensive, taking about 8 ms per step -- and so can add hours to long fits with millions of steps, multiple planets, and/or many transits. This is done by {\tt EXOFAST\_GETTT} and is executed at the end of the fit. It can also be re-executed after the fit has completed (but requires an IDL license).

Separately, to head off additional timing errors, we emphasize that \exofasttwo \ requires \bjdtdb \ as the input timestamp for all data. While it will often give a sensible-looking answer with the data in any timestamp, using the wrong (and especially inconsistent) timestamps can create internal inconsistencies between the models that may be significant, depending on the input data and the actual timestamp used. \exofasttwo \ makes no attempt to convert time standards and always displays the required \bjdtdb \ timestamp as a label in the output table, regardless of what is actually supplied. It is up to the user to ensure the inputs are correct. See \citet{Eastman:2010} for more details about timestamps.

It is, first and foremost, up to every author to clearly state what they mean by ``transit time,'' but we reiterate our recommendation to report $T_T$ for a reliable comparison in the literature. We note that all aggregation sites, like the Exoplanet archive, \url{exoplanets.org} \citep{Wright:2011}, and \url{exoplanet.eu}, necessarily adopt whatever the authors describing the system quote as the transit time to populate their entry. It is issues like these that make the adoption of standard, well-documented, and well-understood codes like \exofasttwo \ extremely important for large-scale comparisons of literature data.

\section{Approximations in derived parameters}
\label{sec:approx}

The transit and occultation durations we report ($T_{FWHM}$, $T_{14}$, $\tau$) are approximations, assuming the velocity at conjunction is constant throughout the transit, and using Eqs 14-16 from \citet{Winn:2010}.

It is important to understand that this is not a deficiency in the model. The orbital path of the model is calculated from Kepler's law and projected using the fitted model parameters as described in \S \ref{sec:planetpath}. There are no explicit parameters for the transit durations -- these are naturally and accurately reproduced from the projected separations as a function of time. While it is certainly possible to calculate them exactly, either analytically \citep[e.g.,][]{Kipping:2008} or numerically, these are expensive calculations with little practical advantage.

The transit times and durations are critical for scheduling, but the maximum difference between the true transit duration and the duration using the \citet{Winn:2010} approximation among all planets in the exoplanet archive with sufficient information to calculate it is 27 seconds (Kepler-105b). This difference is largely irrelevant given typical uncertainties, scheduling constraints, and requirements for out of transit baseline observations. Transit duration variations are important to constrain the dynamics of some systems, but we model those as variations in inclination, not duration.

We also report an approximate form of the primary and secondary impact parameters, $b$ and $b_S$, respectively, using Eqs 7 \& 8 from \citet{Winn:2010}. Again, the exact quantity requires an expensive numerical calculation and the small difference is largely irrelevant. The worst case difference between the quoted value and the approximate form is $3\times10^{-4}$ (HATS-41), well in excess of the typical uncertainty.

\section{Units and Constants}
\label{sec:constants}

For transparency and compatibility, we supply the precise values of all constants used in \exofasttwo \ in Table \ref{tab:constants}. We use the definition of the AU recommended by IAU resolution B2 in 2012\footnote{\url{https://www.iau.org/static/resolutions/IAU2012_English.pdf}}, the nominal units recommended by IAU Resolution B3 \citep{Mamajek:2015}, and the physical constants recommended by \citet{Mohr:2016}. These values are defined once in {\tt MKCONSTANTS} and used consistently throughout the code. They differ slightly but not materially from those used in \exofast. We note that in both codes, we use the equatorial radius of Jupiter as recommended by the IAU, which is significantly (6\%) larger than its polar radius and measurably (2\%) larger than its mean volumetric radius -- a difference that could be a source of systematic error when comparing with results from other codes. While not explicitly stated as a nominal units by \citet{Mamajek:2015}, we derive our own nominal units for the Earth mass and Jupiter mass, \MEnom \ and \MJnom \ respectively, in both solar masses and cgs units. Note that these approximate, derived quantities are never used directly, and are supplied here for reference only. Since all reported masses scale with the stellar mass, we use the ratios of gravitational parameters, which are typically known to 6 significant digits, to convert between solar masses and Jupiter or Earth masses rather than rely directly on the Gravitational constant, $G$, which is only known to about 4 significant digits. $G$ is only used to derive the stellar and planetary densities in cgs units for the output table. Note this substitution is a mild abuse of these nominal units, since we must assume $\GMnom = G\msun$. While to the best of the field's current understanding this is true, the purpose of these nominal units assumes these will diverge as measurements become more precise. We also note a more precise, but statistically inconsistent (4-sigma discrepant) value for the solar radius was measured by SOHO using transits of Mercury \citep{Emilio:2012}. We are unaware of the reason for the difference, but adopt the IAU's standard definition. Regardless, the 0.1\% difference between the two is negligible for our purposes, given the limits of our current theoretical understanding of stellar radii. 

One final word of caution: we rely on external packages that likely use some of the same physical constants. The MIST models, \citet{Chen:2017}, NextGen stellar atmopsheres, \citet{Claret:2011} limb darkening tables. These models may not use the identical values we quote here, but should be statistically consistent.

\begin{deluxetable*}{crcllcc}
\tablecaption{Constants used in EXOFASTv2}
\tablehead{\colhead{Symbol} & \multicolumn{3}{c}{Value} & \multicolumn{1}{l}{Unit} &  \colhead{Description} & \colhead{Source}}

\startdata 
$G$        & 6.67408     & $\times$ & $10^{-8}$ & cm$^3$ g$^{-1}$ s$^{-2}$        & Gravitational Constant              & 1 \\
$c$        & 2.99792458  & $\times$ & $10^{10}$ & cm s$^{-1}$                     & Speed of Light                      & 1 \\
$\sigma_B$ & 5.670367    & $\times$ & $10^{-8}$ & erg s$^{-1}$ cm$^{-2}$ K$^{-4}$ & Steffan Boltzmann Constant          & 1 \\
\Rnom      & 6.957       & $\times$ & $10^{10}$ & cm                              & Solar Radius                        & 2 \\
\Lnom      & 3.828       & $\times$ & $10^{33}$ & erg                             & Solar Luminosity                    & 2 \\
\GMnom     & 1.3271244   & $\times$ & $10^{26}$ & cm$^3$ s$^{-2}$                 & Solar gravitational parameter       & 2 \\
\GMJnom    & 1.2668653   & $\times$ & $10^{23}$ & cm$^3$ s$^{-2}$                 & Jovian gravitational parameter      & 2 \\
\GMEnom    & 3.986004    & $\times$ & $10^{20}$ & cm$^3$ s$^{-2}$                 & Terrestrial gravitational parameter & 2 \\
\MJnom     & 9.5459423   & $\times$ & $10^{-4}$ & \msun                           & Jovian mass$^{4}$                   & 2 \\
\MEnom     & 3.0034893   & $\times$ & $10^{-6}$ & \msun                           & Terrestrial mass$^{4}$              & 2 \\
\MJnom     & 1.89819     & $\times$ & $10^{30}$ & g                               & Jovian mass$^{4}$                   & 1,2 \\
\MEnom     & 5.94236     & $\times$ & $10^{27}$ & g                               & Terrestrial mass$^{4}$              & 1,2 \\
\ReEnom    & 6.3568      & $\times$ & $10^{ 8}$ & cm                              & Equatorial Earth Radius             & 2 \\ 
\ReJnom    & 7.1492      & $\times$ & $10^{ 9}$ & cm                              & Equatorial Jupiter Radius           & 2 \\ 
AU         & 1.495978707 & $\times$ & $10^{13}$ & cm                              & Astronomical Unit                   & 3 
\enddata 
\tablecomments{1 -- \citet{Mohr:2016}, 2 -- \citet{Mamajek:2015}, 3 -- \url{https://www.iau.org/static/resolutions/IAU2012_English.pdf}}
\tablenotetext{4}{This approximate, derived quantity is for reference only and is never used in \exofasttwo. As all masses scale with the stellar mass, the quoted masses are derived directly from the more precise values for \GMnom, \GMJnom, and \GMEnom.}
\label{tab:constants}
\end{deluxetable*}

\section{Runtime}
\label{sec:runtime}
\exofasttwo \ is somewhat slower than the original \exofast. Re-running the original HAT-P-3b fit with \exofast \ on our current computer only took 3.77 minutes to compute the 394,004 steps it required to converge. On the same computer under a similar load, a fit as similar as possible using \exofasttwo \ took 4.75 minutes to compute the 369,356 steps for convergence -- that is, about 25\% longer, due to the additional overheads of bookkeeping for the more complex, generalized fitting framework. This additional overhead is a relatively large fraction of this extremely simple fit, but does not increase much with more complex fits, and so becomes fractionally much smaller.

However, running a fit as similar as possible to the original disables most of the benefits of the newer code. Redoing the HAT-P-3b fit with what we would now consider best practices adds additional parameters with complex covariances, as well as additional models (MIST and the SED) which are computationally expensive. Therefore each step's model takes longer to compute and it takes more steps to converge. We note that including the \gaia \ parallax and SED fit dominates the runtime of this simple model (with a few hundred transit data points and a handful of RVs). While it takes a similar number of steps (372,888), it now takes 37.63 minutes to converge. However, this extra time is well worth the wait, resulting in a 22\% more precise stellar mass, 60\% more precise stellar radius, 55\% more precise planetary radius, and an 8\% more precise planetary mass compared to the original \exofast, largely due to the SED and \gaia \ DR2 parallax.

We have taken great pain to write highly optimized IDL code and the serial performance of the basic routines are within a factor of 2-3 of lower-level languages like C or Fortran \citep{Eastman:2013}. However, IDL's automatic parallelization only applies to a limited number of built-in functions, none of which are a significant contributor to the runtime of \exofasttwo. Therefore, a single \exofasttwo \ fit sees no significant gain from using multiple cores. In principle, MCMC is highly parallelizable and we would be likely to see major (factors of 10-100) runtime gains by porting some or all of our code to parallelized C, Fortran, or CUDA. Currently, we have no plans for such a task.

\section{License-Free Use}
\label{sec:license}

Distributed with the code is {\tt exofastv2.sav}, which is a pre-compiled executable that can be run using an IDL Virtual Machine (VM) without an IDL license. The virtual machine still requires a full (free and legal) installation of IDL and a normal installation of \exofasttwo. Because it is pre-compiled, any arguments to the main program are read from a file. Otherwise, the functionality and use is nearly identical to the full version. There are a few minor limitations of this mode of operation. While the source code is available, the user is unable to tweak it and recompile without a license\footnote{For users with occasional access to a license, {\tt MKEXOFASTV2} can be used to re-create the executables}. Because of this, and the fact that IDL error messages can sometimes be more opaque when using the virtual machine, getting a fit started can be more difficult and provide less feedback. It is not possible to immediately wrap up a fit in progress by setting the {\tt STOPNOW} variable (see \S \ref{sec:convergence}). The \exofasttwo \ distribution comes with a handful of ancillary functions, but only {\tt MKSED} and {\tt MKPRIOR} are currently supported for license-free use. Currently, the precompiled version is known not to work on Windows, though we expect to be able to fix that in the future (it is unknown whether \exofasttwo \ currently works with a license on Windows). Finally, the virtual machine requires the user to click ``OK'' when it starts, with the express intention of forbidding massively parallelized use on supercomputers (see \S \ref{sec:tess}). An example license-free fit of HAT-P-3b is distributed with the code.

Alternatively, we have made many compatibility improvements to \exofasttwo \ such that the HAT-P-3b example now runs to completion using the GNU Data Language (GDL) v0.9.5, an open source compiler for IDL code. However, we caution the user the results are unverified and many features are untested. Limb darkening models outside of the grid are rejected a priori (rather than extrapolated), owing to differences in the implementations of {\tt INTERPOL}. Multi-page postscript files are not supported in GDL and so are broken into many single page files. The covariance corner plot is not generated because the advanced features required to make it are not supported. The GDL-created plots are generally not publication-ready, mostly due to poor font support. Error messages and warnings spam the screen with unknown severity. Finally, GDL was 3.5x slower than IDL in this example. Users wishing to test and improve the behavior and performance with GDL are encouraged to contact the authors.

While cursory tests for both VM and GDL compatibility is part of our suite of tests before each release, these versions are more likely to contain bugs since they are not often used by us. Users should be extra vigilant with these license-free methods, and please notify the authors if problems are encountered.

\section{Under the hood}
\label{sec:underhood}

The core MCMC algorithm is largely similar to the original \exofast, but with a few important enhancements described below. As in the original \exofast, everything about the core MCMC functions ({\tt EXOFAST\_DEMCPT}, {\tt EXOFAST\_GELMANRUBIN}, {\tt GETBURNNDX}, \\ {\tt EXOFAST\_GETMCMCSCALE}, and {\tt EXOFAST\_RANDOM}) are agnostic to the details of the model and could be used for any MCMC optimization problem, similar to the Python MCMC implementation by \citet{ForemanMackey:2012}. While the MCMC algorithm, at its core, is extremely simple, the additional features that make it easy and efficient to use are not trivial. A user wishing to implement their own model may want to use \exofasttwo \ (or \exofast) as a template, but its use is intentionally similar to the way IDL has implemented other optimization algorithms, like {\tt AMOEBA}.

\subsection{Affine Invariant vs Differential Evolution}
\label{sec:affineinvariant}
We investigated the ``Affine Invariant'', ``stretch'' step \citep{Goodman:2010}, popularized by the ``Emcee Hammer'' python code \citep{ForemanMackey:2012}. The difference between the differential evolution step used by \exofast \ and the stretch step used in {\tt EMCEE} turns out to be small in principle and in practice. Both algorithms use a difference between chains (or ``walkers'', in the \citet{ForemanMackey:2012} terminology) to define the step. This is a clever and efficient way to automatically determine and adjust the step size (which should be roughly the 1-sigma uncertainty) and direction (which should follow any covariances), since a snapshot of the accepted steps, after the burn-in, naturally fill out the 1-sigma uncertainties and appropriately map the covariances between parameters. However, the discrete differences between chains may create volumes of parameter space that are inaccessible, potentially leading to nonphysical structure in the posterior distribution. The solution to this problem is the primary difference between the two algorithms. The differential evolution algorithm \textit{adds} a small random uniform deviate to each parameter in the step, whereas the stretch step \textit{scales} the difference between the two chains by a random amount along the stepping vector. An elegant advantage of the affine invariant stretch step is that its random scale is a general multiplicative factor of the step, whereas the differential evolution's additive uniform deviate must be tuned for each parameter such that it is small but not negligible relative to the nominal step.

The biggest difficulty in using both the affine invariant and differential evolution algorithms is in defining the initialization of all chains -- a problem which is not addressed by \citet{terBraak:2006} and is left to the end user in {\tt EMCEE}. \citet{ForemanMackey:2012} recommends initializing the chains in ``a small ball around the a priori preferred position'', but we disagree. While straight-forward to implement, such a scheme is computationally expensive, requiring a longer burn-in than necessary to allow the chains to diffuse into the allowed volume. This makes it far more time consuming to run preliminary fits to get a qualitative feel for the system. More worrisome, it can be difficult to robustly identify the burn-in. A failure to properly exclude the steps before the chains have adequately diffused into the allowed volume will lead to an over-representation around the starting position and severely underestimated uncertainties. Finally, if the starting position was not properly identified as the a priori preferred solution, it is far more difficult for the MCMC algorithm to find it. All of these problems are compounded as the number of fitted parameters increase.

Our solution, integrated in the code and transparent to the user, is to first approximate the 1-sigma uncertainties by assuming Gaussian, uncorrelated uncertainties, and varying each parameter individually about the best fit value until we find the value where $\Delta \chi^2=1$. Of course, real data are rarely Gaussian and uncorrelated, but this need not be precise. It is merely a first order correction to the many orders of magnitude difference in the optimal step sizes between, e.g., period ($\lesssim10^{-6}$) and e.g., \teff \ ($\sim10^{2}$) so that we may get a sufficient scatter in the chains to efficiently seed the MCMC analysis. We then initialize each chain about the best fit plus a random Gaussian deviate times our approximate 1-sigma uncertainty times a scaling factor, $\sqrt{500/N}$, where $N$ is the number of fitted parameters. This yields an average $\Delta \chi^2 = 500$ for each initialized chain, ensuring the spread is large compared to the uncertainty, but also that the $\chi^2$ of the initialized chains is constant as a function of the number of fitted parameters.

This method of initializing the chains then gives us a natural per-parameter scale for the uniform deviate in the differential evolution method described above. In the original \exofast, we simply used a tenth of that value. However, the contribution to the $\chi^2$ due to this additive term scales as the square root of the number of fitted parameters. With the large number of parameters that is typical of \exofasttwo \ fits, even a small uniform deviate dominates the step size, especially since it does not respect the covariance between parameters. In order to avoid this, we now multiply the one tenth additive term by $\gamma$, the factor recommended by \citet{terBraak:2006} for the scaling of the step between two random chains. $\gamma = 2.38/\sqrt{2N}$, where $N$ is the number of fitted parameters.

A minor complication in initializing the chains in this manner is that many chains may step out of bounds (e.g., $e > 1$), surrounded by a uniform landscape of zero likelihood. That means they would have to random walk back to an acceptable value during the MCMC. Such a random walk is rarely a fruitful exercise and never an efficient one. Therefore, if the random deviates make the starting point of the chain out of bounds, we redraw the random normal deviates until it gets a starting value that is in bounds. With many poorly constrained but strictly bounded parameters, it can be exponentially difficult to find an acceptable starting value for the chain when starting as much as 5 times the uncertainty from the best fit in each parameter. To solve this problem, \exofasttwo \ adds an exponential decay to this factor of five with a ``half life'' of 1000 iterations. That is, after each failed attempt to initialize the chain, the random normal deviate gets slightly smaller. After 1000 consecutive failed attempts to find a viable starting value, the random Gaussian deviate is only multiplied by 2.5 times to arrive at the initial starting value for each chain. This ensures that there will be an adequate spread in the initial chains without fear of entering a very long loop of poorly seeded chains.

With the same method to initialize the chains and the algorithms executed as described here, for several inputs, models, and data (but far from an exhaustive combination), the Differential Evolution step converged between 10\% and 300\% faster than the Affine Invariant step. In addition, \citet{Huijser:2015} identified difficult to detect problems with the Affine Invariant sampling for high ($N > 30$) dimensions, which is typical of the more complex fits \exofasttwo \ is designed to do.

We now allow the user to select the affine invariant stretch step if desired, but caution against its general use, both for speed and accuracy. Additional investigation may identify cases where the affine invariant is better, but we continue using the differential evolution step by default.

Both the differential evolution and affine invariant algorithms behave poorly in the presence of non-linear (``banana shaped'') degeneracies. Such degeneracies cause the algorithms to routinely propose steps in the low-likelihood region of parameter space interior to the curve of the covariance, causing the acceptance rate to plummet and the time to convergence to sky-rocket. Short of re-parameterizing the problem or adding priors or assumptions to remove such degeneracies, the only solution is to run the fit significantly longer.

Recently, Hamiltonian Monte Carlo has shown significant promise to dramatically improve convergence times \citep{Neal:2012,Betancourt:2017} and may be the subject of future improvements.

\subsection{Parallel Tempering}
\label{sec:paralleltempering}
Parallel tempering runs some number of parallel chains, each at successively ``hotter'' temperatures, $T$. We compute $C$ as the ratio of $\mathscr{L}$ between the proposed chain ($i$) and the previous ($i-1$) chain, raised to the power of the temperature.

\begin{equation}
    \label{eq:metropolishastings}
    C = \left(\frac{\mathscr{L}_i}{\mathscr{L}_{i-1}}\right)^{T}
\end{equation}

\noindent and draw a uniform random number between 0 and 1. If our random number is less than $C$, we accept the $i$th step. If not, we reject it and make a copy of the $i-1$th step. Therefore, when $T=1$, we recover the standard Metropolis-Hastings algorithm, but when $T>1$, we are more likely to accept a worse model. Then, instead of the usual differential evolution or affine invariant step, with some probability (50\% in our implementation), a step is proposed to jump to the set of parallel parameters in the next hottest chain (and accepted according to the Metropolis-Hastings algorithm). Only the $T=1$ chains can be used for inference (otherwise the looser acceptance criteria would inflate our uncertainites), and chains are never allowed to swap with adjacent chains, which would break their independence and compromise the convergence calculations. This allows the chains to explore regions of parameter space that may cross vast chasms of unlikely volumes, more efficiently exploring complex covariances, sampling multi-modal distributions, or find the optimal global solution despite starting in a worse, locally good solution. 

Parallel tempering is often overkill for the relatively well-behaved likelihood surfaces common for planetary fits. However, complex fits with hundreds of covariant parameters are difficult to optimize. {\tt AMEOBA} does not handle covariances well and often fails unless it starts very close to the global solution. While a differential evolution MCMC often does a better job of optimizing the high dimensional, covariant parameter space, a poor initial optimization can dramatically increase the convergence time or prevent it from finding the best global solution. Utilizing the parallel tempering feature can be far easier than tweaking the starting parameters by hand. In addition, it is useful to explore well beyond the starting assumptions to make sure the user has not overlooked a likely solution.

When widely separated modes exist, (e.g., stellar parameters near the turn off, mass-radius relations near a Jupiter radius, or period aliases), the parallel tempering feature becomes critical to properly sample the allowed volumes. It is not clear how often such a situation might arise -- we have yet to knowingly encounter it for any global planetary models -- but it is a powerful feature to have available by changing a single command line option.

The user can specify the number of chains to run with different temperatures ({\tt NTEMPS}) and the maximum temperature ({\tt TF}). The larger the maximum temperature, the more parameter space it will explore, but the less likely it is to swap between temperatures, and the less efficient it is likely to be. The larger the number of temperatures, the easier it is to swap between chains. However, only the $T=1$ chains represent the true posterior, so only they can be used for inference. It is therefore wasteful to calculate more hot chains than necessary to explore the allowed volume. When parallel tempering is enabled, the real-time ``swap rate'' is displayed -- the acceptance rate of all proposed swaps. The ideal swap rate is the same $\sim 23\%$ as the ideal acceptance rate for any fit. If the swap rate is too low, more temperatures are likely necessary. If it is too high, fewer would likely be more efficient. If parallel tempering is desired, we recommend using 8 temperatures with a maximum temperature of 200, which can robustly find modes separated by 50 sigma. While parallel tempering with 8 temperatures automatically increases the maximum number of steps (model evaluations) by a factor of 8, it rarely increases the runtime by that factor. The correlation length of steps chosen from parallel tempering is generally much shorter, and so convergence generally takes fewer links in each chain. Additionally, many of the hotter chains' proposed steps are out of bounds, which are rejected before computing the full model. The result is that, in many cases, parallel tempering can actually decrease the total runtime.

In the case of the hottest chains, even a deep transit may not be ``significant,'' allowing the chains to accept a flat line model as only slightly worse. Unfortunately, accepting a flat line model is catastrophic for the convergence of the Markov chains, owing to the highly degenerate volume of parameter space that can generate an equally likely, flat-line model. For example, the transit model of planets with any set of parameters (e.g., period, radius, eccentricity) are indisguishable if the transit window falls wholly outside of the data.

We do not a priori impose non-physically motivated boundaries on parameters that may exclude real planets, since we do not want to unnecessarily limit the utility of \exofasttwo. However, this decision formally breaks the assumptions required for convergence of a Markov chain. Because some parameters are unbounded and/or periodic (like $T_C$), the degenerate volume is infinite. Typically, this is irrelevant because, given the strong constraint in the data, the probability of accepting such a model is effectively zero. However, with hot chains (or low SNR transits), the probability of accepting a flat line model becomes much larger -- often even inevitable. 

It is strongly recommended that the user supply realistic, but loose bounds on $T_C$ and Period when using parallel tempering or fitting low signal to noise events. Care should be taken, however, to check the posteriors upon completion to ensure there is no significant probability near the imposed bounds. We also include the option to a priori reject flat line transit models. This is not the default behavior because users may wish to supply data that should not contain a transit to constrain the ephemeris or transit duration. We always reject a transit model that does not transit at all $b > 1+R_P/R_*$. These safeguards help prevent the MCMC from breaking while exploring the highly degenerate, infinite volume of planetary parameter space, but imposes a prior that there is a transit signal in the transit data, which may artificially increase the apparent significance of a marginal detection. Care must be taken to independently evaluate the significance of the transit.

We recommend starting all fits with an initial, relatively short run that enables parallel tempering (and if fitting transits, rejecting flat line transit models), both to ensure the likeliest model is found, and to determine if any parameters are multi-modal (especially \mstar, \rstar, distance, Age, \mplanet, and Period). If any fitted parameters have widely spaced modes, the final fit should be performed with parallel tempering enabled. However, if no multi-modal distributions are found, the user should generate a new prior file based on the best-fit among all MCMC links (using {\tt MKPRIOR}) and restart without parallel tempering to improve the runtime.

A model is just as good a fit to the data if two or more planets are swapped. That is, if planet b were to take on all the properties of planet c and vice versa, the resulting likelihood would be identical. With a typical MCMC sampling, such a swap is effectively disallowed by the vast, unlikely expanse of parameter space that it would have to cross. However, a perfect exploration of the allowed volume (which is more likely with parallel tempering) would find that each planet could take on the properties of any other planet. We now explicitly remove this degeneracy by rejecting steps that scramble the order of the planets relative to their starting order, as defined by their periods. This enables the user to simultaneously explore widely separated, perhaps complicated period aliases without having to worry (much) about a single planet taking on the properties of multiple planets.

\subsection{Burn-in}
\label{sec:burnin}

Often, because of the complex parameter space, the relatively large spread in initial parameters for each chain, sometimes sub-optimal starting position, and the large number of chains, it is relatively easy for one or more chains to get stuck in a sub-optimal solution. While parallel tempering helps tremendously (see \S \ref{sec:paralleltempering}), it is often inefficient and introduces a significant, often unnecessary, overhead. We have modified our burn-in calculation to identify and reject these chains. To calculate the burn-in, \exofasttwo \ finds the maximum likelihood of all links of all chains, then calculates the median likelihood from that point forward in the corresponding chain, which becomes the benchmark for all other chains. The first time each chain's likelihood is above that benchmark denotes its burn-in. The maximum burn-in of all used chains becomes the burn-in.

In order to prevent a bad chain (with a lower-than-typical likelihood) from pushing the burn-in too far out for the rest of the good chains, the chains are sorted by their individual burn-in and the cumulative number of used links is calculated as each chain is added. Since the addition of another chain with a larger burn-in removes links from all previously added chains, adding a chain with a large burn-in may actually reduce the number of total used links. However, subsequent chains with a similar burn-in may increase the number of total links beyond what it was originally. The burn-in is determined to be the burn-in of the chain that maximizes the total number of used links. Chains with a larger burn-in are marked as bad and not used in subsequent analysis. Finally, we set a floor of 10\% of the total number of steps to prevent a poor initial spread about the best fit from biasing the results unfairly toward the best-fit, and a ceiling of 90\% to ensure we have a sufficient number of steps to robustly calculate the convergence statistics and monitor the progress of the fit. If fewer than 3 chains are ``good'', we use all chains (and it is unlikely to be ``well-mixed'').

This functionality is codified in the {\tt GETBURNNDX} function, which calculates both the index of the burn-in and the indices of all ``good'' chains. If users wish to compute their own statistics or figures based on the raw links saved in the output file, it is critical that these bad chains and bad links are discarded first. An example that reads in the output file of saved steps, discards bad links and bad chains, then computes an upper limit, can be seen in the {\tt UPPERLIMIT} code distributed with \exofasttwo.

\subsection{Convergence}
\label{sec:convergence}

A fit is considered ``converged'' or ``well-mixed'' when we deem the PDF to be sufficiently representative of the underlying posterior. It is important to be aware that only non-convergence can be proved. No method can ever prove convergence -- we can simply do our best to convince ourselves that our samples are representative of the underlying posterior.

As in the original, \exofasttwo \ calculates two metrics to judge convergence. The Gelman-Rubin statistic, $Rz$, is a measure of how similar independent chains are to one another, with values that asymptotically approach 1 from above as they become indistinguishable. The rationale is that, the more similar each independent chain is to all others, the more similar they all must be to the underlying posterior. Of course, just as everyone's mother is fond of saying: ``if everyone jumped off a bridge, would you?'', this is not always a good assumption, and gets at the heart of why we can never prove a given run is converged. If all of the chains jumped off a bridge, that doesn't necessarily mean that was the right solution, it may have just been the best solution reasonably accessible given the starting conditions and runtime. This is why we must always be cautious in trusting the results of even a nominally converged chain. The other metric is the number of independent draws, $Tz$, determined by dividing the length of the chains by their correlation length. The larger the better. 

By default, we continue to follow the strict recommendations of \citet{Ford:2006}, which requires $Rz < 1.01$ and $Tz > 1000$ for five successive evaluations. It is not uncommon for complex fits with large data sets to take weeks to meet these criteria -- the most complicated system we have modeled to date, K2-266, a system with 80 days of \ktwo \ long cadence data, 6 planet candidates, an extremely short duration transit that required a higher density of model evaluations, TTVs, and hundreds of free parameters described in \citet{Rodriguez:2018b} took 60 days to converge.

We note that \citet{Brooks:1998} recommend a far less conservative value of $Rz < 1.2$, with no explicit requirement on $Tz$, and many in the literature have followed this advice \citep[e.g.][]{Vanderburg:2016}, leading to significantly faster runtimes. We now let the user change both the Gelman-Rubin Statistic and the number of independent draws that define the convergence.

While the Gelman of \citet{Brooks:1998} and of the Gelman-Rubin statistic is the same, well-respected statistician whose knowledge of MCMC certainly exceeds that of any astronomer, there are two reasons we generally care about the convergence criteria, and this recommendation assumes the user has already confidently excluded the most important -- and difficult to identify -- problem of convergence.

\begin{figure}[!htbp]
  \begin{center}
    \includegraphics[width=3.5in]{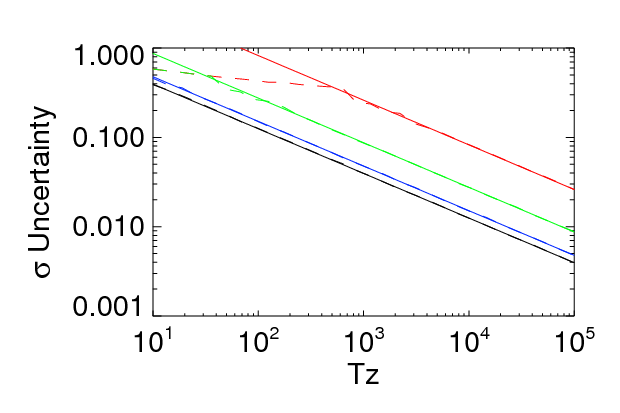} 
    \caption{The uncertainty in the median (black), 68\% confidence interval (blue), 95\% confidence interval (green), and 99.7\% confidence interval (red) reported by MCMC sampling as a function of the number of independent draws, $Tz$, as simulated by drawing random Gaussian numbers (dashed lines). We see that the uncertainty in all is proportional to $1/\sqrt{Tz}$ (solid lines) once $Tz$ is large enough to appropriately determine the desired range.}
    \label{fig:tz}
  \end{center}
\end{figure}

First, the stricter the convergence criteria, the more accurately the PDF represents the true posterior. That is, there is an error on the quoted median values and confidence intervals that shrinks as the MCMC runs longer. When we have $T_Z=100$, we can think of this as drawing 100 Gaussian random numbers to approximate a Gaussian. Since the underlying Gaussian has a median of 0 and a 68\% confidence interval of 1, the difference between our approximate Gaussian and the known underlying distribution can be computed. In Figure \ref{fig:tz}, we plot this error, averaged over 10,000 trials at each $Tz$ to reduce the noise, as a function of $Tz$. Since we determine our confidence intervals by using the range of values in a given interval, for low $Tz$, there is a quantization error in the rounding of the ranges. When computing the best-fit line, we exclude $Tz < 1000$ for this reason.

From Figure \ref{fig:tz}, we can see that with a $Tz$ of 100, the median and 68\% confidence intervals have errors of $< 0.15\sigma$. For Gaussian distributions, this scales as $1/\sqrt{Tz}$, so that our default threshhold of $Tz > 1000$ should produce median values and 68\% confidence intervals that are accurate to $0.05\sigma$. Accurately computing limits or wider confidence intervals requires accurately measuring the position along a less well populated region of parameter space, which has significantly higher fractional uncertainties for a given $Tz$. The uncertainty in the 95\% and 99.7\% confidence intervals, as a function of $Tz$, are also shown in Figure \ref{fig:tz}, and show the same $1/\sqrt{Tz}$ scaling, but with a larger zero point. The uncertainties in the median values and confidence intervals as a function of $Tz$ are summarized in Equation \ref{eq:errerr}:

\begin{equation}
    \label{eq:errerr}
\begin{tabular}{lcl}
$\sigma_{Median}$  & $\sim$ & $0.13\sigma/\sqrt{Tz/100}$\\
$\sigma_{1\sigma}$ & $\sim$ & $0.15\sigma/\sqrt{Tz/100}$\\
$\sigma_{2\sigma}$ & $\sim$ & $0.28\sigma/\sqrt{Tz/100}$\\
$\sigma_{3\sigma}$ & $\sim$ & $0.82\sigma/\sqrt{Tz/100}$\\
\end{tabular}
\end{equation}

In reality, our distributions are not generally Gaussian, so the true uncertainties and scaling may differ slightly for each parameter, but these rough values give a good sense for the impact of $Tz$. Importantly, even relatively small values of $Tz$ give reasonably precise answers, which is likely why \citet{Brooks:1998} recommends less strict criteria, and this general advice is repeated by \citet{ForemanMackey:2012}.

However, the reason we use such strict convergence criteria by default and advise caution when relaxing them is that the stricter the convergence criteria, the more steps it must take. The more steps it takes, the more likely it is to be able to jump from an isolated local solution to the optimal global solution, the more likely it is to properly explore complex covariances, and the less impact improperly scattered initial chains or a poorly trimmed burn-in might have. A fit that has not sampled the optimal global solution or a fit that has not properly diffused into the allowed parameter space could yield a catastrophically incorrect solution that is not representative of the true underlying posterior.

This concern is largely eliminated if we are confident the fit has sampled the global solution with an adequate spread in the chains, the covariances are well-behaved, and the burn-in is properly trimmed, but such assurances are difficult to obtain. Parallel tempering may help dramatically with this, but we note that with the standard MCMC (without parallel tempering), we have performed many fits that would have passed the less strict convergence criteria ($Rz < 1.2$) before it actually found the correct global solution, particularly for transit-only, eccentric fits or those with many highly correlated parameters that are difficult to optimize in the initial solution.

As a reasonable compromise, we typically perform several relatively short runs to evaluate the significance of various assumptions. We use those results to determine which data sets and priors are self-consistent, and begin writing a draft while we run a longer version that formally passes our strict convergence criteria. With such a procedure, the long runtimes of \exofasttwo \ using strict convergence criteria are rarely the rate-limiting step in publishing a paper.

In addition to setting {\tt MAXSTEPS}, there are two new ways a fit can be stopped before convergence while retaining all results. The user can pause the fit (control + C), set the {\tt STOPNOW} global variable to 1, then continue the fit. Or, the user can supply a time limit, {\tt MAXTIME}, on the MCMC portion of the code, which will automatically set the {\tt STOPNOW} global variable to 1 once the runtime exceeds {\tt MAXTIME}. When {\tt STOPNOW} is 1, the MCMC code will finish its current loops over each chain, the thinning, and the number of temperatures (if parallel tempering is enabled), then halt the fit and summarize the results as they stand. Currently, a fit cannot be restarted once stopped, but that functionality may be added in the future. Adding checkpoints and more detailed status of running fits may also be added in future updates.

\section{Walkthrough}
\label{sec:walkthrough}

The original \exofast \ required minimal human inputs, deriving its own global starting point using a lomb-scargle periodogram of the RVs, then using the best-fit RV solution to predict the transit time. For the multi-planet systems \exofasttwo \ now fits, this initial step can require a significant amount of human judgment and expertise that we have not attempted to codify. Things like how many planets, their periods ($P$), and phases ($T_C$) are usually required inputs. Their radii ($R_P/R_*$), masses ($K$), inclinations, and eccentricities are sometimes helpful.

It is therefore necessary that the user has some idea of the properties of the system and will use those results to start the global fit relatively close to its final values. That, coupled with its additional complexity, makes \exofasttwo \ somewhat less forgiving than its predecessor, and failures can often be opaque to a casual user with a limited understanding of how certain parameters may influence the model. However, \exofasttwo \ is significantly more tolerant of poor starting parameters with Parallel tempering enabled (see \S \ref{sec:paralleltempering}), and we believe the example described here (and others distributed with the code) will provide sufficient templates to allow most users to model most systems sensibly. Note that we intend to add all fits from our team to the examples distributed with the code to help broaden this base of useful templates.

Here we walk through the same fit of HAT-P-3b we did in \citet{Eastman:2013}, but using the most advanced features of \exofasttwo. The input RV and transit data files, {\tt HAT-3b.HIRES.rv} and {\tt n20070428.Sloani.KepCam.dat}, respectively, are identical to the original \exofast. Note that the filenames must follow a specific format to encode the date, telescope, and filter used in the output tables, figures, and for the limb darkening. While any extra columns in the RV data file are currently ignored, we are considering a future improvement that will use those columns to detrend like we do with transit data sets. Therefore, to future proof current fits, we recommend that users only have three columns in the RV data files (\bjdtdb, RV, $\sigma_{\rm RV}$).

Because HAT-P-3 is a sun-like star, the default MIST models should be reliable. In addition, we use {\tt MKSED} to generate an SED input file, {\tt hat3.sed}, which contains broadband photometry from available, trusted catalogs with appropriate systematic error floors applied. These values are summarized in Table \ref{tab:LitProps}.

\begin{table}
\scriptsize
\centering
\caption{Literature Properties for HAT-P-3}
\begin{tabular}{llcc}
\hline
\hline
Parameter & Description & Value & Source\\
\hline 
$\alpha_{J2000}$\dotfill	&Right Ascension (RA)\dotfill & 13:44:22.56340 & 1	\\
$\delta_{J2000}$\dotfill	&Declination (Dec)\dotfill &  +48:01:42.8341 & 1	\\
\\
$B_T$\dotfill	& Tycho B$_T$ mag.\dotfill & 12.677 $\pm$ 0.201		& 2	\\
$V_T$\dotfill	& Tycho V$_T$ mag.\dotfill & 11.932 $\pm$ 0.147		& 2	\\
$B$\tablenotemark{a}\dotfill		
                & APASS Johnson $B$ mag.\dotfill	& 12.287 $\pm$	0.06& 3	\\
$V$\dotfill		& APASS Johnson $V$ mag.\dotfill	& 11.470 $\pm$	0.04& 3	\\
$g'$\dotfill	& APASS Sloan $g'$ mag.\dotfill	& 11.745$\pm$0.08	& 3	\\
$r'$\dotfill	& APASS Sloan $r'$ mag.\dotfill	& 11.183$\pm$0.05	& 3	\\
$i'$\dotfill	& APASS Sloan $i'$ mag.\dotfill	& 10.994$\pm$0.04 & 3	\\
\\
$J$\dotfill		& 2MASS $J$ mag.\dotfill & 9.936  $\pm$ 0.02	& 4	\\
$H$\dotfill		& 2MASS $H$ mag.\dotfill & 9.542 $\pm$ 0.03	    & 4	\\
$K_S$\dotfill	& 2MASS $K_S$ mag.\dotfill & 9.448 $\pm$ 0.03& 4	\\
\\
\textit{WISE1}\dotfill		& \textit{WISE1} mag.\dotfill & 9.377 $\pm$ 0.030 	& 5	\\
\textit{WISE2}\dotfill		& \textit{WISE2} mag.\dotfill & 9.451 $\pm$ 0.030 	& 5 \\
\textit{WISE3}\dotfill		& \textit{WISE3} mag.\dotfill & 9.389 $\pm$ 0.035 	& 5	\\
\\
$\pi$\dotfill & Parallax (mas) \dotfill & 7.403 $\pm$ 0.026 & 1 \\
\hline
\\[-6ex]
\end{tabular}
\\
\begin{flushleft} 

\tablecomments{
    \footnotesize{References are:
    $^1$\citet{Gaia:2018};
    $^2$\citet{Hog:2000};
    $^3$\citet{Henden:2016};
    $^4$\citet{Cutri:2003};
    $^5$\citet{Cutri:2014};
    }
}
\end{flushleft}
\label{tab:LitProps}
\end{table}

In the original \exofast, the starting values and priors were specified in an array that was opaque and error prone to define. Now, they are specified in a configuration file by name (see \S \ref{sec:priors}). As with all \exofasttwo \ fits, we must specify a starting value for Period and $T_C$. Note that $T_C$ becomes highly covariant with the period if it is many epochs from the data, which is difficult for {\tt AMOEBA} to optimize and unnecessarily increases the uncertainty of $T_C$. If retrieving a $T_C$ from the literature, it is good practice to propagate it to the epoch of the fitted data. In this fit, because there is very little RV data and only one transit, we include a Gaussian penalty on the period determined from the (unfit) discovery lightcurve quoted in \citet{Torres:2007}. For most fits, a Gaussian penalty on the period should not be used. If one is supplied, it is extremely important that this prior come from data not being fit, so as not to double count the constraint from the data and underestimate the uncertainties.

Because we fit an SED, we include a prior on parallax from \gaia \ DR2 (adding 82 $\mu as$ to the reported value and adding 33 $\mu as$ in quadrature to the reported error, following the recommendation of \citet{Stassun:2018}), and an upper limit on the V-band extinction using the maximum value from \citet{Schlafly:2011} using their web interface\footnote{\url{https://irsa.ipac.caltech.edu/applications/DUST/}}. Note their quoted values are B-V extinction, and so must be multiplied by 3.1 to convert to the V-band extinction required by \exofasttwo. While the fit would work without a parallax prior, the SED would not constrain the stellar radius -- the stellar radius (determined less precisely from MIST, the transit, and the spectroscopic priors), would determine a photometric parallax, largely negating the utility of including the SED in the global fit. The upper limit on the V-band extinction is also not required to run, but the uncertainties would be larger.

Previously, we required priors on \teff \ and \feh. The SED and MIST models can now independently constrain them, so they are no longer required. However, if they are available and trusted, spectroscopic priors can often reduce the uncertainties in many fitted and derived parameters. We use values from spectroscopy quoted in \citet{Torres:2007}. 

Because the period is so short, the planet is likely tidally circularized. Because of that, and because we have so few RVs, we fix the eccentricity to zero. This is not strictly necessary. This could be done in the prior file or with the circular flag. We use the circular flag.

The starting guess for the RV zero point and semi-amplitude is the mean and $\sqrt{2}$RMS of the RVs, respectively, and is usually good enough for single-planet fits. The \citet{Chen:2017} mass-radius relation is used to seed the guess for $R_P/R_*$. If no RVs are supplied, the code defaults to a Hot-Jupiter-like value of $R_P/R_*=0.1$.

So, with a relatively minimal configuration file, {\tt hat3.priors}, reproduced below, we can start the fit. 

\begin{verbatim}
teff 5185 80
feh 0.27 0.08
tc 2454218.76016
period 2.899703 0.000054
parallax 7.485 0.042
av 0 -1 0 0.07409
\end{verbatim}

The default starting values for the star are sun-like, which is marginal for the HAT-P-3 system with published values of $\mstar=0.904\msun$ and $\rstar=0.818\rsun$. As we can see in the diagnostic plots in \S \ref{sec:plots}, supplying better starting values for the stellar parameters would help {\tt AMOEBA} find the correct global solution, which it failed to do in this case. If {\tt AMOEBA} does not find the correct global solution, the acceptance rate and overall efficiency can be dramatically impacted due to a poor spread in the initial chains and a sub-optimal random addition to each parameter at each step required by the DEMC algorithm (see \S \ref{sec:underhood}). Further, the burn-in period is likely to be longer as it must optimize the global solution. In this case, the fit converged 3 times slower than the same fit with optimized starting values, taking just under 2 hours instead of 40 minutes (see \S \ref{sec:runtime}). If the starting values are very far off, it can lead to catastrophic failures where it never finds or explores the true globally optimal solution.

However, in this case, optimizing the starting values is not strictly required to achieve a good fit. If the host star were very different from a Sun-like star, the planet were small and RVs were not supplied, the planet were highly grazing, or the system had multiple planets, the user may need to refine the starting values for \mstar, \rstar, EEP, $R_P/R_*$, $i$, or $K$, and/or enable the parallel tempering option so the fit is able to explore a larger volume of parameter space around the starting point. For most fits, no other parameters will need to be changed from their starting values to work, but the closer the starting values are to the optimal values, the faster and more robust the fit will be. {\tt MKPRIOR} can take the results from a previous run and regenerate this configuration file to specify all fitted parameters, starting as close to the best-fit of the previous run as possible. Because {\tt MKPRIOR} retains the Gaussian priors imposed on the original fit, the starting fit may not be exactly equal to the best-fit, but it will be well within the range for {\tt AMOEBA} to easily optimize.

Finally, we simply call \exofasttwo \ from IDL to use all of the files described above like this:

\begin{verbatim}
exofastv2, nplanets=1, circular=[1],$
    tranpath='n20070428.Sloani.KepCam.dat',$
    rvpath='HAT-3b.HIRES.rv',$
    fluxfile='hat3.sed',$
    priorfile='hat3.priors',$
    prefix='fitresults/HAT-3b.',$
    maxsteps=25000, nthin=2
\end{verbatim}

The outputs are explained in the following section. This and other examples for a variety of different types of fits are distributed with the code, in the {\tt \$EXOFAST\_PATH/examples} directory. A tutorial with exercises and walkthroughs with other types of fits and systems is linked in the {\tt README} supplied with the code.

\section{Explanation of outputs}
\label{sec:explain}

\exofasttwo \ creates a number of outputs. Some are intended to be used directly for publication, many are diagnostic, and some allow the user to create figures to their own aesthetic tastes without having to recreate the global model.

One of the most important outputs is the latex source code for a table summarizing the median and 68\% confidence intervals for all parameters. While this table is large and somewhat unwieldy, we strongly urge users to publish it in its entirety, not to pick and choose values most relevant to the current work. There is significant value in having a uniformly reported set of parameters, particularly for ingestion into aggregation sites like the exoplanet archive \citep{Akeson:2013} and future comparative analyses. In addition, many derived parameters are useful for other follow up work, like the secondary eclipse parameters. You may refer the readers to Table \ref{tab:explain} of this paper for a detailed explanation of each parameter and the underlying assumptions behind them. The first two columns are the symbol and short explanation that is included as the output of every fit. The ``variable'' column refers to the name one would use to reference it in the prior file. The ``bounds'' column refers to the most conservative, explicit bounds we have imposed on the parameter, as well as the propagated bounds on derived parameters. In many cases, stricter bounds are imposed with certain options or combinations of data sets, explained in the last column. The final column is the detailed explanation of the corresponding parameter. This table includes additional parameters that are never displayed in the output table, but which the user may wish to modify with the prior file, as well as a few parameters that are only displayed in an ancillary table. If we add, change, or clarify parameters in the future, the file {\tt \$EXOFAST\_PATH/explain.tex}, distributed with the code, contains an updated version of Table \ref{tab:explain}. Users are welcome to publish any part of this or the updated table they feel is necessary to explain the \exofasttwo \ fit.

\input{explain.tex}

\subsection{Plots}
\label{sec:plots}

\begin{figure*}
  \begin{center}
    \includegraphics[width=6.5in]{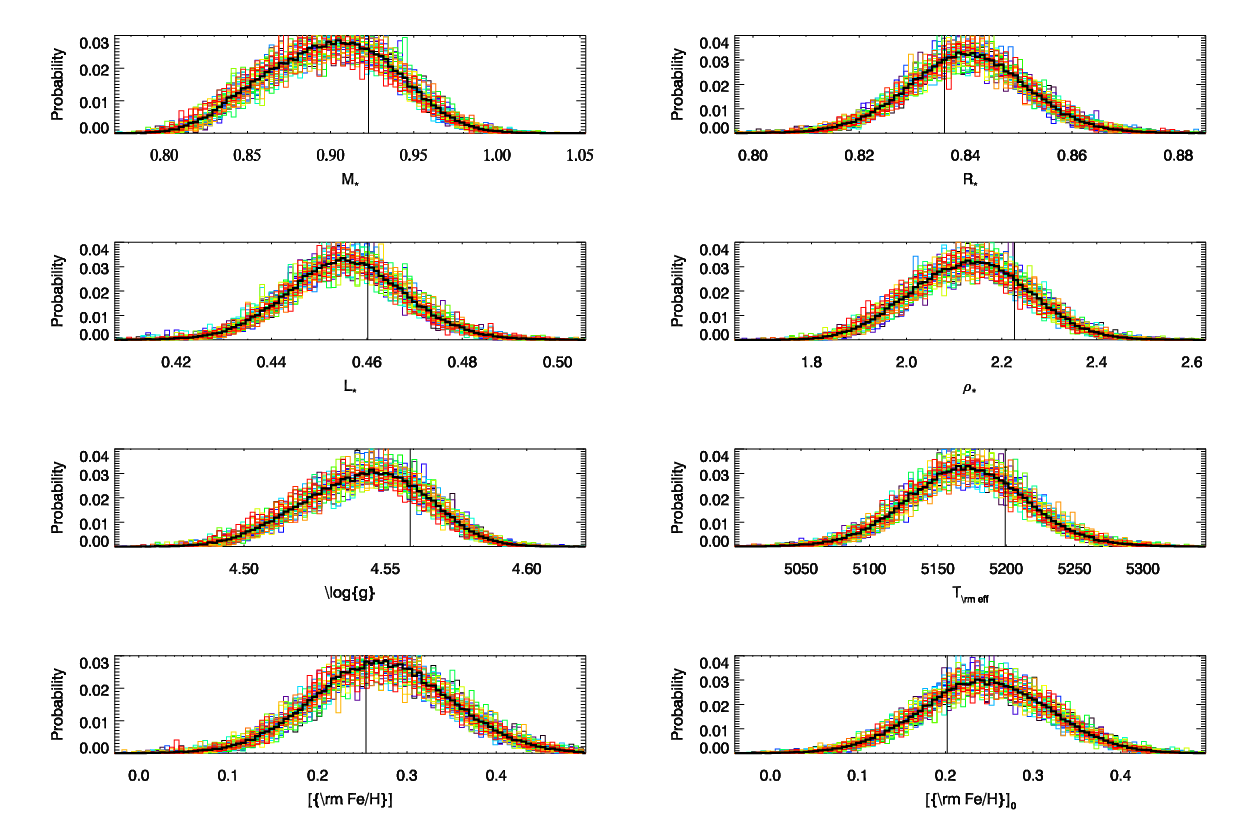}
    \caption{The first page of the Probability Distribution Function (PDF) diagnostic plots. Each plot shows the PDF for a parameter. Each color represents a chain and the thick black PDF is the average of all chains. The thin black vertical line denotes the best-fit value. A similar plot is made for each fitted and derived parameter. Some parameters are poorly behaved (like EEP, which often shows discontinuities), but generally, we are looking for smoothly varying Gaussian like plots with all chains in agreement with one another, as we see here.}
    \label{fig:pdf}
  \end{center}
\end{figure*}

\begin{figure*}
  \begin{center}
    \includegraphics[width=6.5in]{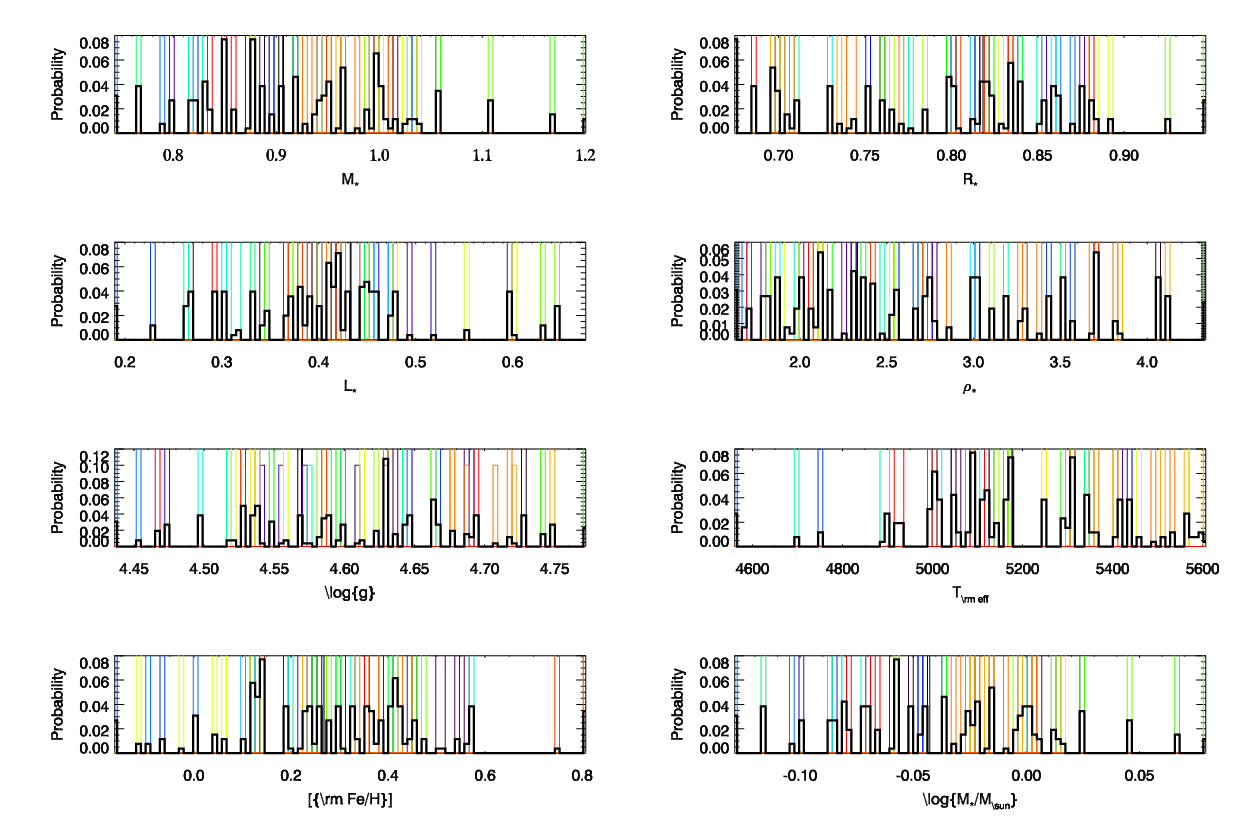}
    \caption{Same as Figure \ref{fig:pdf}, but showing an example of a poorly mixed chain that should not be used for inference. Note the multi-modal distributions, with each mode is dominated by a separate chain, suggesting they are still near their initialized values.}
    \label{fig:pdfunmixed}
  \end{center}
\end{figure*}

\begin{figure*}
  \begin{center}
    \includegraphics[width=6.5in]{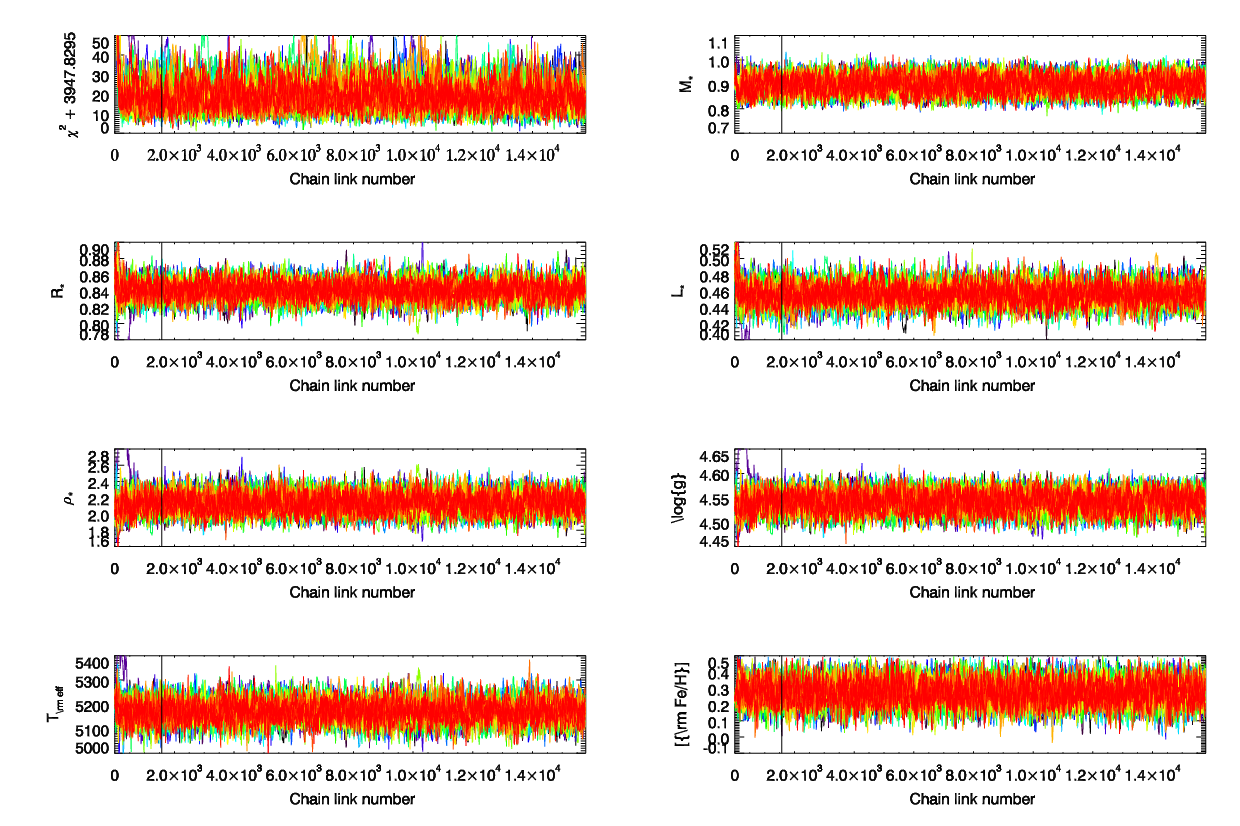}
    \caption{The first page of the diagnostic chain plots. Each plot shows the parameter value as a function of the link number. Each color represents a chain. The thin vertical line denotes the calculated burn-in (see \S \ref{sec:burnin}). Only links to the right are kept and used for inference. A similar plot is made for the $\chi^2$ and each fitted and derived parameter. Generally, we are looking for a lack of long-term evolution to the right of the black line (especially in the $\chi^2$), as well as agreement among all chains, as we see here.}
    \label{fig:chain}
  \end{center}
\end{figure*}

\begin{figure*}
  \begin{center}
    \includegraphics[width=6.5in]{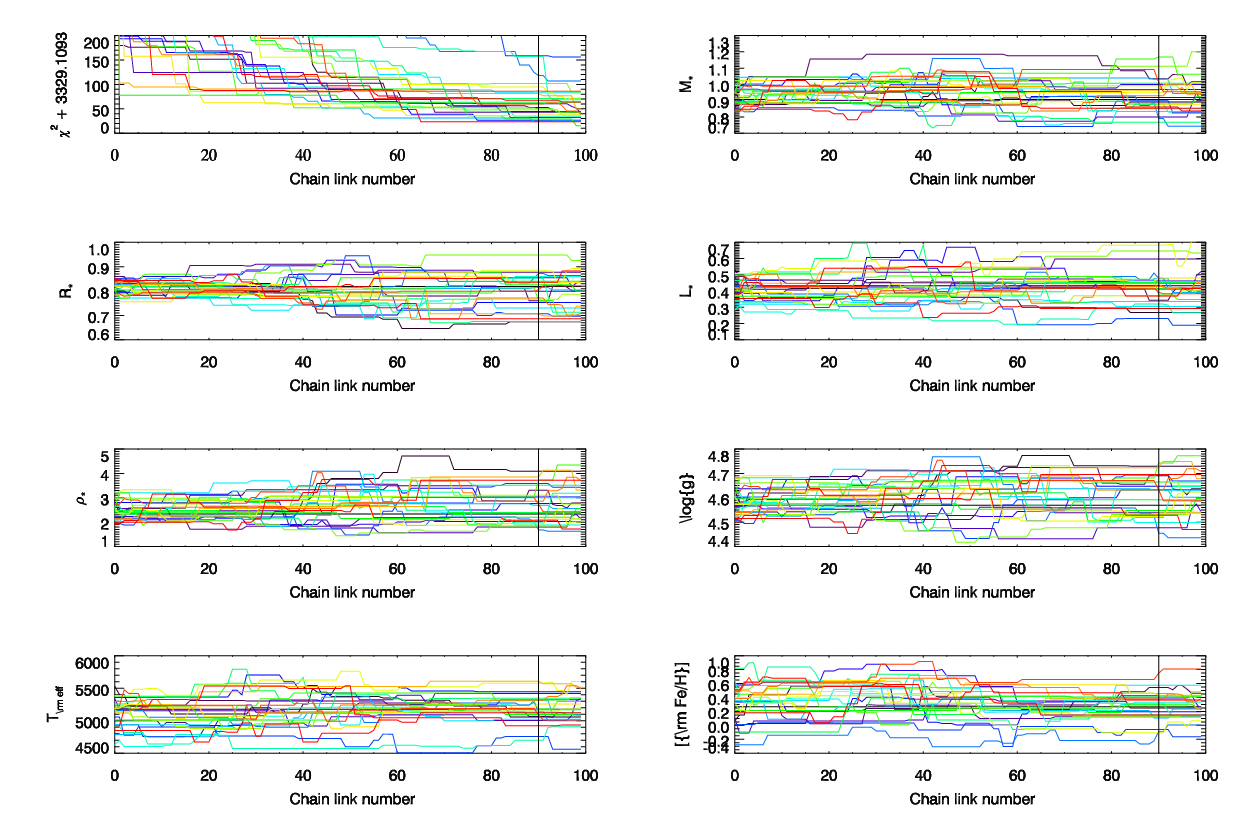}
    \caption{Same as Figure \ref{fig:chain}, but showing an example of a poorly mixed chain that should not be used for inference. Note the $\chi^2$ still trending downward, many parameters still expanding into their allowed volume and systematically drifting, and the burn-in pegged at its maximum (90\% of the total number of steps) each of which are major red flags individually, casting serious doubt on the robustness of this fit. Users should not trust the results that come out of a fit like this. They should increase MAXSTEPS and/or NTHIN and rerun the fit.}
    \label{fig:chainunmixed}
  \end{center}
\end{figure*}

We now generate improved versions of the diagnostic output plots. The Probability Distribution Function (PDF) plots now include the PDF of each chain as well as the average of all chains and the best fit value. Figure \ref{fig:pdf} shows an example for our canonical HAT-P-3 fit that is well-behaved, while Figure \ref{fig:pdfunmixed} shows a much shorter run to demonstrate what one might look for as an example of a poorly behaved fit. We also add a new diagnostic plot -- a plot of the $\chi^2$ and each parameter vs chain link for each chain and a line marking the burn in. These plots will aid the user in identifying runs that may be poorly mixed despite passing the convergence criteria, may help the user gain confidence in the results even though they may not have passed our strict convergence criteria, or may help the user identify problems with the runs (e.g., a chain stuck in a local minimum, a poor starting value). An example for our canonical HAT-P-3 fit is shown in Figure \ref{fig:chain} for a while behaved run, while Figure \ref{fig:chainunmixed} shows the same fit that is not yet well-mixed.

\begin{figure*}
  \begin{center}
    \includegraphics[width=6.5in, trim={2.5cm 1cm 6.5cm 0cm}]{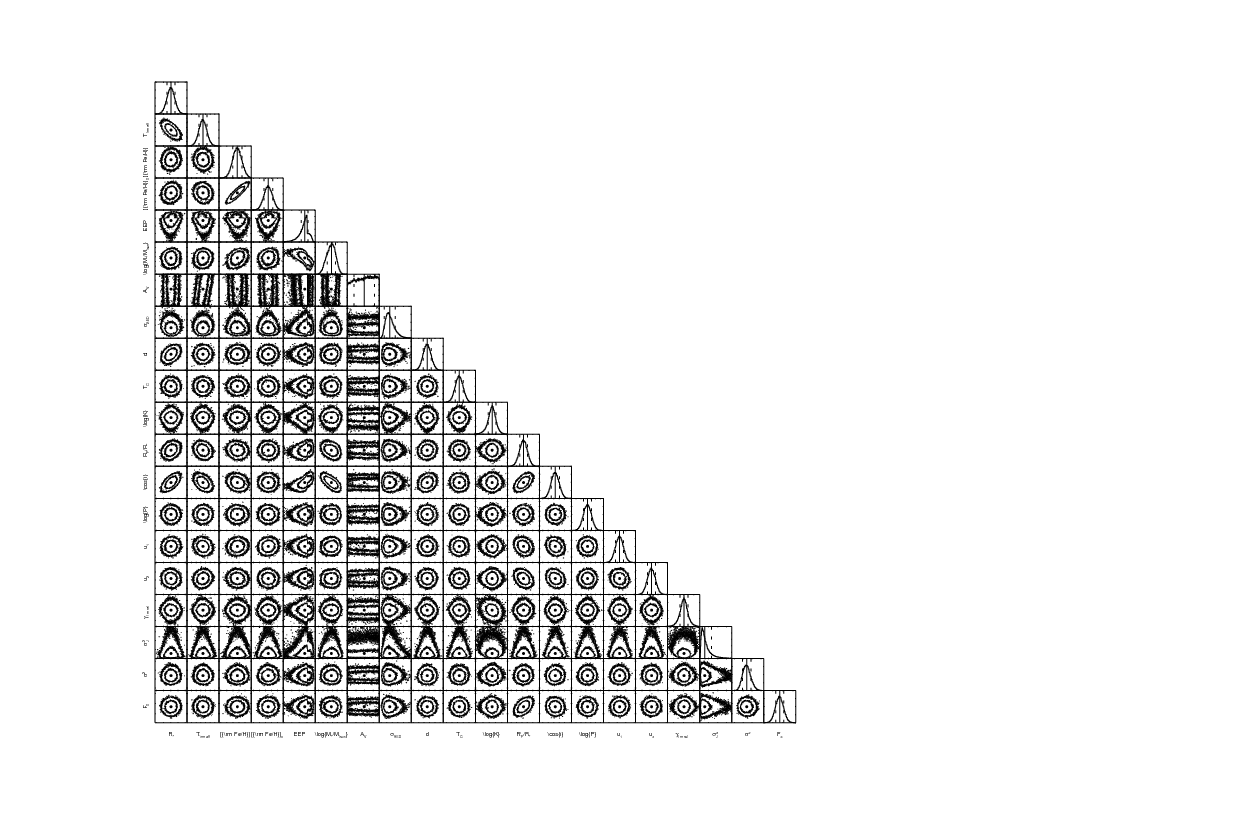}
    \caption{The corner plot showing the covariances for all fitted parameters in the HAT-P-3b global fit. The contours show the 68\% and 95\% confidence intervals, while the black circle marks the best-fit model parameter. The top of the corner plot shows the probability distribution function for each parameter, with the vertical line denoting the best-fit value. This plot is not intended for publication, but is useful for identifying diabolical covariances and building intuition about the nature of which data constrain what parameters, or what data sets would be most helpful to improve. With eccentric orbits, additional planets, or more data sets, the number of fitted parameters can be very large and this corner plot can get extremely tight.}
    \label{fig:covar}
  \end{center}
\end{figure*}

The covariance plots are now, by default, limited to just the fitted parameters and are plotted in a much easier to interpret, corner plot format. An example for HAT-P-3b fit is shown in Figure \ref{fig:covar}. For complex models, the number of derived parameters is large, and plotting each of those against all others made it time consuming to compute (sometimes hours) and the files too large to manage (many GBs) and difficult to navigate. 

With an arbitrary number of planets, transits, RV instruments, and wavelengths, the model plots can be complex. We have not handled all permutations to produce publication-quality figures, and typically multi-planet fits or fits that mix ground-based and space-based light curves will require some additional effort on the user's part. This is a source of ongoing improvement, but we supply the user with several ascii files with residuals and models to allow the user to create their own figures based on the best-fit model without having to recreate it. In the future, we may provide ancillary codes that use these outputs and generate more specialized publication quality figures.

\begin{figure*}
\centering
\includegraphics[height=2in, trim={0.5cm 0cm 0cm 0.5cm}]{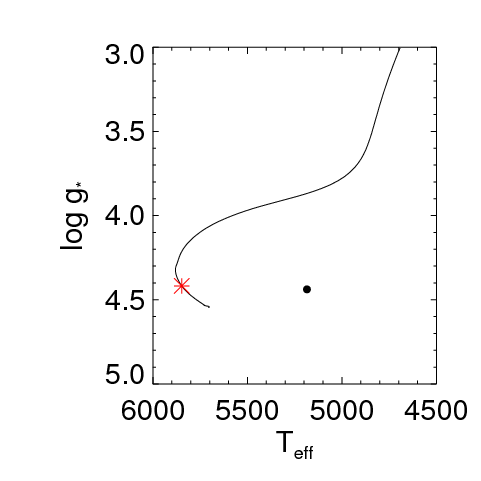}\label{fig:startmist}
\includegraphics[height=2in, trim={0.5cm 0cm 0cm 0.5cm}]{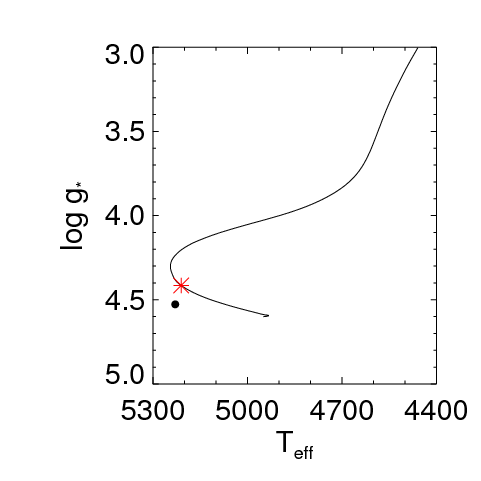}\label{fig:amoebamist}
\includegraphics[height=2in, trim={0.5cm 0cm 0cm 0.5cm}]{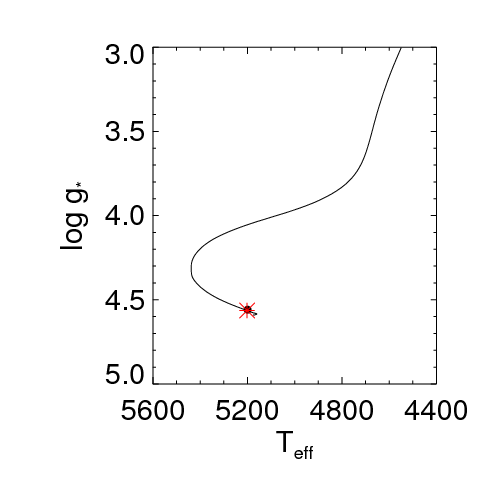}\label{fig:mcmcmist}
\caption{(a) -- The starting model, before any optimization, showing the MIST stellar track for HAT-P-3 fit. The black line represents the mass track interpolated at the model values for \mstar, \feh$_0$, and $EEP$. The black circle is at the model value for \teff \ and \logg. The red asterisk corresponds to the model value $EEP$ (age) along the track. Perfect consistency among all components of the global model would have the black circle perfectly overlap with the red asterisk. The internal inconsistency of the starting values, due to the default solar values for the stellar mass and radius, but using the spectroscopic values for \teff \ and \feh \ (a much cooler star) is relatively large and is likely to impact performance. The user could improve performance by supplying better values for \mstar, \rstar, and $EEP$, and should be concerned about the potential for a catastrophic failure to find the optimal global solution, due to the rough likelihood surface of the MIST models, but could proceed with caution. This diagnostic plot is not intended for publication. (b) -- Same as (a), but with {\tt AMOEBA}-optimized global model. Ideally, there would be perfect agreement between the red asterisk and black point, and would be indistinguishable from panel (c). Discrepancies within the uncertainties are not necessarily cause for alarm, but that we find a better fit in panel (c) means the MCMC could have been more efficient, spending extra time finding the global solution and using sub-optimal steps to define the initial spread in chains and the random offset in each step. We could improve performance by adjusting the starting parameters for the star. This diagnostic plot is not intended for publication. (c) Same as (a), but using the best-fit model among all links of all chains. That there is good agreement between black point and red asterisk is a good indication that the MCMC likely found the correct global solution. Subsequent fits would be more efficient if the prior file were regenerated based on this best-fit using {\tt MKPRIOR}. This publication-quality figure is a direct output of \exofasttwo \ when MIST models are used to constrain the star and should be published alongside the other figures of the global model.}
\label{fig:mist}
\end{figure*}

\begin{figure*}
\centering
\includegraphics[height=1.5in, trim={0.5cm 0cm 0cm 0.5cm}]{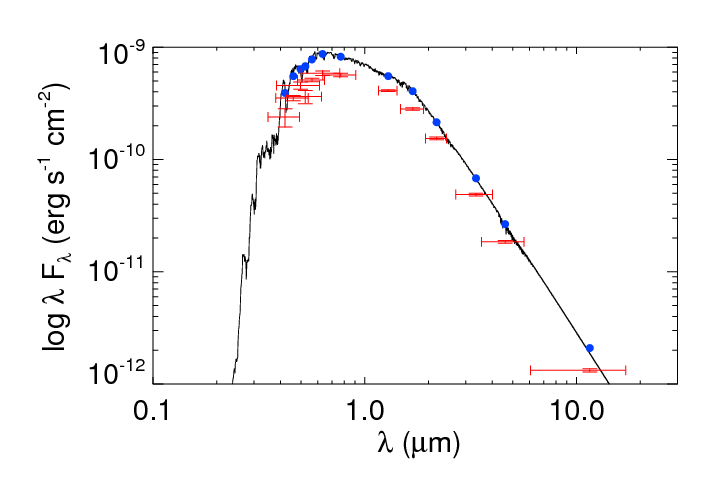}\label{fig:startsed}
\includegraphics[height=1.5in, trim={0.5cm 0cm 0cm 0.5cm}]{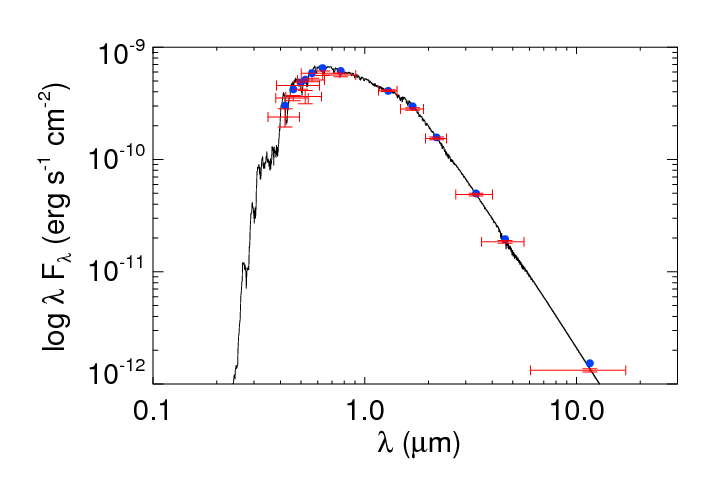}\label{fig:amoebased}
\includegraphics[height=1.5in, trim={0.5cm 0cm 0cm 0.5cm}]{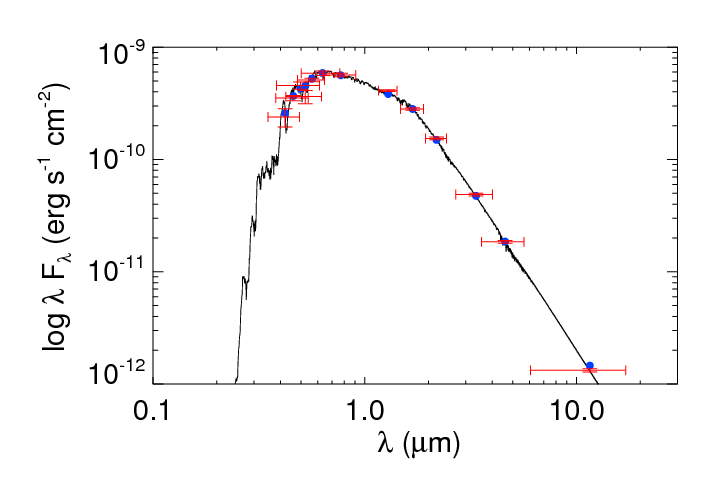}\label{fig:mcmcsed}
\caption{(a) -- The starting model, before any optimization, showing the model Spectral Energy Distribution (black) for HAT-P-3, with broad band averages (blue circles) and broad band measurements (red) from Table \ref{tab:LitProps}. The error bars in wavelength denote the bandwidth of the corresponding filter and the error bars in flux denote the measurement uncertainty. Perfect consistency among all components of the global model would have the blue circles perfectly overlap with the red data. With a robust prior on parallax from \gaia, the offset in normalization is due to the stellar radius. The SED model has a smooth likelihood surface and is easily optimized in the next step, but adjusting \rstar \ in the prior file so the normalization starts closer is a good way to help optimize the MIST model. This diagnostic plot is not intended for publication. (b) -- Same as (a), but with the SED model from the {\tt AMOEBA}-optimized global fit. The good agreement between the data and model, and with panel (c) is exactly what we want to see. This diagnostic plot is not intended for publication. (c) Same as (a), but using the best-fit model among all links of all chains. This publication-quality figure is a direct output of \exofasttwo \ when the SED model is used to constrain the star and should be published alongside the other figures of the global model.}
\label{fig:sed}
\end{figure*}

\begin{figure*}
\centering
\includegraphics[height=1.5in, trim={0.5cm 0cm 0cm 0.5cm}]{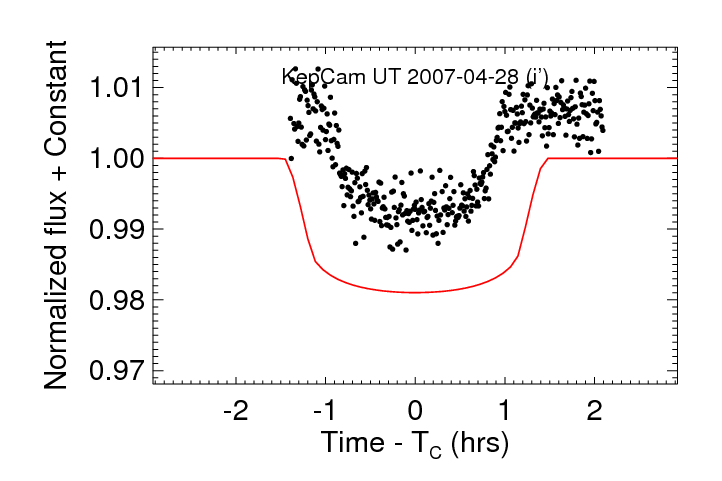}\label{fig:starttransit}
\includegraphics[height=1.5in, trim={0.5cm 0cm 0cm 0.5cm}]{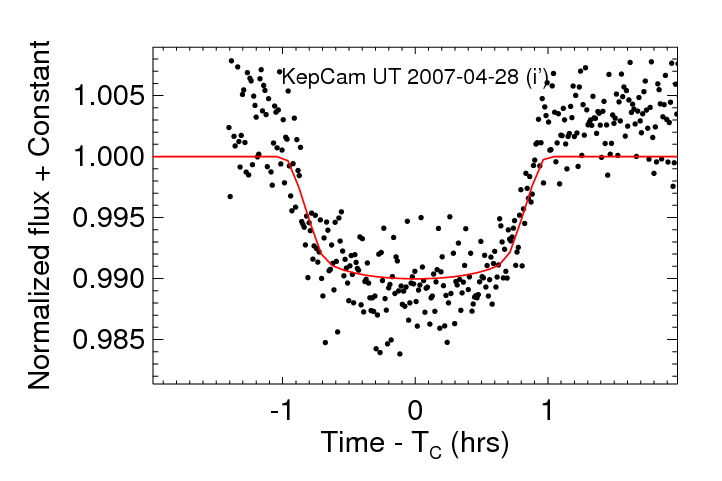}\label{fig:amoebatransit}
\includegraphics[height=1.5in, trim={0.5cm 0cm 0cm 0.5cm}]{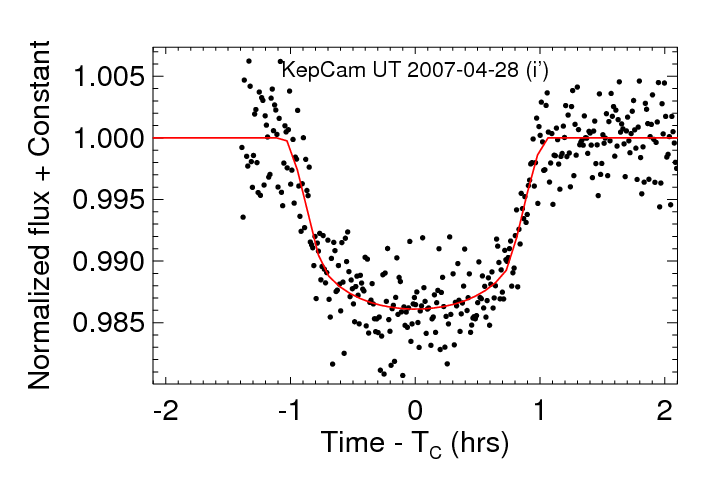}\label{fig:mcmctransit}
\caption{(a) -- The starting model, before any optimization, showing the model transit (red) over the data (black circles) for HAT-P-3b. The input light curve should be normalized so the out of transit data is 1, but it is $\sim$1\% off. The transit model has a relatively smooth likelihood surface and is usually easily optimized in the next step, but adjusting $F_0$ in the prior file or renormalizing the lightcurve may help improve performance. Adjusting \mstar \ and \rstar \ (\rhostar) will also improve the agreement with the transit duration. This diagnostic plot is not intended for publication. (b) -- Same as (a), but with transit model from the {\tt AMOEBA}-optimized global fit. Ideally, this figure would be indistinguishable from panel (c), but the poor match between the model and the out-of-transit normalization means {\tt AMOEBA} did not find the correct global solution. Large discrepancies are likely to lead to catastrophic failures. Small discrepancies (like we see here) are likely to lead to a performance hit. We could improve performance by adjusting the starting parameters for $F_0$, \mstar, and \rstar \ to better match the out of transit flux and transit duration. This diagnostic plot is not intended for publication. (c) Same as (a), but using the best-fit model among all links of all chains. That there is good agreement between the data and model is a good indication that the global fit likely found the optimal global solution. Subsequent fits would be more efficient if the prior file were regenerated based on this fit using {\tt MKPRIOR}. This publication-quality figure is a direct output of \exofasttwo \ when the transit is fit and should be published alongside the other figures of the global model.}
\label{fig:transit}
\end{figure*}

\begin{figure*}
\centering
\includegraphics[height=1.5in, trim={0.5cm 0cm 0cm 0.5cm}]{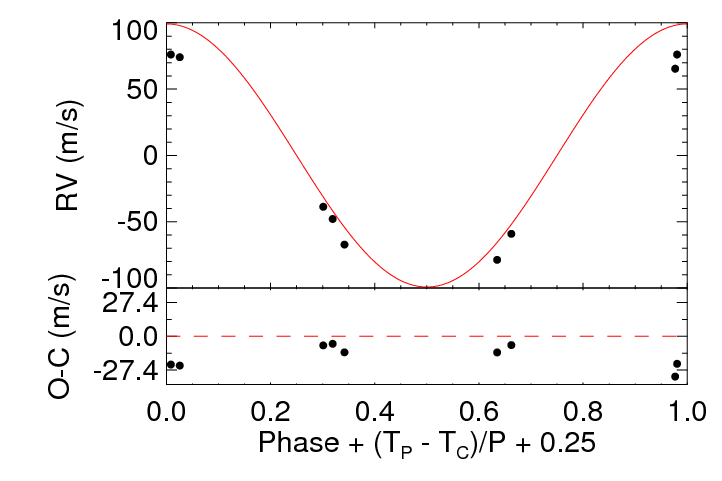}\label{fig:startrv}
\includegraphics[height=1.5in, trim={0.5cm 0cm 0cm 0.5cm}]{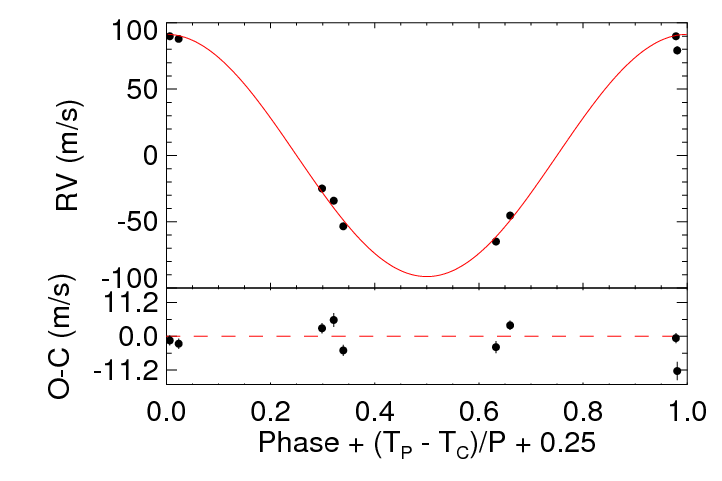}\label{fig:amoebarv}
\includegraphics[height=1.5in, trim={0.5cm 0cm 0cm 0.5cm}]{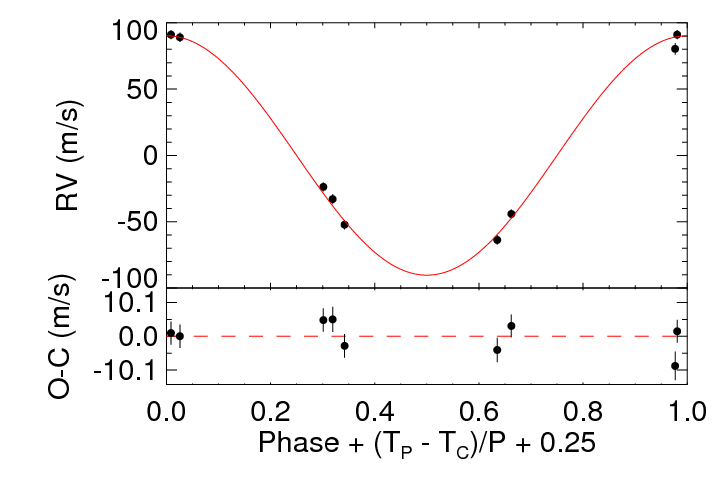}\label{fig:mcmcrv}
\caption{(a) -- The starting model, before any optimization, showing the model radial velocity (red) over the data (black circles) for HAT-P-3b. There is good agreement between the data and model, so no further adjustments are necessary. This diagnostic plot is not intended for publication. (b) -- Same as (a), but with {\tt AMOEBA}-optimized global model. The slight improvement over panel (a) and that it is indistinguishable from panel (c) is exactly what we want to see. This diagnostic plot is not intended for publication. (c) Same as (a), but using the best-fit model among all links of all chains. That there is good agreement between the data and model is a good indication that the global fit likely found the optimal global solution. This publication-quality figure is a direct output of \exofasttwo \ when the radial velocity is fit and should be published alongside the other figures of the global model.}
\label{fig:rv}
\end{figure*}

Three versions of each of the model plots are created. The first version is created at the very beginning of the fit and shows the starting model overplotted on the data. Another version shows the best-fit found by the {\tt AMOEBA} optimization. It is essential that the user inspect all of these plots, especially when starting a new fit. If the starting model is way off, subsequent steps of the fit are likely to fail and the error messages may be opaque. If the {\tt AMOEBA} model is a poor fit to the data, the MCMC is likely to fail or take much longer than it otherwise would. In either case, the user should change the starting values in the prior file to get better results. If the user is unable or unwilling to manually get better results, parallel tempering is compute-time expensive, but often an effective way to find a good, global solution. Subsequent runs can be started with a prior file created by {\tt MKPRIOR}. The final version of these plots show the best-fit among all links of all chains overlaid on the data, and is generally expected to be published. Examples of the MIST, SED, transit, and RV, model plots generated at all three steps for the HAT-P-3b fit are shown in figures \ref{fig:mist}, \ref{fig:sed}, \ref{fig:transit}, and \ref{fig:rv}, respectively.

\section{Troubleshooting}
\label{sec:troubleshooting}
The flexibility of \exofasttwo \ is tremendous, but it is easy to create a combination of inputs that is nonphysical and fails. We have provided a number of example fits that should serve as a template for a wide variety of potential use cases, but the number of combinations is large and difficult to anticipate and handle with graceful failures and clear error messages in every case. 

If the supplied examples fail, it is likely that it was not installed correctly. The most common reasons for this are incorrectly set environment variables, missing libraries, or versions of critical codes used by \exofasttwo \ that behave differently with a higher precedence in the IDL PATH. When version problems with library routines are identified, we rename the working versions and include copies with our distribution.

We will continue to improve the code and push updates to github, but a user unfamiliar with nuances of IDL, \exofasttwo, or global fitting who ventures too far from the examples is likely to encounter failures with limited feedback and perhaps even misleading error messages. Most often, the starting conditions (as defined in the prior file) will be too far from the best fit or a critical parameter will be unconstrained (either by an explicit prior or missing data set) and \exofasttwo \ will be unable to begin sensibly. In such cases, using the {\tt DEBUG} and {\tt VERBOSE} flags can be extremely helpful. {\tt DEBUG} will overplot the model on top of the data to the screen, whereas {\tt VERBOSE} will print out the penalties for each component of the fit and additional information like when a step encounters a boundary. The user can then adjust relevant starting values via the prior file and restart. Or, the user can continue and the code and it will generate those same outputs at each step as it settles into the best fit and begins the MCMC, which often quickly reveals the problem (but is much much slower, so should not be used once problems are resolved).

Another common failure is an unconstrained parameter due to an absent (or low-quality) critical data set or prior. Ignoring these warnings is likely to end poorly. For example, setting the {\tt NOMIST} keyword to disable the MIST evolutionary models without providing a stellar mass and radius prior as an alternative will result in problems. The user must read the documentation carefully and make sure to have the proper constraints.

Finally, if the reader is stumped, do not hesitate to ask the author for help. The documentation is not perfect and the code will not work sensibly in every case, especially in the beginning. Bringing failures to our attention allows us to improve the documentation or code for everyone.

\section{Fitting TESS TOIs}
\label{sec:tess}
The flexibility of \exofasttwo \ allows us to batch automate large numbers of fits on a supercomputer. We are now fitting all the \tess \ Objects of Interest (TOIs) and uploading the results to ExoFOP-TESS to aide the vetting, characterization, and publication of candidates. For example, the duration of some transits requires an unlikely eccentricity when around the presumed host star. Or, some are so V-shaped the only solutions allowed are a large, grazing body. Therefore, such events can be de-prioritized as a likely false positive.

We automate the creation of the SED with {\tt MKSED}, which also grabs the parallax from \gaia \ DR2. We use {\tt GETAVPRIOR} to determine the upper limit on $A_V$ from the \citet{Schlafly:2011} dust maps. Because the stellar parameters in the TIC were often wrong, and difficult to correct in an automated fashion, we start all fits with a Sun-like host star. Therefore, in order to adequately explore the stellar parameter space, we use parallel tempering with 8 temperatures and a maximum temperature of 200.

The dilution is already accounted for in the lightcurves, but we fit it, with a prior of $0 \pm 10\%$ of the already-applied correction to propagate any uncertainty in the correction. Note that the dilution in the TIC is given as a contrast ratio, $C$, so we must correct it as described in \S \ref{sec:dilute}. The accuracy of the contamination reported in the TIC has not been thoroughly evaluated, but fails catastrophically when the 2MASS does not resolve close companions ($\lesssim2\farcs0$), and so the TIC is unaware of their contribution. This will be improved in TIC8, which will be based on \gaia \ positions, but unidentified companions will always be cause for concern when modeling SEDs and can lead to catastrophic failures in the automated fits we upload to ExoFOP-TESS. These can often be identified by poor SED, MIST, or transit fits.

All available sectors of TESS data are modeled simultaneously. Multi-planet candidates are modeled as such and we constrain the eccentricity by requiring their orbits not to cross. We use starting points from the TOI releases for $T_C$, $Period$, and transit depth for each candidate. We trim the lightcurves to only include points within twice the duration determined in the alert around each transit, except for models with single-transit events, which include all points to constrain its period. We bound $T_C \pm P/3$ (or $\pm 3$ days for single-transit planets) and $P \pm 10\%$ (or $P=30$ days with no limits for single-transit planets) and reject flat transit models to prevent the parallel tempering from running away. We do not impose priors on \teff, \feh, or \logg. The provenance for those parameters in the TIC is not programmatically available and often originate from an SED and/or isochrones. Instead, we fit the SED and MIST stellar tracks to constrain \teff, \feh, and \logg. Typically, more precise results could be obtained by applying priors on \teff \ and \feh \ from high resolution spectroscopy.

If the data are from the Quick Look Pipeline (QLP), there is no need to flatten the lightcurve, and we set {\tt EXPTIME=30, NINTERP=10}, to generate 10 points over the 30 minute exposure for each data point to integrate the model. If the data are from the SPOC, we flatten the light curve by dividing out a spline with automatically identified breakpoints using the procedure and code by \citet{Vanderburg:2014}. We set limits on the runtime of 2.5 days, 7,500 steps (thinned by 30), or until all parameters are converged ($Tz > 1000$, $Rz < 1.01$). In practice, most fits reach the 2.5 day runtime limit, and so do not pass our strict convergence tests, but barring catastrophic failures, the reported results are usually accurate to $< 0.5\sigma$.

In addition to the results, we upload a refined prior file (to start the fit at the best fit we found) and all the startup files we used to run the fit. This makes it almost trivial for \exofasttwo \ users to re-run a longer or customized fit, adding additional follow-up data or spectroscopic priors, if available.

Because these fits are done in an automated fashion with limited error checking, the user would be wise to inspect the fit as they would any of their own fits before blindly trusting the results. The results are usually reasonably accurate, but rarely pass our strict convergence criteria and could almost always be improved by running longer. Some 5-10\% of fits fail catastrophically, mostly due to blended SED photometry and the resulting bias in the stellar radius. Some additional fits fail catastrophically due to poor flattening of the transit light curve, which we hope to improve in the future.

One complication to running \exofasttwo \ on a supercomputer is the issue of IDL licenses. While \exofasttwo \ can be run without a license (see \S \ref{sec:license}), starting the virtual machine requires the user to click ``OK'', with the intention of forbidding this type of use case. We considered automating the graphical interaction, but found a more robust work around. IDL uses one license per display. Because IDL is not highly multi-threaded, we can run a fit on each core of a given node with only a minor impact on the serial runtime and use only one license per node. Odyssey, the supercomputer at Harvard we are using for these fits, has 62 IDL licenses, and many 32-core machines. We wish to leave some licenses for others, but we have never seen more than 10 licenses in use by others. Therefore, we can run 32 fits on each of fifty 32-core machines at once, or a total of 1600 simultaneous fits -- far exceeding the current number of TOIs. We include an example sbatch file that implements this scheme, {\tt \$EXOFAST\_PATH/fit.sbatch}, to submit jobs to a supercomputer using the SLURM workload manager, should others wish to batch automate large numbers of fits with \exofasttwo.

\section{Acknowledgements}
\label{sec:acknowledgement}

If you use \exofasttwo \ in your work, please acknowledge the work of the many people whose work it is built from by using the following text as a guideline.

This work made use of \exofasttwo \ \citep{Eastman:2013,Eastman:2017}, which is built on the work of many. The transit model is generated using \citet{Mandel:2002,Agol:2019} with limb darkening parameters constrained by \citet{Claret:2011} and \citet{Claret:2017}. The exoplanet mass radius relation from \citet{Chen:2017} is used to estimate the mass or radius (and all relevant derived parameters) of the exoplanet, in the absence of a RV data set or transit, respectively. The stellar physics is constrained from either the empirical relations laid out by \citet{Torres:2010}, the Yonsie Yale stellar evolutionary models \citet{Yi:2001}, or the MIST evolutionary models \citep{Choi:2016, Dotter:2016}, which itself is built using MESA \citep{Paxton:2011,Paxton:2013,Paxton:2015}. Stellar atmospheric models from NextGen \citep{Allard:2012}, ATLAS \citep{Kurucz:1979}, and PHOENIX \citep{Hauschildt:1997} underlie several aspects of the code. \exofasttwo \ also makes use of the IDL astronomy library \citep{Landsman:1993}.

\exofasttwo's SED modeling code relies on catalogs from Galex \citep{Bianchi:2011}, Tycho-2 \citep{Hog:2000}, UCAC4 \citep{Zacharias:2012}, APASS \citep{Henden:2016}, 2MASS \citep{Cutri:2003}, WISE \citep{Cutri:2013}, Gaia \citep{Gaia:2016}, the Kepler INT Survey \citep{Greiss:2012}, the UBV Photoelectric Catalog \citep{Mermilliod:1994}, and the Stroemgren-Crawford uvby$\beta$ \ photometry catalog \citep{Paunzen:2015}, as well as extinction from \citet{Schlegel:1998} and \citet{Schlafly:2011} and parallax from \gaia \ DR2 \citep{Gaia:2018}.

An updated version of this acknowledgement, along with the bibtex entries for each citation, is maintained and distributed with the code as \\ {\tt \$EXOFAST\_PATH/acknowledgements.tex}.

\acknowledgements
We would like to thank the many users of \exofast \ and early adopters of \exofasttwo \ who have asked questions, reported bugs, and suggested features, which have dramatically improved the code and documentation. We thank the Sagan Summer Workshop for providing a platform to teach students how to use \exofasttwo. Work by J.D.E. was funded by the National Science Foundation (grants 1516242 and 1608203) and NASA ADAP 80NSSC19K1014. Work by K.G.S. was partially funded by NASA grant 17-XRP17 2-0024.  Work by E.A.\ was supported by NSF grant AST-1615315 and NASA grants NNA13AA93A and 80NSSC18K0829.

Computations in this paper were run on the Odyssey cluster supported by the FAS Division of Science, Research Computing Group at Harvard University.


\end{document}

%% file: explain.tex
\newcommand{\specialcell}[2][c]{%
  \begin{tabular}[#1]{@{}c@{}}#2\end{tabular}}

\startlongtable
\begin{deluxetable*}{llcc}
\tablecaption{Explanation of \exofasttwo \ output parameters.}
\label{tab:explain}
\tablehead{\colhead{~~~Parameter} & \colhead{Units} & \colhead{Variable} & \colhead{Bounds}}
\startdata
\smallskip\\\multicolumn{2}{l}{Stellar Parameters:}&\smallskip\\
\hline
~~~~$\log{ \left(M_*/M_{\sun}\right) }$\dotfill & Mass\dotfill & logmstar & $0.001 <  M_* \leq 500$ \\
\multicolumn{4}{p{17cm}}{The log of the stellar mass in solar  units. The MIST grid spans $0.1 \leq M_* \leq 300$, beyond which models are rejected (not extrapolated). Known systematic errors exist for stars on the low mass end of their range. The YY grid spans $0.4 < M_* < 5$. The span of the Torres data are $0.2 < M_* < 30$, though it is sparsely populated at the extremes. All stellar models should be disabled and replaced with direct priors for extreme stars.}\\
\hline
~~~~$M_*$\dotfill & Mass (\msun)\dotfill & mstar & $0.001 <  M_* \leq 500$  \\
\multicolumn{4}{p{17cm}}{Same as above, but in linear solar units.} \\
\hline
~~~~$R_*$\dotfill & Radius (\rsun)\dotfill & rstar & $10^{-6} < R_* < 2000$ \\
\multicolumn{4}{p{17cm}}{The stellar radius in solar units. The span of the MIST models is $ 0.008 < R_* < 1432$. 
The span of the Torres relation is $0.2 < R_* < 30$.}\\
\hline
~~~~$L_*$\dotfill & Luminosity (\lsun)\dotfill &lstar & None \\
\multicolumn{4}{p{17cm}}{The stellar luminosity, $\left(\frac{R_*}{R_\sun}\right)^2\left(\frac{T_{\rm eff}}{T_{\sun}}\right)^4$, in solar units. There are no explicit limits on this derived parameter, but the bounds on $T_{\rm eff}$ and $R_*$ limit its range.}\\
\hline
~~~~$\rho_*$\dotfill & Density (cgs)\dotfill & rhostar & None \\
\multicolumn{4}{p{17cm}}{The stellar density, $\frac{3M_*}{4\pi R_*^3}$, in g~cm$^{-3}$. There are no explicit limits on this derived parameter, but the bounds on $M_*$ and $R_*$ limit its range.}\\
\hline
~~~~$\log{g}$\dotfill & Surface gravity (cgs)\dotfill & logg & None \\
\multicolumn{4}{p{17cm}}{The log$_{10}$ of the stellar surface gravity, $\log_{10}\left({\frac{GM_*}{R_*^2}}\right)$, in cgs units. The span of MIST is $-1 \lesssim \log{g} \lesssim 10$. 
}\\
\hline
~~~~$T_{\rm eff}$\dotfill & Effective Temperature (K)\dotfill & teff & $100 < T_{\rm eff} < 250000$ \\
\multicolumn{4}{p{17cm}}{The stellar effective temperature in Kelvin. Note that the stellar evolution and stellar atmosphere grids are not applicable in this entire range. The span of the MIST models is $2200 < T_{\rm eff} < 400000$, and we would advise caution fitting stars cooler than 3500 K or hotter than 10000 K. 
The span of the Torres relation is $3000 < T_{\rm eff} < 40000$. 
} \\
\hline
~~~~$[{\rm Fe/H}]$\dotfill & Metallicity (dex)\dotfill & feh & $-10 < [{\rm Fe/H}] < 2$ \\
\multicolumn{4}{p{17cm}}{The stellar surface iron abundance. The MIST-derived surface iron abundances are $[{\rm Fe/H}] \gtrsim -5$, though for most stars, it tracks $[{\rm Fe/H}]_{0}$ relatively closely, which spans $-4 < [{\rm Fe/H}]_{0} \leq 0.5$. The YY grid spans from $-3.29 < [{\rm Fe/H}] < 0.78$. The metalicity of only 21 stars in the Torres sample was known, and span a relatively narrow range around solar metalicity: $-0.6 < [{\rm Fe/H}] < 0.4$. 
}\\
\hline
~~~~$[{\rm Fe/H}]_{0}$\dotfill & Initial Metallicity \dotfill & initfeh & $-5 \leq [{\rm Fe/H}]_{0} \leq 0.5$ \\
\multicolumn{4}{p{17cm}}{The initial stellar surface iron abundance at Age=0 that define the grid points for the MIST stellar tracks. See \citet{Dotter:2016}. Only displayed for MIST fits. The MIST grid spans $-5 < [{\rm Fe/H}]_{0} \leq 0.5$.} \\
\hline
~~~~$Age$\dotfill & Age (Gyr)\dotfill & age & $0 < Age < 13.82$ \\
\multicolumn{4}{p{17cm}}{The stellar age, in billions of years. Only displayed for YY or MIST fits. While the MIST grid spans much older stars, we exclude stars who's age exceeds that of the universe. It is fit when an age prior is supplied or for YY fits. The Torres relation does not constrain the age.}\\
\hline
~~~~$EEP$\dotfill & Equal Evolutionary Phase \dotfill & eep & $0 \leq EEP \leq 1709$ \\
\multicolumn{4}{p{17cm}}{The ``Equal Evolutionary Phase'', which indexes common evolutionary phases (e.g., the turn off = 454) and defines the grid points for the MIST stellar tracks, which is essentially a proxy for age. See \citet{Dotter:2016} and \S \ref{sec:mist} for a detailed explanation. Only displayed for MIST fits.}\\
\hline
~~~~$vsinI_*$\dotfill & Projected rotational velocity (m/s)  \dotfill& vsini & None \\
\multicolumn{4}{p{17cm}}{The projected rotational velocity of the star, in m/s. Only displayed for DT or RM fits. See \S \ref{sec:dt}.}\\
\hline
~~~~$V_{\rm line}$\dotfill & Unbroadened line width (m/s)  \dotfill& vline & None \\
\multicolumn{4}{p{17 cm}}{The average line-width for the star without any rotational broadening, in units of m/s. It includes effects from macroturbulence, thermal, and pressure broadening and is typically 2,000 to 10,000 m/s. Only displayed for DT fits.
}\\
\hline
~~~~$A_V$\dotfill & V-band extinction (mag)\dotfill & av & $0 \leq A_V \leq 100$ \\
\multicolumn{4}{p{17cm}}{The V-band extinction for the SED model. Only displayed for SED fits.}\\
\hline
~~~~$\sigma_{SED}$\dotfill & SED photometry error scaling \dotfill & errscale & $0.01 < \sigma_{SED} < 100$ \\
\multicolumn{4}{p{17cm}}{The multiplicative factor to scale the supplied broadband SED photometric errors to ensure they are consistent with the model. If these conservative bounds are encountered, the error bars should be checked and scaled manually.}\\
\hline
~~~~$\mu_{\alpha}$\dotfill & RA Proper Motion (mas/yr)  \dotfill& pmra & $-20000 < \mu_{\alpha} < 20000$ \\
\multicolumn{4}{p{17cm}}{The proper motion in Right Ascension, in milli arcseconds per year. Includes the $\cos{\delta}$ term (i.e., what \gaia \ reports). Only displayed for astrometry fits. The bound is twice Barnard's star, the star with the highest known proper motion.}\\
\hline
~~~~$\mu_{\delta}$\dotfill & Dec Proper Motion (mas/yr) \dotfill & pmdec & $-20000 < \mu_{\delta} < 20000$ \\
\multicolumn{4}{p{17cm}}{The proper motion in Declination, in milli arcseconds per year (i.e., what \gaia \ reports). Only displayed for astrometry fits. The bound is twice Barnard's star, the star with the highest known proper motion.}\\
\hline
~~~~$\gamma_{\rm abs}$\dotfill & Absolute RV (m/s)  \dotfill & rvabs & $-c < \gamma_{\rm abs} < c$ \\
\multicolumn{4}{p{17cm}}{The Absolute RV zero point, in m/s. Only displayed for astrometry fits. It is bounded by the speed of light.}\\
\hline
~~~~$\varpi$\dotfill & Parallax (mas)\dotfill & parallax & None \\
\multicolumn{4}{p{17cm}}{The parallax in milli arcseconds. Only displayed for SED or astrometry fits. There are no explicit bounds, but the bound on distance limits its range.}\\
\hline
~~~~$d$\dotfill & Distance (pc)\dotfill & distance & $0.9 AU < d < 3\times10^{10}$ pc \\
\multicolumn{4}{p{17cm}}{The distance in parsecs. Only displayed for SED fits. The bounds include the Sun to the size of the observable universe.}\\
\hline
~~~~$\dot{\gamma}$\dotfill & RV slope (m/s/day) \dotfill& slope & $\mid \dot{\gamma} \mid < 10000$ \\
\multicolumn{4}{p{17cm}}{A linear trend in RV, in m/s/day, referenced to the midpoint of all supplied RV data from all instruments ($RV += \dot{\gamma}(t-t_m)$, where $t_m = max(t) + min(t)/2$). Intended for long period planets where the data are insufficient to constrain a full Keplerian orbit. Only displayed if {\tt FITSLOPE} flag set.}\\
\hline
~~~~$\ddot{\gamma}$\dotfill & RV quadratic term (m/s/day$^2$)\dotfill & quad & $\mid \ddot{\gamma} \mid < 10000$ \\
\multicolumn{4}{p{17cm}}{A quadratic trend in RV, in m/s/day$^2$, referenced to the midpoint of all supplied RV data from all instruments ($RV += \ddot{\gamma}(t-t_m)^2$, where $t_m = max(t) + min(t)/2$).  Intended for long period planets where the data are insufficient to constrain a full Keplerian orbit. Only displayed if {\tt FITQUAD} flag set. This should rarely be used without also fitting for a slope in the RV (i.e., setting the {\tt FITSLOPE} flag).}\\
\smallskip\\\multicolumn{2}{l}{Planetary Parameters:}\smallskip\\
\hline
~~~~$P$\dotfill &Period (days)\dotfill & period & $0.1 < P < 10^{13}$ \\
\multicolumn{4}{p{17cm}}{The period of the planet, in days. Note this period is in the solar system barycentric frame and does not account for the small (but often statistically significant) difference due to the absolute radial velocity of the planet. A tighter, user-supplied uniform bound is highly recommended, especially when using parallel tempering. The upper bound is roughly a Hubble time.}\\
\hline
~~~~$R_P$\dotfill &Radius (\rj)\dotfill & rp & None \\
\multicolumn{4}{p{17cm}}{The radius of the planet, in Jupiter radii. Note that for RV-only fits, this quantity is derived from the \citet{Chen:2017} exoplanet mass-radius relations. There is no internal bound on this quantity and it may be negative. Negative values produce a bump in the transit lightcurve instead of a dip. When \citet{Chen:2017} is used, it must be positive.} \\
\hline
~~~~$R_P$\dotfill &Radius (\re)\dotfill & rpearth & None \\
\multicolumn{4}{p{17cm}}{Same as above, but in Earth radii. Only displayed when {\tt EARTH} flag is set.}\\
\hline
~~~~$T_C$\dotfill &Time of conjunction (\bjdtdb)\dotfill & tc & $\mid T_C-T_{C,0} \mid < P/2$ \\
\multicolumn{4}{p{17cm}}{The time of conjunction that is closest to the starting value supplied in the prior file, which is typically a good proxy for the time of transit. See \S \ref{sec:ttvs} for a more detailed explanation. This value is not allowed to stray more than $\pm P/2$ from its starting value, though note that the period used for this bound is from the current step and may be much larger than the initial period.} \\
\hline
~~~~$T_0$\dotfill &Optimal conjunction Time (\bjdtdb)\dotfill & t0 & None \\
\multicolumn{4}{p{17cm}}{The time of conjunction that minimimzes the covariance with Period and therefore has the smallest uncertainty. Practically, it will never be more than 1 period outside of the span of the input data.}\\
\hline
~~~~$T_T$\dotfill &Transit time (\bjdtdb)\dotfill & tt & N/A \\
\multicolumn{4}{p{17cm}}{The time of minimum projected separation between the star and planet, as seen by the observer, including any modeled TTVs. This is the often assumed, but rarely used, meaning of ``transit time''. This value is reported in a separate table, never in the primary output table, but is described here for reference only. This quantity is not computed during the MCMC, and no bounds can be applied based on it.}\\
\hline
~~~~$a$\dotfill &Semi-major axis (AU)\dotfill & a & None \\
\multicolumn{4}{p{17cm}}{The semi-major axis of the planetary orbit in AU. There are no explicit limits, but it is bounded by the stellar and planetary masses, and planetary period through Kepler's law.}\\
\hline
~~~~$i$\dotfill &Inclination (Degrees)\dotfill & ideg & $0 \leq i \leq 180$ \\
\multicolumn{4}{p{17cm}}{The inclination of the orbit in degrees. When only transits and/or RVs are fit, $i \leq 90^\circ$. When only RVs are fit, there is no constraint and we marginalize over $0 \leq \cos{i} \leq 1$.}\\
\hline
~~~~$i$\dotfill &Inclination (Radians)\dotfill & i & $0 \leq i \leq \pi$ \\
\multicolumn{4}{p{17cm}}{Same as above, but in radians.}\\
\hline
~~~~$\cos{i}$\dotfill & Cos of Inclination \dotfill & cosi & $-1 \leq \cos{i} \leq 1$ \\
\multicolumn{4}{p{17cm}}{The cosine of the inclination. When only transits and/or RVs are fit, $\cos{i} \geq 0$. When only RVs are fit, there is no constraint and we marginalize over $0 \leq \cos{i} \leq 1$. This fitted parameter is never displayed in the output table.} \\
\hline
~~~~$e$\dotfill &Eccentricity \dotfill & e & $0 \leq e \leq 1-\frac{a+R_p}{R_*}$ \\
\multicolumn{4}{p{17cm}}{The eccentricity of the planet, with a uniform eccentricity prior. The upper bound rejects models where the star and planet would collide during periastron. For multi-planet systems, orbits that cross into each other's hill spheres are also excluded. If {\tt TIDES} flag is set, models are also rejected when $1-3a/R_* < e < 1-\frac{a+R_p}{R_*}$, which is an empirical limit on eccentricity justified by tidal circularization. Because of the Lucy-Sweeney bias due to hard boundary at 0, eccentricities that are non-zero with significance of 2.3 $\sigma$ or less should be considered consistent with circular.}\\
\hline
~~~~$\omega_*$\dotfill &Argument of Periastron (Degrees)\dotfill & omegadeg & $-180 \leq \omega_* < 180$ \\
\multicolumn{4}{p{17cm}}{The argument of periastron of the star's orbit due to the planet, in degrees, as measured in the standard, left-handed coordinate system. This is the standard value to report because that is the measured quantity for RV orbits. It differs from $\omega_P$ by 180 degrees. Its reported confidence interval may cross the stated bounds since the distribution is re-centered around the mode.}\\
\hline
~~~~$\omega_*$\dotfill &Argument of Periastron (Radians)\dotfill & omega & $-\pi \leq \omega_* < \pi$ \\
\multicolumn{4}{p{17cm}}{Same as above, but in radians. This fitted parameter is never displayed in the output table.} \\
\hline
~~~~$\Omega_*$\dotfill &Longitude of ascending node (Deg)\dotfill & bigomegadeg & $0 \leq \Omega \leq 360$ \\
\multicolumn{4}{p{17cm}}{The longitude of the ascending node, as measured in the standard, left-handed coordinate system. Only displayed for astrometry fits. If only astrometry is fit, this value is restricted from 0 to 180 and it is the longitude of an unknown node. Adding transits or RV break this degeneracy, and we report the longitude of ascending node from 0 to 360 degrees. Its reported confidence interval may cross the stated bounds since the distribution is re-centered around the mode.}\\
\hline
~~~~$\Omega_*$\dotfill &Longitude of ascending node (Rad)\dotfill & bigomega & $0 \leq \Omega \leq 2\pi$ \\
\multicolumn{4}{p{17cm}}{Same as above, but in radians. This fitted parameter is never displayed in the output table.} \\
\hline
~~~~$T_{eq}$\dotfill &Equilibrium temperature (K)\dotfill & teq & None \\ 
\multicolumn{4}{p{17cm}}{The equilibrium temperature of the planet, in Kelvin, calculated according to Eq. 1 of Hansen \& Barman, 2007, $T_{eq} = T_{\rm eff}\sqrt{\frac{R_*}{2a}}$, which assumes no albedo and perfect redistribution. The quoted statistical error is likely severely underestimated relative to the systematic error inherent in this assumption. There are no explicit limits on this derived parameter, but it is bounded by the limits on $T_{\rm eff}$, $R_*$, and $a$.}\\
\hline
~~~~$\tau_{\rm circ}$\dotfill &Tidal circularization timescale (Gyr)\dotfill & tcirc & $\tau_{\rm circ}$ > 0 \\
\multicolumn{4}{p{17cm}}{The tidal circularization timescale, using Equation 3 from \citet{Adams:2006}, $\tau_{\rm circ} = 1.6 {\rm Gyr} \frac{Q_P}{10^6}\frac{m_P}{m_J}\frac{M_*}{M_\sun}^{-3/2}\frac{R_P}{R_J}^{-5}\frac{a}{0.05 AU}^{-13/2}$, and assuming Q=10$^6$. The quoted uncertainty only propagates the uncertainty in the stellar mass and the planetary mass, radius, and semi-major axis. However, it will typically be dominated by the orders of magnitude uncertainty in Q. Because the timescale is linearly dependent on Q, the uncertainty should be at least an order of magnitude in either direction.}\\
\hline
~~~~$M_P$\dotfill &Mass (\mj)\dotfill & mp & $\mid M_P \mid \lesssim 100 M_\sun$ \\
\multicolumn{4}{p{17cm}}{The mass of the planet, in Jupiter masses. Note that for transit-only fits, this quantity is derived from the \citet{Chen:2017} exoplanet mass-radius relations. This value may be negative if the {\tt LINEARK} flag is set, which flips the RV curve. That introduces a degeneracy for RV-only fits. In RV-only fits, this is the true mass with large uncertainties to reflect the marginalization over $\cos{i}$, though users may wish to quote $M_P\sin{i}$, which is more standard and more precisely known. The upper bound is derived from a conservative limit on $\log{K} < 5$. Also note the boundary $M_P + M_* > 0$.} \\
\hline
~~~~$M_P$\dotfill &Mass (\me)\dotfill & mpearth & $\mid M_P\mid \lesssim 100 M_\sun$ \\
\multicolumn{4}{p{17cm}}{Same as above, but in Earth masses. Only displayed when {\tt EARTH} flag is set.} \\
\hline
~~~~$K$\dotfill &RV semi-amplitude (m/s)\dotfill & k & $\mid K \mid < 10^5$ \\
\multicolumn{4}{p{17cm}}{The semi-amplitude of the RV signal in m/s. Note that for transit-only fits, this quantity is derived from the \citet{Chen:2017} exoplanet mass-radius relations. This value may be negative if the {\tt LINEARK} flag is set and \citet{Chen:2017} not used, which flips the RV curve.}\\
\hline
~~~~$\log{K}$\dotfill &Log of RV semi-amplitude \dotfill & logk & $\mid \log{K} \mid < 5$ \\
\multicolumn{4}{p{17cm}}{The log$_{10}$ of the RV-semi-amplitude in m/s. If {\tt LINEARK} flag is set, it will exclude all negative values of $K$ before deriving this quantity and there is no lower limit.}\\
\hline
~~~~$R_P/R_*$\dotfill &Radius of planet in stellar radii \dotfill & p & None \\
\multicolumn{4}{p{17cm}}{The radius of the planet in stellar radii. Negative values produce a bump in the transit lightcurve instead of a dip. When \citet{Chen:2017} is used, it must be positive.}\\
\hline
~~~~$a/R_*$\dotfill &Semi-major axis in stellar radii \dotfill & ar & $a/R_* > 0$ \\
\multicolumn{4}{p{17cm}}{The semi-major axis of the planetary orbit in stellar radii.}\\
\hline
~~~~$\delta$\dotfill &Transit depth (fraction)\dotfill & delta & None \\
\multicolumn{4}{p{17cm}}{$ \left(  R_P / R_* \right) ^{2}$, which is the transit depth for non-grazing transits in the absence of limb darkening.}\\
\hline
~~~~$Depth$\dotfill &Flux decrement at mid transit \dotfill & depth & None \\
\multicolumn{4}{p{17cm}}{The depth of the primary transit at $T_T$, including grazing geometries, band-specific limb darkening, and any modeled $T \delta V$s. This value is reported in a separate table, never in the primary output table, but is described here for reference only. Prior to 2019-07-22, this was reported in the primary output table as the depth of the primary transit at $T_C$, not including limb darkening.}\\
\hline
~~~~$\tau$\dotfill &Ingress/egress transit duration (days)\dotfill & tau & $\tau \geq 0$ \\
\multicolumn{4}{p{17cm}}{The ingress/egress primary transit duration (first to second or third to fourth contact), in days, approximated with \citet{Winn:2010}, Eqs 14-16.} \\
\hline
~~~~$T_{14}$\dotfill &Total transit duration (days)\dotfill & t14 & $T_{14} \geq 0$ \\
\multicolumn{4}{p{17cm}}{The total primary transit duration (first to fourth contact), in days, approximated with \citet{Winn:2010}, Eqs 14 \& 16.} \\
\hline
~~~~$T_{FWHM}$\dotfill &FWHM transit duration (days)\dotfill & tfwhm & $T_{FWHM} \geq 0$ \\
\multicolumn{4}{p{17cm}}{The full width at half maximum primary transit duration (1.5 to 3.5 contact), in days, approximated with \citet{Winn:2010}, Eqs 14-16.} \\
\hline
~~~~$b$\dotfill &Transit Impact parameter \dotfill & b & None \\
\multicolumn{4}{p{17cm}}{The approximate minimum projected separation at the time of transit, in stellar radii, $\frac{a\cos{i}}{R_*}\frac{1-e^2}{1+e\sin{\omega_*}}$, from Eq 7 of \citet{Winn:2010}. See \S \ref{sec:ttvs}. For RV only fits, $b \geq 0$. When astrometry is fit, $b$ may be negative. When a transit model is fit, a bound $\mid b \mid < 1+R_P/R_*$ is imposed to prevent the fit from exploring the infinite volume of non-transiting transit models (flat lines). This bound may artificially increase the significance of marginal transit detections.} \\
\hline
~~~~$b_S$\dotfill &Eclipse impact parameter \dotfill & bs & None \\
\multicolumn{4}{p{17cm}}{The approximate minimum projected separation at the time of secondary occultation, in stellar radii, $\frac{a\cos{i}}{R_*}\frac{1-e^2}{1-e\sin{\omega_*}}$, from Eq 8 of \citet{Winn:2010}. See \S \ref{sec:ttvs}. Note that while values $b_S > 1+R_P/R_*$ show no secondary occultation, we only a priori exclude values based on the primary impact parameter when a transit model is computed. This will be an obvious problem if fitting eccentric systems where there is a secondary occultation but not a primary transit and will be fixed if we ever encounter it.}\\
\hline
~~~~$\tau_S$\dotfill &Ingress/egress eclipse duration (days)\dotfill & taus & $\tau_S \geq 0$ \\
\multicolumn{4}{p{17cm}}{The ingress/egress secondary eclipse duration (first to second or third to fourth contact), in days, approximated with \citet{Winn:2010}, Eqs 14-16.} \\
\hline
~~~~$T_{S,14}$\dotfill &Total eclipse duration (days)\dotfill & t14s & $T_{S,14} \geq 0$ \\
\multicolumn{4}{p{17cm}}{The total secondary eclipse duration (first to fourth contact), in days, approximated with \citet{Winn:2010}, Eqs 14 \& 16.}\\
\hline
~~~~$T_{S,FWHM}$\dotfill &FWHM eclipse duration (days)\dotfill & tfwhms & $T_{S,FWHM} \geq 0$ \\
\multicolumn{4}{p{17cm}}{The full width at half maximum primary transit duration (1.5 to 3.5 contact), in days, approximated with \citet{Winn:2010}, Eqs 14-16.}\\
\hline
~~~~$\delta_{S,3.6\mu m}$\dotfill &BB eclipse depth at 3.6$\mu$m (ppm)\dotfill & eclipsedepth36\dotfill & None \\
\multicolumn{4}{p{17cm}}{The predicted secondary occultation depth at 3.6$\mu$m using a black-body approximation of the stellar flux, $F_{*}$, at \teff, and of the planetary flux, $F_{P}$, at T$_{\rm eq}$. Equal to $\frac{(R_P/R_*)^2}{(R_P/R_*)^2 + F_*/F_P}$.}\\
\hline
~~~~$\delta_{S,4.5\mu m}$\dotfill &BB eclipse depth at 4.5$\mu$m (ppm)\dotfill & eclipsedepth45 & None \\
\multicolumn{4}{p{17cm}}{Same as above for 4.5$\mu$m.} \\
\hline
~~~~$\rho_P$\dotfill &Density (cgs)\dotfill & rhop & None \\
\multicolumn{4}{p{17cm}}{The density of the planet, $\frac{3M_P}{4\pi R_P^3}$, in g~cm$^{-3}$.}\\
\hline
~~~~$logg_P$\dotfill &Surface gravity (cgs) \dotfill & loggp & None \\
\multicolumn{4}{p{17cm}}{The log$_{10}$ of the planetary surface gravity, $\log_{10}\left({\frac{GM_*}{R_*^2}}\right)$, in cgs units.}\\
\hline
~~~~$\lambda$\dotfill &Projected Spin-orbit alignment (deg)  \dotfill & lambdadeg & $-180 < \lambda < 180$ \\
\multicolumn{4}{p{17cm}}{The projected alignment between the spin axis of the star and the orbital axis of the planet, in degrees. Only displayed for DT or RM fits. Its reported confidence interval may cross the stated bounds since the distribution is re-centered around the mode.}\\
\hline
~~~~$\lambda$\dotfill &Projected Spin-orbit alignment (Rad)  \dotfill & lambda & $-\pi < \lambda < \pi$ \\
\multicolumn{4}{p{17cm}}{Same as above, but in radians. This fitted parameter is never displayed in the output table.} \\
\hline
~~~~$\Theta$\dotfill &Safronov Number \dotfill & safronov & None \\
\multicolumn{4}{p{17cm}}{The Safronov Number is calculated using Eq 2 from \citet{Hansen:2007}. $\Theta = \frac{1}{2}\left(\frac{V_{esc}}{V_{orb}}\right)^2 = \frac{a}{R_P}\frac{M_P}{M_*}$.}\\
\hline
~~~~$\fave$\dotfill &Incident Flux (\fluxcgs)\dotfill & fave & None \\
\multicolumn{4}{p{17cm}}{The orbit-averaged flux incident on the planet, in 10$^9$ erg~s~cm$^{-2}$. $\fave = \sigma_B\teff^4\left(\frac{R_*}{a}(1-e^2/2)\right)^2.$ While there are no explicit bounds on this derived quantity, it is limited by the bounds on \teff, $R_*$, $a$, and $e$.}\\
\hline
~~~~$T_P$\dotfill &Time of Periastron (\bjdtdb)\dotfill & tp & None \\
\multicolumn{4}{p{17cm}}{The time of periastron of the orbit, in \bjdtdb. While there are no explicit bounds on this derived quantity, it should be within one period of $T_C$.} \\
\hline
~~~~$T_S$\dotfill &Time of superior conjunction (\bjdtdb)\dotfill & ts & None \\
\multicolumn{4}{p{17cm}}{The time of superior conjunction, in \bjdtdb. $T_S$ is analagous to $T_C$ for the secondary occultation. The caveats about the difference between $T_T$ and $T_C$ detailed in \S \ref{sec:tc} apply to the difference between $T_E$ and $T_S$. While there are no explicit bounds on this derived quantity, it should be within one period of $T_C$.} \\
\hline
~~~~$T_E$\dotfill &Time of eclipse (\bjdtdb)\dotfill & ts & None \\
\multicolumn{4}{p{17cm}}{The time of the minimum projected separation during the secondary occultation. $T_E$ is analagous to $T_T$ for the secondary occultation. The caveats about the difference between $T_T$ and $T_C$ detailed in \S \ref{sec:tc} apply to the difference between $T_E$ and $T_S$. While there are no explicit bounds on this derived quantity, it should be within one period of $T_C$.} \\
\hline
~~~~$T_A$\dotfill &Time of Ascending Node (\bjdtdb)\dotfill & ta & None \\
\multicolumn{4}{p{17cm}}{The time of the Ascending node (RV$_*$ minimum) in \bjdtdb. While there are no explicit bounds on this derived quantity, it should be within one period of $T_C$.}\\
\hline
~~~~$T_D$\dotfill &Time of Descending Node (\bjdtdb)\dotfill & td & None \\
\multicolumn{4}{p{17cm}}{The time of the Descending node (RV$_*$ maximum) in \bjdtdb. While there are no explicit bounds on this derived quantity, it should be within one period of $T_C$.}\\
\hline
~~~~$ecos{\omega_*}$\dotfill & \dotfill & ecosw & $-1 \lesssim e\cos{\omega_*} \lesssim 1$ \\
\multicolumn{4}{p{17cm}}{The eccentricity times the cosine of the argument of periastron of the stellar orbit due to the planet. The true bounds are somewhat stricter based on the physical constraints on $e$.}\\
\hline
~~~~$esin{\omega_*}$\dotfill & \dotfill & esinw & $-1 \lesssim e\sin{\omega_*} \lesssim 1$ \\
\multicolumn{4}{p{17cm}}{The eccentricity times the sine of the argument of periastron of the stellar orbit due to the planet. The true bounds are somewhat stricter based on the physical constraints on $e$.}\\
\hline
~~~~$M_P\sin i$\dotfill &Minimum mass (\mj)\dotfill & msini & None \\
\multicolumn{4}{p{17cm}}{The minimum mass of the planet, typically quoted for RV-only fits, in Jupiter masses.  Note that for transit-only fits, this quantity is derived from the \citet{Chen:2017} exoplanet mass-radius relations. This value may be negative if the {\tt LINEARK} flag is set.}\\ 
\hline
~~~~$M_P\sin i$\dotfill &Minimum mass (\me)\dotfill & msiniearth & None \\
\multicolumn{4}{p{17cm}}{Same as above, but in Earth masses. Only displayed when {\tt EARTH} flag is set.}\\
\hline
~~~~$M_P/M_*$\dotfill &Mass ratio \dotfill & q & None \\
\multicolumn{4}{p{17cm}}{The mass of the planet divided by the mass of the star. Note that for transit-only fits, this quantity is derived from the \citet{Chen:2017} exoplanet mass-radius relations. This value may be negative if the {\tt LINEARK} flag is set.}\\
\hline
~~~~$d/R_*$\dotfill &Separation at mid transit \dotfill & dr & $d/R_* > 0$ \\
\multicolumn{4}{p{17cm}}{The separation between the star and planet at the time of inferior conjunction.}\\
\hline
~~~~$P_T$\dotfill &A priori non-grazing transit prob \dotfill & pt & $P_T > 0$ \\
\multicolumn{4}{p{17cm}}{The a priori probability that the transit would be seen as non-grazing ($b \leq 1-R_P/R_*$). This is useful in searching for transits of RV planets or for correcting for the observational bias of transiting planets. The reciprocal of this number is the number of similar planets that would go undetected in a transit survey for each planet like this detected. Note: to estimate the a posteriori transit probability, see if $b \leq 1-R_P/R_*$.} \\
\hline
~~~~$P_{T,G}$\dotfill &A priori transit prob \dotfill & ptg & $P_{T,G} > 0$ \\
\multicolumn{4}{p{17cm}}{Same as above, but allowing for grazing transits ($b \leq 1+R_P/R_*$). Note: for the a posteriori transit probability, see if $b \leq 1+R_P/R_*$.} \\
\hline
~~~~$P_S$\dotfill &A priori non-grazing eclipse prob \dotfill & ps & $P_S > 0$ \\
\multicolumn{4}{p{17cm}}{The a priori probability that the secondary eclispe would be seen as non-grazing ($b \leq 1-R_P/R_*$). This is useful in searching for eclipses of RV-only planets or for correcting for the observational bias of transiting planets. The reciprocal of this number is the number of similar planets that would go undetected in an eclipse survey for each planet like this detected. Note: to estimate the a posteriori eclipse probability, see if $b_s \leq 1-R_P/R_*$.} \\
\hline
~~~~$P_{S,G}$\dotfill &A priori eclipse prob \dotfill & psg & $P_{S,G} > 0$ \\
\multicolumn{4}{p{17cm}}{Same as above, but allowing for grazing eclipses. Note: for the a posteriori eclipse probability, see if $b_s \leq 1+R_P/R_*$.}\\
\smallskip\\\multicolumn{2}{l}{Wavelength Parameters:}\smallskip\\
\hline
~~~~$u_{1}$\dotfill &linear limb-darkening coeff \dotfill & u1 & $0 \lesssim u_1 \lesssim 2$ \\
\multicolumn{4}{p{17cm}}{The linear limb darkening coefficient for the quandratic limb darkening law. The bounds are actually on combinations of $u_1$ and $u_2$ from \citet{Kipping:2013}. $u_1 + u_2 < 1$, $u_1 > 0$, and $u_1 + 2u_2 > 0$.}\\
\hline
~~~~$u_{2}$\dotfill &quadratic limb-darkening coeff \dotfill & u2 & $-1 \lesssim u_2 \lesssim 1$ \\
\multicolumn{4}{p{17cm}}{The quadratic limb darkening coefficient for the quadratic limb darkening law. The bounds are actually on combinations of $u_1$ and $u_2$ from \citet{Kipping:2013}. $u_1 + u_2 < 1$, $u_1 > 0$, and $u_1 + 2u_2 > 0$.}\\
\hline
~~~~$A_T$\dotfill & Planetary Thermal emission (ppm) \dotfill & thermal & None \\
\multicolumn{4}{p{17cm}}{The amount of thermal emission from the planet, in ppm, modeled as an offset that disappears during secondary eclipse. Only displayed when {\tt FITTHERMAL} includes the corresponding band.}\\
\hline
~~~~$A_D$\dotfill &Dilution from neighboring stars \dotfill & dilute & $-1 < A_D < 1$ \\
\multicolumn{4}{p{17cm}}{The fractional dilution, $F_2/(F_{1}+F_{2})$, where $F_1$ is the flux of the host star and $F_2$ is the combined flux of all blended stars. Only displayed when {\tt FITDILUTE} includes the corresponding band. Allowed to be negative to account for over-corrected dilution.}\\
\hline
~~~~$A_R$\dotfill & Reflection from the planet (ppm)\dotfill & reflect & None \\
\multicolumn{4}{p{17cm}}{The amount of reflected light from the planet, in ppm, modeled as a phase curve added to the baseline transit model that disappears during secondary eclipse. Only displayed when {\tt FITREFLECT} includes the corresponding band.}\\
\smallskip\\\multicolumn{2}{l}{Telescope Parameters:}\smallskip\\
\hline
~~~~$\gamma_{\rm rel}$\dotfill &Relative RV Offset (m/s)\dotfill & gamma & $-c < \gamma_{\rm rel} < c$ \\
\multicolumn{4}{p{17cm}}{The arbitrary instrumental zero point offset in the radial velocities, in m/s. It is bounded by the speed of light, though no relativistic effects are included.}\\
\hline
~~~~$\sigma_J$\dotfill &RV Jitter (m/s)\dotfill & jitter & $0 \leq \sigma_J < c$ \\
\multicolumn{4}{p{17cm}}{The RV jitter, in m/s. If jitter variance is negative, the values are set to zero here. Thus, the reported value for this parameter may be biased if jittervar is negative.}\\
\hline
~~~~$\sigma_J^2$\dotfill &RV Jitter Variance \dotfill & jittervar & $-c^2 < \sigma_J^2 < c^2$ \\
\multicolumn{4}{p{17cm}}{The RV jitter variance, in (m/s)$^2$. This fitted quantity can be negative to correct for over-estimated RV errors, but can never be more negative than the smallest user-supplied error squared. That is, models where min$(\sigma_{\rm RV})^2 + \sigma_J^2 \leq 0$ are rejected.} \\
\smallskip\\\multicolumn{2}{l}{Transit Parameters:}\smallskip\\
\hline
~~~~$\sigma^{2}$\dotfill &Added Variance \dotfill & variance & None \\
\multicolumn{4}{p{17cm}}{The added variance for the lightcurve. This quantity can be negative to correct for over-estimated photometric errors, but can never be more negative than the smallest user-supplied error squared. That is, models where min$(\sigma_{\rm Tran})^2 + \sigma^2 \leq 0$ are rejected}\\
\hline
~~~~$TTV$\dotfill & Transit Timing Variation (days)\dotfill & ttv & $-P/2 < TTV < P/2$ \\
\multicolumn{4}{p{17cm}}{The difference between the modeled time of conjunction and the best-fit time of conjunction for this transit, in days. For a TTV analysis, the user is likely to prefer the ancillary TTV table.} \\
\hline
~~~~$TiV$\dotfill & Transit Inclination Variation (Radians)\dotfill & tiv & None \\
\multicolumn{4}{p{17cm}}{The difference between the modeled inclination and the best-fit inclination for this transit, in radians. This cannot make the total inclination violate the bounds stated above.}\\
\hline
~~~~$T\delta V$\dotfill & Transit Depth Variation \dotfill & tdeltav & None \\
\multicolumn{4}{p{17cm}}{The difference between the modeled $\left(R_P/R_*\right)^2$ and the best-fit $\left(R_P/R_*\right)^2$ for this transit.}\\
\hline
~~~~$F_0$\dotfill & Baseline flux \dotfill & f0 & None \\
\multicolumn{4}{p{17cm}}{The baseline flux of the transit. When secondary eclipses are fit, this is the baseline contribution from just the star and should be normalized to 1.}\\
\smallskip\\\multicolumn{2}{l}{Doppler Tomography Parameters:}\smallskip\\
\hline
~~~~$\sigma_{DT}$\dotfill & Doppler Tomography Error scaling  \dotfill & dtscale & $\sigma_{DT} > 0$ \\
\multicolumn{4}{p{17cm}}{The multiplicative factor to scale the supplied DT errors to ensure they are consistent with the model. Only displayed for DT fits.}\\
\smallskip\\\multicolumn{2}{l}{Astrometry Parameters:}\smallskip\\
\hline
~~~~$\sigma_{Astrom}$\dotfill & Astrometric Error Scaling \dotfill & astromscale & $\sigma_{Astrom} > 0$ \\
\multicolumn{4}{p{17cm}}{The multiplicative factor to scale the supplied astrometric errors to ensure they are consistent with the model. Only displayed for astrometric fits.}\\
\hline
~~~~$\alpha_{\rm ICRS}$\dotfill & Right Ascension (deg)  \dotfill & ra & $0 \leq \alpha_{\rm ICRS} \leq 360$ \\
\multicolumn{4}{p{17cm}}{The ICRS right ascension, in degrees. Only displayed for astrometry fits.}\\
\hline
~~~~$\delta_{\rm ICRS}$\dotfill & Declination (deg) \dotfill & dec & $-90 \leq \delta_{\rm ICRS} \leq 90$ \\
\multicolumn{4}{p{17cm}}{The ICRS right ascension, in degrees. Only displayed for astrometry fits.}\\
\enddata
\end{deluxetable*}